\documentclass[a4paper,twocolumn, 10pt]{article}

\usepackage{times} 

\usepackage[margin=1in]{geometry} 

\usepackage{amsmath}
\usepackage{amsthm}
\usepackage{amssymb}

\usepackage[utf8]{inputenc}
\usepackage{hyperref}

\usepackage[sort&compress,numbers,square]{natbib}

\usepackage{graphicx, color}
\graphicspath{{fig/}}
\usepackage{subcaption}
\usepackage{tikz}
\usepackage{gensymb}

\usepackage{algorithm, algpseudocode} 
\usepackage{mathrsfs} 

\usepackage{lipsum}  

\usepackage{wrapfig}

\usepackage{caption}

\usepackage{dblfloatfix}

\title{Integrated Circuit Architecture for Real-Time Sensing with Embedded Microbial Whole-Cell Sensors}
\author{Amritha Janardanan$^{1,2}$ \and Soner Sonmezoglu$^{3,4}$ \and Stefano Sonedda$^{3,5,6}$ \and Tom J. Zajdel $^{3,7,8}$ \and James L. Flewellen$^{1,2}$ \and Meera Lester$^3$ \and Behzad Rad$^{7,10}$ \and Michel M. Maharbiz $^{3,9,11}$ \and Teuta Pilizota$^{1, 2, 12}$}

\date{
	$^1$School of Biological Sciences, University of Edinburgh,        Edinburgh, UK.
	$^2$K\=ahu SiliconBio, Edinburgh, UK.
    $^3$Department of Electrical Engineering and Computer Sciences, University of California Berkeley, Berkeley,          California, USA.
    $^4$ Department of Electrical and Computer Engineering, Northeastern University, Boston, Massachusetts, USA.
    $^5$Department of Electrical and Electronic Engineering, University of Cagliari, Italy
    $^6$Corticale Srl, Genoa, Italy
    $^7$Molecular Foundry, Lawrence Berkeley National Laboratory, Berkeley, California, USA.
    $^8$ Department of Electrical and Computer Engineering, Carnegie Mellon University, Pittsburgh, Pennsylvania, USA.
    $^9$Department of Bioengineering, University of California, Berkeley, California, USA.
    $^{10}$Molecular Biophysics and Integrated Bioimaging Division, Lawrence Berkeley National Laboratory, Berkeley, California, USA.
    $^{11}$Chan Zuckerberg Biohub, San Francisco, California, USA. 
    $^{12}$Department of Physics, University of Cambridge, Cambridge, UK.
}

\begin{document}

    \maketitle
    
    \begin{abstract}

    Bacteria sense a diverse range of environmental analytes with high sensitivity and temporal resolution. Engineering and synthetic biology approaches enabled harnessing this capability through development of whole-cell biosensors that respond to specific molecules of interest.  However, converting these responses into electrical signals in real time, across different environmental conditions, in miniaturized, field-deployable microelectronic devices, remains challenging. Here we present a bioelectronic platform that directly couples engineered bacteria to an integrated circuit (IC) chip through custom on-chip microelectrodes, enabling real-time, electronic readout of analyte sensing through bacterial flagellar motor dynamics. Using non-Faradaic electrochemical impedance measurements the device resolves both the direction and speed of motor rotation with a signal-to-noise ratio (SNR) of 15 dB. The IC is further integrated with a microfluidic system that enables controlled delivery and removal of analytes, nutrients and bacteria. When combined with whole-cell biosensors engineered to detect specific analytes, this work provides a miniature, portable platform for continuos monitoring in a range of liquid environments.
    
    \end{abstract}

    \vspace{0.5em}
    \noindent\textbf{Keywords:} bioelectrical interfaces, electrochemical impedance sensing, micron sized electrodes, custom designed ICs,  bacterial flagellar motor, chemotaxis, whole-cell biosensors.
        
    \section*{Introduction}
    \label{sec:intro}

    Electronic signal transduction has fundamentally transformed the way humans access and communicate information encoded in the physical world \cite{ZookSchroeder2008}. Modern electronic systems harness electrons to efficiently process, store, and transmit massive quantities of digital data across the globe. Coupling electronics with other information processing  modalities, such as converting it into electromagnetic and photonic signals, continues to drive transformative global advances \cite{Li2025Photodetectors,Zhou2025WearableUltrasound,Luo2024MagnetoelectricIoT, Mehrotra2019EMWaveBiosensors}. 
    Living cells are an intriguing source of novel and complex information processing capabilities. Bacteria, in particular, evolved over billions of years for life in every corner of our planet, developing an extraordinary array of sensing capabilities. They sense, respond to, and transform a range of environmental cues with remarkable efficiency and speed \cite{Bi2018Bacterialchemotaxis,Dufrene2020Mechanomicrobiology,VanderHorst2007Photosensing}. Capitalising on our increasing ability to synthesize and insert desired genetic information into living cells, the fields of synthetic and engineering biology aim to exploit this capability to provide a new range of sensing modalities. Whole-cell based biosensors, which are engineered living cells designed to report on specific analytes, offer a powerful platform for detecting a wide range of different targets: from toxins \cite{Dou2025envpollutants,Roy2021}, pathogens \cite{Jeon2022pathogens} and explosives \cite{Shemer2021explosives} to human-health related markers \cite{Barger2021}. Although various bacterial whole-cell biosensors have been reported, their integration with electronic systems, effectively bridging the signal gap between biology and technology, is needed to gain unprecedented access to, and control over, their rich information processing capabilities \cite{Rivnay2025}.  
 
Most reported whole-cell biosensors rely on expression of fluorescent or bioluminescent reporters, where signal intensity reflects the analyte concentration \cite{Meer2010, Moraskie2021}. Because transcription and translation of reporter proteins takes $\sim$30~min \cite{Meer2010}, this approach is inherently not real-time signalling. Fluorescence-based whole cell biosensors with faster response times are used for detection of intracellular molecules, rather than environmental sensing, and rely on, e.g. Förster Resonance Energy Transfer and, thus, complex optics for detection \cite{Shi2018}. Similar is also true for most whole-cell biosensors that rely on protein expression, where interfacing with an optoelectronic transducer required to bridge the biology to semiconductor-based technology gap, often remains bulky and laboratory based \cite{Hicks2020}. Notable recent efforts to change this include transduction of optical signal into an electronic output via a low-power, miniature photodetector circuits \cite{Mimee2018Ingestiblebiooptoelectronicsensor,Tsai2015waterpollutionCCD,IndiaWebb2023}. Miniaturisation of the optoelectronic transducer is limited by the photon budget, thus placing strong requirements on optical shielding, and the need for long integration time. Optoelectronic readout is therefore subject to a trade-off between power consumption, bandwidth and SNR. Therefore, the relevant comparison with the electrical readout is SNR at a defined bandwidth, under realistic power and area constraints.  In this regime, electrical readout sustains higher bandwidth at similar device scale without prolonged optical integration, making it better suited for monitoring of bacterial responses in real time.

Another approach is inspired by bacteria that output electrons naturally, e.g. \textit{Geobacter} and \textit{Shewanella} \cite{Gralnick2023}. The intracellular metabolic reactions that produces the electrons can be engineered to sense specific molecules, enabling transfer of electrons to the electrodes \cite{Zhu2023,Greenman2021}. However, genetically engineering these bacteria remains difficult, hampering progress \cite{Li2024}. A recent solution synthetically introduced the extracellular electron-transfer pathway from \textit{Shewanella oneidensis} into \textit{Escherichia coli} \cite{Jensen2010, Atkinson2022}, where genetic manipulation is straightforward and well-established. Albeit measured using bulky electrochemical reactors \cite{Atkinson2022,Zhang2025}, an electron current was, thus, obtained as a quantitative readout of down to 12.5~$\mu$M analyte concentration, in $\approx$ 2-10 min, with speed limited by mass transport \cite{Atkinson2022}.

Collectively, current approaches highlight the need for a bioelectronic transduction technology that leverages the miniaturisation and low power of semiconductor microelectronics, without compromising the speed or sensitivity of the bacterial analyte detection circuits. To achieve this, we looked to the canonical bacterial signalling network -- chemotaxis \cite{Sourjik2012, Wadhams2004}. With the help of protein `antennas', called chemoreceptors, the chemotactic network senses favourable and unfavourable inputs in the environment across a wide dynamic range, down to nanomolar concentrations \cite{Jasuja1999, Kim2002, Hansen2008}. The signal is processed by a series of protein feedback loops and sent to a mechanical output, in the form of a rotary rotary motor protein complex called the bacterial flagellar motor \cite{Sowa2008}. The motor is roughly 50 nm in diameter, spans the entire cell envelope and connects to a thin, up to $\sim$10 $\mu$m long filament via a flexible hook joint \cite{Sowa2008, Tan2024, Santiveri2020}. The motor also rapidly changes its rotational direction in response to the network inputs \cite{Bai2010}. These so-called switching events last hundreds of milliseconds and reset quickly, preserving both high sensitivity and temporal resolution of the network \cite{Sowa2008, Bai2010}. Relaying the information processed by the chemotactic network while keeping its sensitivity, speed, and specificity requires detecting the rotational information of the flagellar motor in real-time.

To detect the motor rotation electrically, we developed a biohybrid IC architecture that integrates micron-sized electrode sensors in a custom-designed IC. \textit{E. coli} cells were positioned close to the electrode sensors of carefully selected sizes. The cells have genetically engineered filaments, so that they readily attach to micron sized polystyrene spheres (beads) \cite{Ryu2000}. The chip performs electrochemical impedance measurements of motor-driven beads positioned over the electrode sensor at MHz frequencies, thus allowing us to detect the motor rotation with signal-to-noise ratio (SNR) of up to $\sim$15~dB. Our IC-based sensor can not only detect rotation speed but changes in direction of rotation, thereby accessing the output of the chemosensory network in real-time. To our knowledge, this is the first realisation of impedance monitoring that tracks rotational motion at 0.5-10$\mu$m spatial resolution. Apart from increasing the frequencies across which we can perform spectroscopy into the MHz range, the IC design allows future fabrication of hundreds of thousands of sensing arrays onto one chip, giving more robust sensing and the ability to collect large data sets. Furthermore, the chemotactic network is one of the best-studied regulatory networks in living cells \cite{Sourjik2012,Wadhams2004,Huynh2025}, facilitating the design of various different biosensors whose output is the motor rotation, e.g., by engineering the chemoreceptors \cite{Bi2016, Bergman2025}. The speed with which sensors for novel analytes can be designed, when only techniques required to genetically alter \textit{E. coli} are needed, combined with the small size, and ease of integration into existing electronics, can prove essential to meet the global challenges of monitoring various liquid environments.

    \section*{Results}
\begin{figure*}[htb!] 
    \centering    \includegraphics[width=0.95\textwidth]{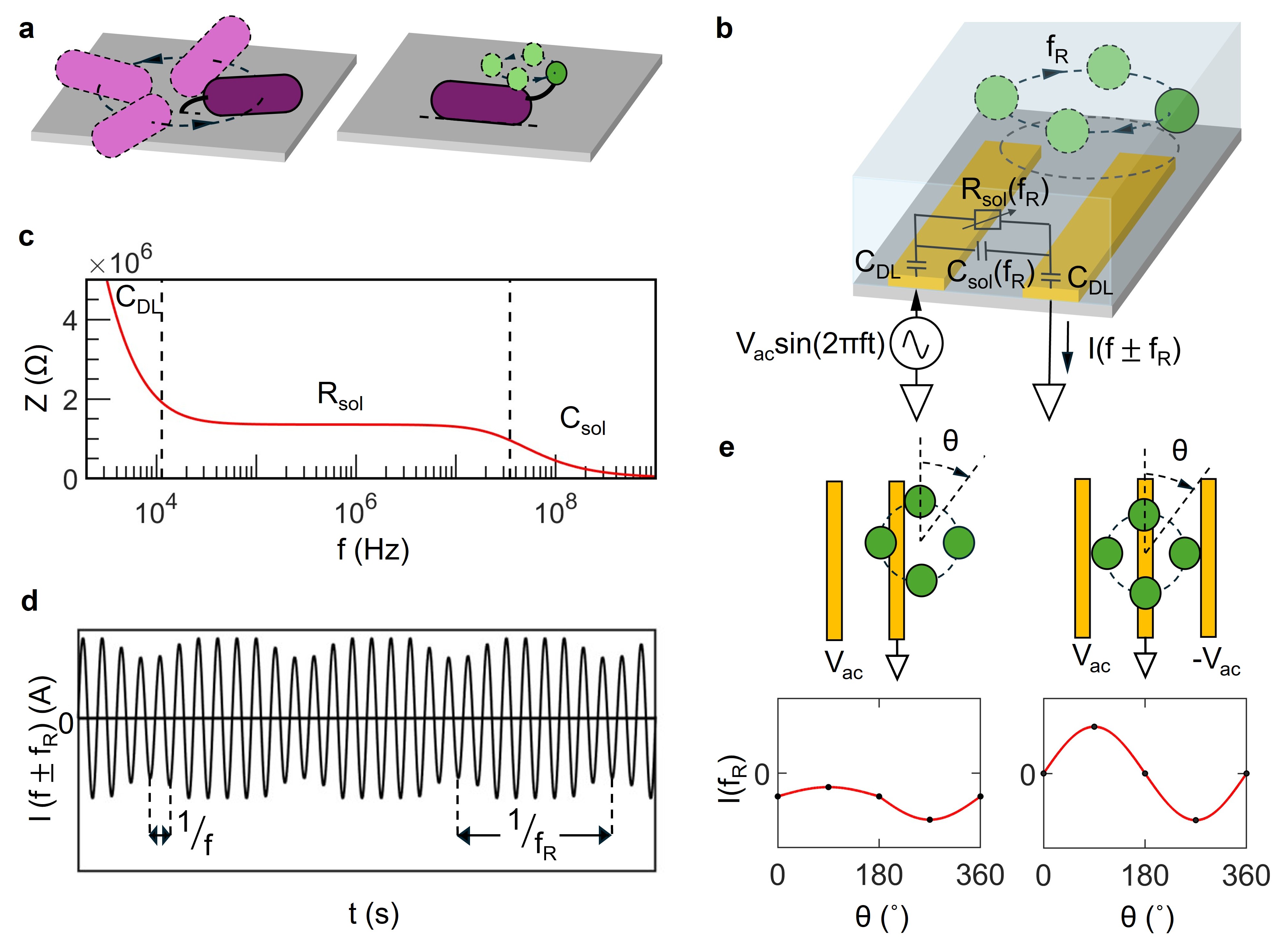} 
    \caption{\textbf{Illustration of impedance-based detection of flagellar motor rotation.} \textbf{a.} Schematic of (left) tethered cell and (right) bead assay. On the left, the sheared flagellum (black) is attached to a surface, causing the cell body (purple, $\sim$3~$\mu$m in length, and $\sim$1~$\mu$m in width, \cite{Taheri-araghi2015} and \textit{Methods}), to rotate about the attachment point. On the right, the cell body is attached to the surface and a polystyrene bead (green, $\sim$0.2-10~$\mu$m in diameter) is attached to the sheared flagellum, causing this bead to rotate. 
    \textbf{b.} A microscale sphere rotating at frequency $f_\mathrm{R}$ above a pair of microelectrodes (yellow rectangles) in solution, with an AC sinusoidal voltage of frequency $f$ applied across the electrodes. The equivalent circuit model of the impedance between the electrodes is shown, comprising $C_\mathrm{DL}$, $R_\mathrm{sol}$ and $C_\mathrm{sol}$. Rotation of the sphere modulates $R_\mathrm{sol}$ and $C_\mathrm{sol}$ at $f_\mathrm{R}$, resulting in current components $I$ at $f\pm f_\mathrm{R}$.
    \textbf{c.} Representative impedance spectrum calculated for an electrode sensor pair with length 10~$\mu$m, width 0.75~$\mu$m, and interelectrode distance 4~$\mu$m (see \textit{Supplementary text}). With dashed vertical lines we indicate frequency windows where either $C_\mathrm{DL}$, $R_\mathrm{sol}$ or $C_\mathrm{sol}$ is highest.
    \textbf{d.} Schematic representation of the current $I$ flowing through the electrodes, when $Z \sim R_\mathrm{sol}$. The input frequency $f$ and marker rotation frequency $f_\mathrm{R}$ amplitude modulating this current are shown. 
    \textbf{e.} Schematic representation of $I(f_\mathrm{R})$ for the two-electrode (left) and three-electrode (right) sensor configurations, with the marker positioned to maximise the signal in each case. In the two-electrode configuration, the current exhibits a peak negative amplitude at $\theta = 270^\circ$, when the marker is in between the electrodes. The current remains negative at other phases with smaller amplitudes. In the three-electrode configuration, with the marker rotating about the central electrode, the current varies near-sinusoidally with rotation phase, showing four distinct amplitudes: zero, positive, zero and negative, corresponding to $90^\circ$ phase increments.}    
    \label{fig:1} 
    \end{figure*}

    \textbf{Bacterial analyte sensing can be transduced into electrical signals via impedance detection of flagellar motor rotation.} To detect the motor rotation electrically we wish to position a marker of its rotation between microscale electrodes and measure impedance changes between them. To do so, we rely on the previously used tethered cell, and bead assays, Fig.~\ref{fig:1}a \cite{Krasnopeeva2021,Sowa2008}. Both are made possible by mutations in the flagellar filament that render it sticky to glass and polystyrene beads (see also \textit{Methods}) \cite{Ryu2000,Krasnopeeva2021}. In the tethered cell assay, the flagellar filament is stuck to a surface, so the motor drives the rotation of the cell body. In the bead assay, the cell body is attached to the surface, and the motor drives the rotation of the bead, stuck to the flagellar filament. Positioning either marker so that a portion of its rotational path passes sufficiently close above and between the microelectrodes, periodically modulates the impedance ($Z$) between them. The latter can be detected as the corresponding change in the current output, Fig.~\ref{fig:1}b.
    
The impedance between electrodes has significant contributions from the electrode-solution interfacial double layer capacitance ($C_\mathrm{DL}$) and the bulk solution impedance components ($R_\mathrm{sol}$ and $C_\mathrm{sol}$), where each component dominates at different applied frequencies, Fig.~\ref{fig:1}b and c. As an example, Fig.~\ref{fig:1}c is the spectrum calculated for one representative electrode sensor configuration (see \textit{Supplementary Text}), with dashed lines indicating where contribution from each of the components becomes highest. $C_\mathrm{DL}$ will not be influenced by a marker rotating several $\mu$m above the microelectrodes because the interfacial layer extends to a maximum of a few nm above it \cite{Schott2024}. We wish to operate in the frequency regime where the bulk solution impedance components are not just bigger but dominate \cite{Hong2005}. At the same time we wish to keep the power requirements to a minimum, which increase with increasing frequencies \cite{Razavi}. In the illustrative example shown in Fig.~\ref{fig:1}c, the sweet spot is the frequency window of 0.1–10~MHz. It is characterised by periodic changes in $R_\mathrm{sol}$ caused by marker rotation at frequency $f_\mathrm{R}$, which amplitude-modulate the output current at $f_\mathrm{R}$, Fig. \ref{fig:1}d. This component at $I(f_\mathrm{R})$ can then be extracted.

In the two-electrode sensor configuration only the rotational speed of the marker is detected through the changes in electric field strength when it is closest and furthest from the electrode sensor, respectively ($\theta = 270^\circ$ and $90^\circ$ in Fig.~\ref{fig:1}e). This is because, to first order, the change in $R_\mathrm{sol}$ follows the spatial variation of the electric field strength $\mathbf{E}$ along the marker trajectory (although the finite size of the marker means the perturbation reflects the local field distribution across its extent). 

 In a symmetric three-electrode sensor configuration, apart from the rotational frequency, the complete phase information can be measured as well, Fig.~\ref{fig:1}e. This is relevant, because the dominant direction of rotation in bacterial whole cell biosensor can be genetically encoded, and thus serve as the signal output \cite{Wadhams2004}. With equal electrode size and spacing, equal and opposite AC voltages can be applied to the two outer electrodes. When the centre of rotation of the marker is positioned over the middle electrode, the sensor can distinguish between four phase angles along the trajectory ($\theta = 0^\circ, 90^\circ, 180^\circ, 270^\circ$) as the local electric field on the central electrode transitions from zero to positive, to zero, and then to negative. The resulting current output can appear approximately sinusoidal due to an inhomogeneous $\mathbf{E}$-field between these microscale electrodes and a finite marker size. This configuration significantly improves SNR, as the differential measurement suppresses baseline impedance contributions, and the current flowing from the centre electrode reflects only asymmetry of the marker relative to the electrode centre.

\begin{figure*}[htb!] 
    \centering
    \includegraphics[width=0.95\textwidth]{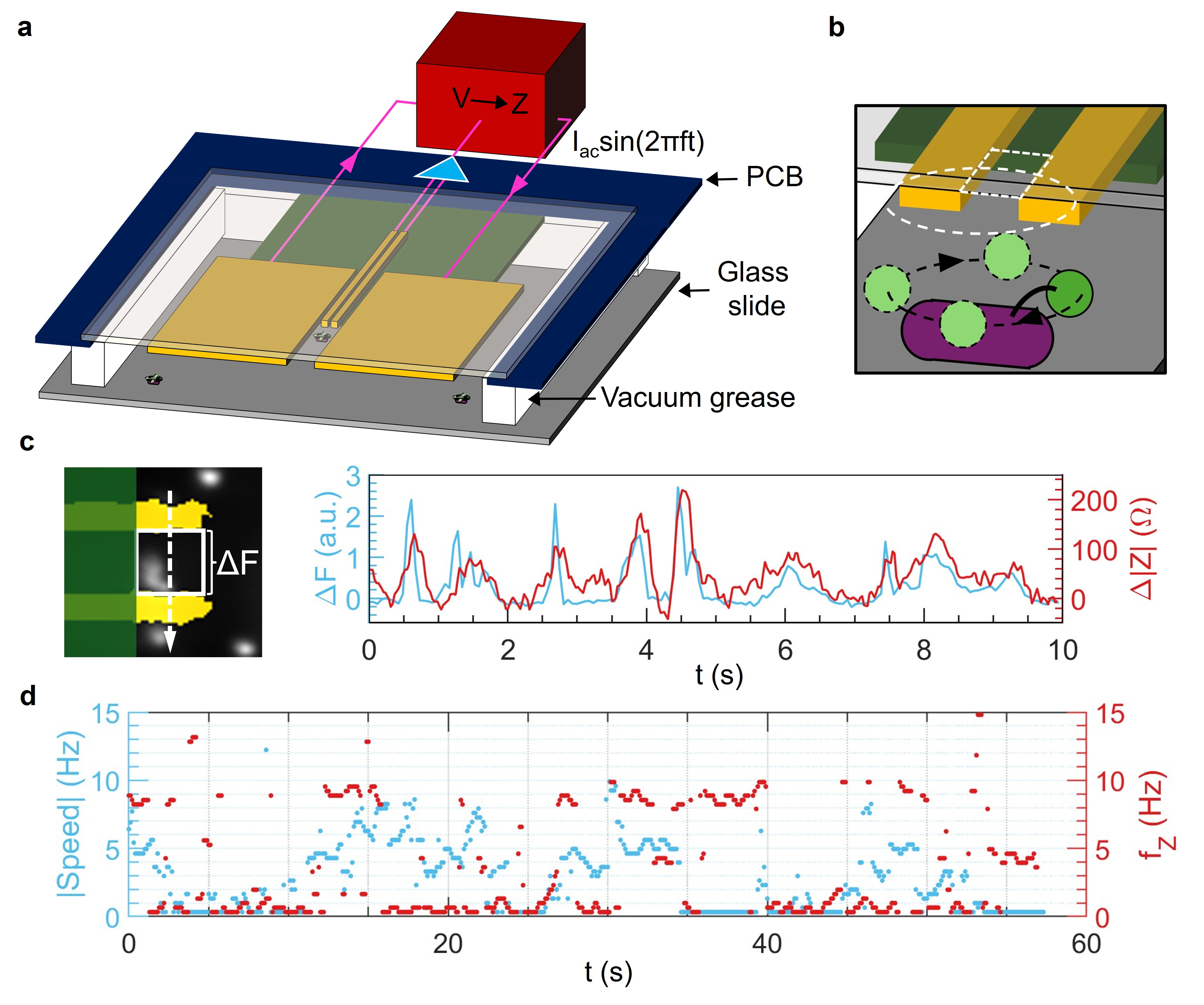} 
    \caption{\textbf{Cleanroom fabricated four-electrode sensors were not able to detect the rotation of the bacterial flagellar motor.}
    \textbf{a.} A four-electrode array (dark yellow) was fabricated on a transparent silica substrate (transparent white) with an insulating layer (dark green) protecting the electrical wiring (pink), and defining the sense electrode dimensions (zoomed in b.). The electrodes on the substrate with the insulating layer were mounted on a PCB (dark blue), which electrically connects them to an impedance analyser (red). A microfluidic channel is formed beneath the electrodes using a glass slide (dark grey) and a vacuum grease spacer. Bacterial cells with markers of the flagellar motor rotation are attached to the surface of such channels (three small dots on the glass slide), and one of the cells positioned near the sense electrodes.
    \textbf{b.} Magnified view of the sensing region showing the bead assay (purple bacterium with attached light green bead) and sense electrodes. The bacterium is immobilized on the glass slide, and the projected rotational trajectory of the bead (dashed white circle) passes through the sensing area (dashed white square) between the sense electrodes.
    \textbf{c.} Left: Microscope image used to measure the change in fluorescence intensity in the sensing area, $\Delta F$. The fabricated sense electrodes are shown in yellow, with insulation (transparent green) defining the exposed electrode area. The white arrow indicates the flow direction of individual fluorescent 2~$\mu$m beads, and the white box marks the region used to extract the total intensity ($F$) and its changes with respect to baseline fluorescence $\Delta F$. Bright spots in the field of view correspond to beads stuck on the surface. Right: $\Delta F$ (cyan) and $\Delta Z$ (red) (\textit{Methods}), recorded simultaneously as the beads transited the sensing region show overlap between the two signals. 
    \textbf{d.} Rotational speed of the flagellar motor (cyan) and impedance frequency $f_\mathrm{Z}$ (red), extracted from the simultaneously recorded images of marker rotation and impedance, are plotted as a function of time. There is no visible overlap between the two signals. 
    } 
    \label{fig:2}  
    \end{figure*}

 \textbf{Cleanroom fabricated microelectrode sensors do not provide sufficient signal-to-noise to detect flagellar motor rotation.} We first wished to detect bacterial flagellar motor rotation electrically with microelectrodes fabricated using standard lithography techniques in a cleanroom \cite{Fruncillo2024}. A four-electrode sensor array consisting of a pair of 4~$\mu$m $\times$ 4~$\mu$m sense electrodes spaced 4~$\mu$m apart, and a pair of 100~$\mu$m $\times$ 100~$\mu$m current-injection electrodes spaced 20~$\mu$m apart, was mounted on a printed circuit board (PCB), Fig.~\ref{fig:2}a and \textit{Methods}.  An impedance analyser, interfaced with the electrode array via the PCB, sends the current via the current-injection electrodes, and measures the voltage between the sense electrodes, Supplementary Fig.~1. PCB-induced parasitic impedances favoured measurements below 80~kHz, where electrode interfacial capacitances become significant. Therefore, a four-point configuration separating current injection from voltage sensing was used, which, together with high-conductivity polymer-coated current-injection electrodes, enabled operation at 10 kHz (see \textit{Supplementary Text} for further details) \cite{Linderholm2006,Smits1958}.
 
To enable placement of the bacterium with the motor driven marker, a $\sim$40~$\mu$m high microfluidic channel was created under the electrode sensor using a glass slide and vacuum grease. This configuration enabled delivery of bacteria while allowing the movement of the electrode sensor with respect to the slide (see \textit{Methods} and Supplementary Fig.~3) \cite{Saxl2007}. The bacteria were immobilised on the glass slide and 2~$\mu$m diameter polystyrene beads introduced to form the bead-assay configuration, Fig.~\ref{fig:1}a, Fig.~\ref{fig:2}a, b and \textit{Methods}. Optical imaging was done from above. A micromanipulator was used to move the sense electrodes and bring them as close as possible to the bacterial cell with the rotating bead, Supplementary Fig.~3b.
 
    We next confirmed that a single bead of the same size (2~$\mu$m in diameter) can be detected when flown across the electrode sensor. For the purpose, and given we here did not require electrodes to be moved with respect to the opposite channel surface, we used PDMS channels whose height ranged between 15-20~$\mu$m, but was well defined and verified by profilometry of the mould used for all channels (\textit{Methods}). The beads were fluorescent, enabling simultaneous optical recording and impedance measurement in the 4~$\mu$m $\times$ 4~$\mu$m sensing area (\textit{Methods}). Fig.~\ref{fig:2}c shows that beads passing over and in between the sense electrodes coincided with $\Delta Z$ peaks that denote the change in impedance of the electrode sensor relative to baseline, confirming detection. Beads passing near, and not over, the sensing area produced lower-amplitude $\Delta Z$ peaks (e.g., at 2.2, 3.25, and 5.75~s). $\Delta Z$ changes for beads passing over the sensor had SNRs of $\sim 8–22$~dB ($\Delta Z \sim 43–219~\Omega, \sigma_{\mathrm{Noise}} 
    \sim 16.5~\Omega$ and see \textit{Methods}), and depended on the degree of bead overlap with the sensing area, and on bead height above the electrodes (estimated to be within $10~\mu$m from the surface as a better estimate could not be obtained with our long-working distance objective, see \textit{Methods}). Beads passing outside the sensor area resulted in SNRs of approximately 11~dB ($\Delta Z \sim 60~\Omega$). In \textit{Supplementary Text} and Supplementary Fig.~4 we analyse these effects in more detail, whereas our previous publication gives more examples of similar data recordings \cite{Zajdel2017}.

  Having confirmed we can detect single 2~$\mu$m beads, we simultaneously recorded the electrochemical impedance and optically imaged the beads rotated by the flagellar motor. Fig. ~\ref{fig:2}d (also see \textit{Supplementary Text}) shows that the frequency peaks obtained from the impedance data spectra did not match those obtained from optical images, and no correlation was observed between impedance variation and bead displacement from its centre of rotation, Supplementary Fig.~6. The vertical distance of the rotating beads from the electrodes was $\sim$30-40~$\mu$m, a combination of channel height and the position of the motor, and hence the marker, along the cell body (there are $\sim$~5 flagellar motors randomly distributed along the cell body \cite{Honda2022}). The radius of rotation of the marker was $\sim$1-2~$\mu$m. While in principle the length of the filament, and thus the radius of rotation, can be varied with the amount of shearing, this is rarely tightly controlled (\textit{Methods}). Given the lack of precision control over the vertical positioning of the marker with respect to the electrodes, we conclude that cleanroom fabricated microelectrodes connected to benchtop instrumentation are not sufficient to detect the rotation of the bacterial flagellar motor.

    \textbf{Microelectrodes were integrated with electronics on an integrated circuit to increase signal-to-noise-ratio.} Having been unsuccessful in detecting flagellar motor rotation with cleanroom-fabricated, relatively large microelectrodes, we hypothesized that scaling down the electrodes and sensing area to the sub-$\mu$m regime, together with tighter confinement of the electric field, would increase the SNR and enable detection of the motor rotation. This miniaturization, however, increases impedance contribution from $C_\mathrm{DL}$, and requires operation at higher frequencies. We therefore turned to a custom application specific integrated circuit (ASIC) that minimises parasitic capacitance at the interface between the electrodes and the electronics, reducing noise and enabling measurements at higher frequencies.
    
\begin{figure*}[htb!] 
    \centering \includegraphics[width=0.95\textwidth]{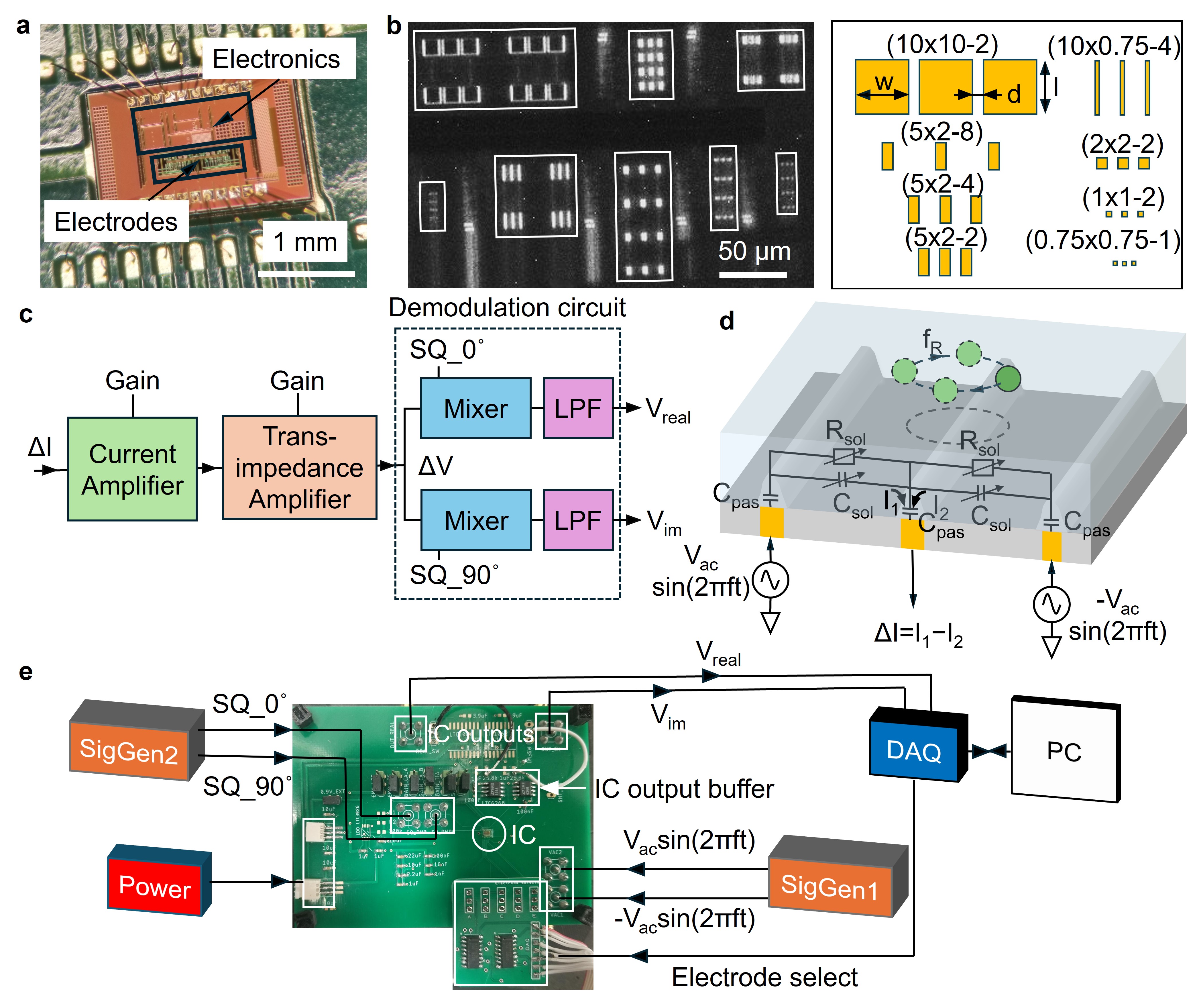} 
    \caption{\textbf{Design of the microelectrode sensors on an Integrated Circuit.} \textbf{a.} Microscope image of the foundry-fabricated IC (orange) connected to the PCB (green) with the visible gold wire bonds. The areas on the IC containing the electrode sensors and the processing electronics are shown.  \textbf{b.} Left: Microscope image showing eight electrode sensor configurations (outlined in white), each comprising four identical three-electrode sensors. Right: Schematic of the eight electrode sensor configurations shown in the microscope image on the left, drawn to scale. Electrode dimensions are indicated above each electrode. 
    \textbf{c.} Signal processing circuits on the IC are given from left to right. First a low-noise current amplifier that amplifies the differential current $\Delta I$, followed by a transimpedance amplifier that converts this current to an amplified differential voltage $\Delta V$.  Finally, $\Delta V$ is demodulated at the input frequency $f$ using two mixers driven by digital signals at $f$, with phase offset $0^\circ$ and $90^\circ$ relative to the input. Each mixer output is low-pass filtered (LPF) to yield voltages representing the in-phase (V$_{real}$) and quadrature (V$_{im}$) components of the modulating impedance, respectively (see Supplementary Fig.~7 for details).
    \textbf{d.} Schematic of the electrode sensor on the IC drawn to scale for the $l \times w - d = (10 \times 0.75 - 4)\,\mu\mathrm{m}$ configuration. A surface passivation layer of SiO$_2$(0.6~$\mu$m)/Si$_3$N$_4$(0.6~$\mu$m) is present over the electrodes as part of the standard foundry process, thus forming a ridged 3D interface between the electrodes and the solution (see Supplementary Fig.~8). Two out-of-phase sinusoidal signals of frequency $f$ are applied to the outer electrodes, and the difference in thus generated currents, $\Delta I = I_1 - I_2$, flowing into the centre electrode is measured. The marker modulates $I_1$ and $I_2$ described in Fig.~\ref{fig:1}b as it rotates over electrodes. \textbf{e.} The PCB integrating the IC with input supply and output processing units is shown. Two signal generators provide sensor excitation and signal demodulation inputs, while power supplies feed the IC and amplifiers. The IC outputs are buffered and sent to a data-acquisition (DAQ) interface, which converts them to digital signals for processing by a PC. Electrode sensor selection is controlled by the PC through the DAQ and the addendum PCB (small green board) connected to the main PCB. See Supplementary Fig.~10-12 for PCB schematics and instrumentation setup.
    }
    \label{fig:3}  
    \end{figure*}

    Our ASIC integrates an array of 64 electrode sensors arranged in eight different configurations, with the electronics required to amplify and demodulate the measured signals, Fig.~\ref{fig:3}a-c. Each electrode configuration varies in length ($l$), width ($w$) and interelectrode-spacing ($d$) from the sub-micron dimensions to 10~$\mu$m, and we thus refer to them as ($l \times w – d$) given in $\mu$m. We adopted the three-electrode configuration of Fig.~\ref{fig:1}e, which enables increased SNR and detection of the direction of marker rotation as well as its speed. 
    
    The electrode sensor configuration and amplifier gains could be digitally selected via the IC. At an operating frequency of 10~MHz, the solution provides a predominantly conductive pathway and can be approximated by the solution resistance ($R_\mathrm{sol}$), Fig.~1c. A passivation stack comprising silicon dioxide (SiO$_2$) and silicon nitride (Si$_3$N$_4$) forms a ridged three-dimensional capacitive interface between the electrodes and the solution, Fig.~\ref{fig:3}d and Supplementary Fig.~8, with the passivation capacitance element adding in series to the electrode-solution equivalent circuit model (compare Fig.~\ref{fig:1}b and Fig.~\ref{fig:3}d; see \textit{Supplementary Text} for details). We retain the passivation layer because it replaces direct coupling between the metal and electrolyte, whose capacitance is strongly influenced by ionic strength, surface chemistry and adsorption, with a geometry-defined dielectric coupling. Furthermore, the sub-$\mu$m size of some of the electrodes and the standard TSMC process, makes passivation removal a complex, multi-step workflow that requires careful optimisation, (\textit{Methods} \cite{Moser2019,Li2024_electroless}.
    The $E$-field distribution in this configuration, with symmetry broken by the rotating marker produces a differential current at the central electrode (Fig.~\ref{fig:3}d). 
    The amplifier circuits amplify the current and convert it to voltage, while the demodulation circuit provides two output voltages representing the in-phase (V$_{real}$) and out-of-phase (V$_{im}$) components of the impedance variation caused by the rotating marker, Fig.~\ref{fig:3}c (see \textit{Supplementary Text} for details of the electronic circuit).
    
    To interface the IC with the electronic inputs and power supplies, as well as the output processing unit, we again used a PCB, Fig.~\ref{fig:3}e, which receives the electrode sensor excitation signals and demodulation reference inputs from two synchronized signal generators. The outputs from the IC (V$_{real}$ and V$_{im}$) are routed through a buffer amplifier to a data acquisition (DAQ) board for analog-to-digital conversion, and digital data is subsequently processed on a computer. The DAQ also relays electrode sensor selection signals from the computer back to the IC. See \textit{Methods} and Supplementary Fig.~10-12 for further details on the electronic interfacing setup.

\begin{figure*}[!tb] 
    \centering
    \includegraphics[width=0.85\textwidth]{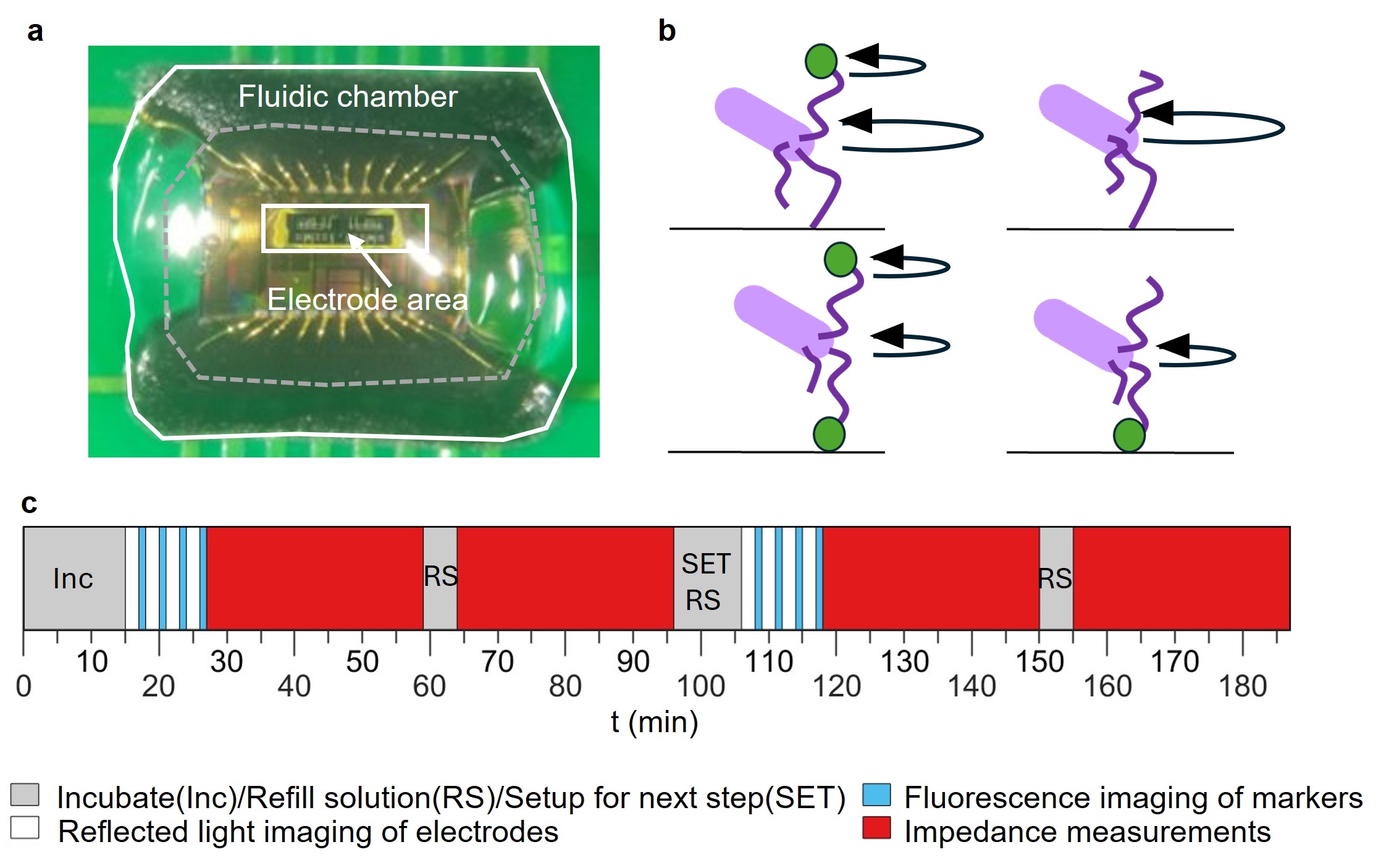} 
    \caption{
    \textbf{Interfacing bacteria with electrode sensors on the IC.} 
    \textbf{a.} A top-view microscope image of the IC with a fluidic chamber that protects the wire bonds and its surface, while leaving the electrode sensor area exposed to interface with the bacteria--bead suspension. The area between the white line and the grey dashed line marks the raised chamber walls, while the area within the grey line holds the bacteria-bead solution, and the white rectangle indicates the exposed electrode area. 
    \textbf{b.} Different configurations in which bacteria and marker beads can attach to the IC surface. Both sticky flagella and beads may bind to the surface, resulting in combinations of bead and tethered-cell assays (see Fig.~\ref{fig:1}a). The flagella were not sheared, leading to a range of rotation radii.
    \textbf{c.} A typical experimental timeline showing the sequence of steps after the bacteria--bead suspension was introduced into the fluidic chamber. Grey indicates one of the following: the incubation period at the beginning of the experiment, time to refill the chamber with buffer, or time to setup instrumentation for the next step. White indicates the reflected light imaging of the electrode sensor area, cyan fluorescence imaging of markers in solution, and, finally, red the impedance measurement. 
  }
     \label{fig:4}  
    \end{figure*}

    Finally, we designed a simple fluidic chamber for delivery of the bacteria to the sensors while protecting the surrounding electronics, Fig.~\ref{fig:4}a and \textit{Methods}. The bacterial solution,
   consisting of fluorescently labelled \textit{E.\ coli} mixed with fluorescent 0.5~$\mu$m polystyrene beads pre-attached to their sticky flagella, was placed in the chamber, which was then covered with a glass coverslip. See \textit{Methods} for further details on assay preparation and delivery. While the standard flagellar motor assays shown in Fig.~\ref{fig:1}a use a glass surface, the surface of the IC is the passivating layer of SiO$_2$/Si$_3$N$_4$. Therefore, we did not prepare a single specific assay configuration, but allowed multiple possible bead–flagellum attachment geometries, Fig.~\ref{fig:4}b and \textit{Methods}. To further increase the likelihood of the rotating marker passing near the electrode sensors we wished to use not only individual beads, but also their clusters of various sizes. For the purpose, we tested various salt concentrations \cite{Pilizota2007} and selected 130~mM NaCl, which provided a balance between forming differently sized bead clusters (0.5–5~$\mu$m in size) and minimizing excessive bead attachment to the IC surface. At this salt concentration and our operating frequency of 10~MHz, the current path in the solution is strongly conduction dominated (see \textit{Supplementary Text} on impedance spectrum calculations), and the increased conductivity was also expected to enhance the conductivity contrast between insulating bead markers and the surrounding medium, thereby improving the SNR.

    \textbf{Flagellar motor rotational speed and direction can be detected using the IC}
   To demonstrate that our IC based electrode sensors can detect the motor rotation we followed the data acquisition sequence depicted in Fig.~\ref{fig:4}c. It consisted of optical imaging of the IC surface, to identify rotating markers over or near the electrode sensors, and electronic probing to measure impedance changes. Reflected white-light microscopy was used to identify the electrode sensors, and fluorescence microscopy to visualize the rotation markers, see also \textit{Methods}. Optical imaging was not performed simultaneously with longer impedance measurements because the fluorescence excitation light can generate light-dependent impedance artefacts (photocarrier generation and surface charge accumulation \cite{Shen2021}) and cause photodamage to the bacteria \cite{Krasnopeeva2019}. Typically, the experiment consisted of: (i) 15-20~min incubation period that allowed bacteria with markers to settle to the IC surface, (ii) imaging of different regions of the electrode sensor area in sequential 15-20~s intervals, upon which (iii) the output signals from all 64 electrode sensors were recorded for 60~s each. This procedure was repeated until actively rotating markers were no longer observed. Done this way,
  the presence of any actively rotating markers near a given sensor was confirmed only during subsequent image analysis. The chamber was replenished with fresh MB$_\mathrm{NaCl}$ (see \textit{Methods} for media definition) every 60~min to compensate for evaporation; the resulting effects on the IC DC voltage output are depicted in Supplementary Fig.~13. Occasionally, during the initial 15-20~s imaging interval it was immediately obvious that a marker is located close to given electrode sensors. Optical imaging was extended for those cases, in order to better record its rotation, and electrical measurements of only that sensor were performed.

\begin{figure*}[!tb] 
    \centering   \includegraphics[width=0.75\textwidth]{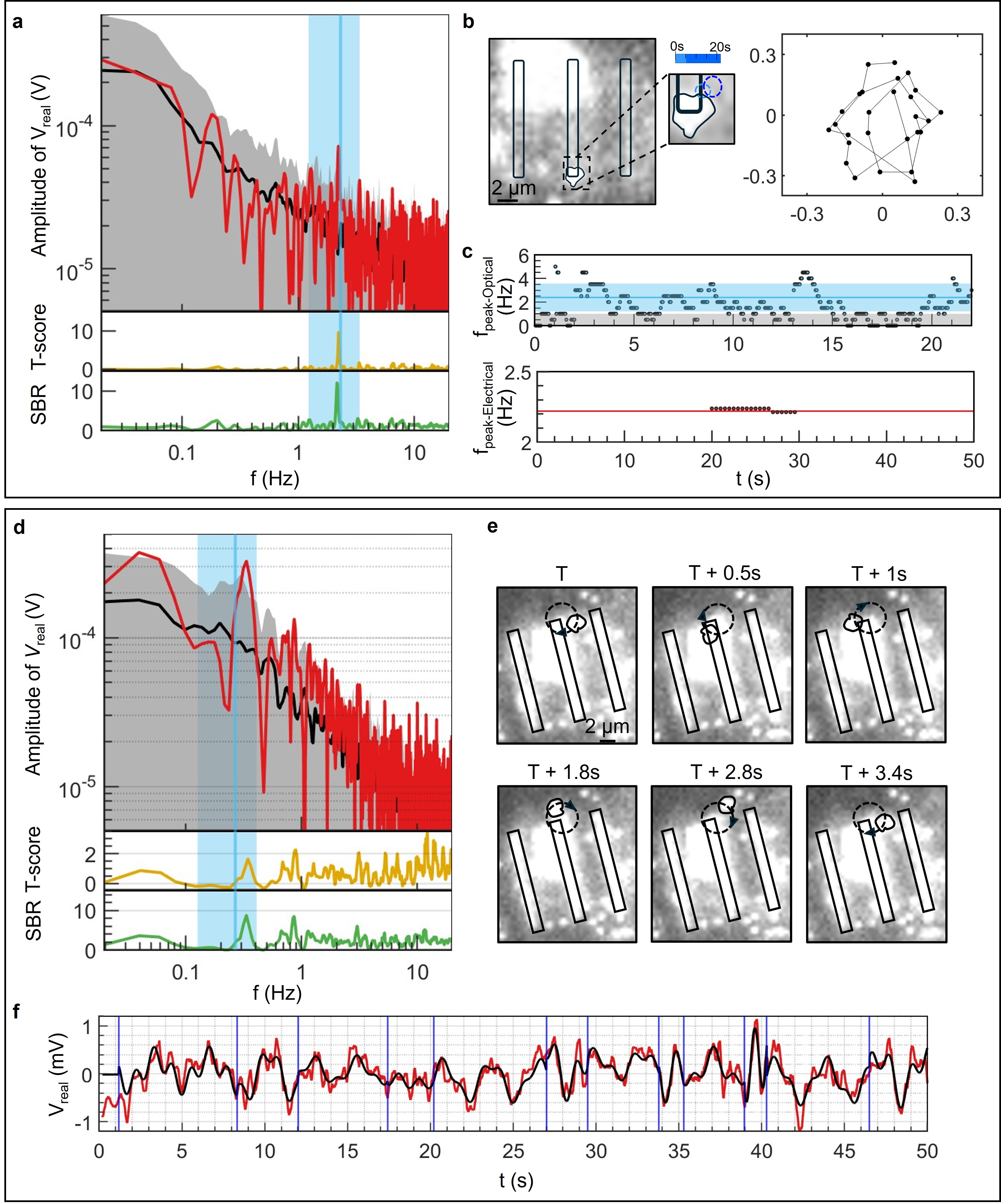} 
    \caption{
    \textbf{Flagellar motor rotation detected with two different electrode sensors on the IC.} Panels (a–c) correspond to one sensor and (d–f) to another, both with $l \times w - d = (10 \times 0.75 - 4)~\mu\mathrm{m}$.  
    \textbf{a,d}: Top: Amplitude spectrum of the voltage output $V_{\mathrm{real}}$ for the entire $\sim$50~s-long recording (red), representing the in-phase component of the impedance. The mean amplitude spectrum of control measurements (obtained without bacteria) is shown in black, with the grey region indicating $\pm$ 2 standard deviations. The corresponding \textit{t}-score and SBR spectra are shown below (see \textit{Methods} for details on data analysis).  The mean optically measured rotation frequency is indicated by a cyan vertical line, with the shaded region indicating one standard deviation.
    \textbf{b.} Left: Fluorescence image showing the bead-cluster marker relative to the electrode sensor; both are outlined for clarity. The zoomed-in region shows the rotational trajectory of the marker (blue) over a 22~s recording, with the colour scale indicating temporal progression. Right: Rotational trajectory extracted from the optical images (see \textit{Methods}).
    \textbf{c.} Marker rotation frequency obtained from fluorescence images ($f_\mathrm{peak-Optical}$) and from the sensor output ($f_\mathrm{peak-Electrical}$) in sliding windows as a function of time (black dots). Details of the data analysis are given in \textit{Methods}. The cyan horizontal line and shaded region depict the mean and standard deviation of all speeds (used in panel~(a)), excluding speeds obtained during pauses in rotation (grey region; see \textit{Methods} for how these were identified). The red horizontal line indicates the peak-frequency obtained from the full duration spectrum in panel~(a).
    \textbf{e.} Fluorescence image sequence showing a full rotation cycle of the bead-cluster marker over the electrode sensor during a 3.4~s period.
   \textbf{f.} Mean-filtered time-domain sensor output V$_{\mathrm{real}}$ (red) compared with its piecewise least-squares fit model consisting of a sum of two sinusoids (black), where the frequency of those was varied in a small window close to 0.33 and 0.86~Hz as described in \textit{Methods}. The dark blue vertical lines indicate the edges of the manually selected time windows over which the fits were performed. See \textit{Methods} and Supplementary Fig.~16 for details of the analysis.
   }
    \label{fig:5}  
    \end{figure*}

    Fig.~\ref{fig:5}a-f shows two cases where the rotation of the marker located on the central electrode of a $l \times w - d = (10 \times 0.75 - 4)$~$\mu$m sensor was detected. The exact acquisition sequences for these two cases is given in Supplementary Fig.~14. Using fluorescence microscopy we measured both markers to be a bead cluster  $\sim2\mu$~m in diameter, traversing both sides of the central electrode. In the first case, Fig.~\ref{fig:5}a-c, the cluster was rotating at (2.30$\pm$1.08)~Hz with radius $\sim$0.2~$\mu$m, and an offset of $\sim$0.4~$\mu$m from the electrode centre, while in the second, Fig.~\ref{fig:5}d-f, it was rotating at (0.27$\pm$0.14)~Hz with a $\sim$2.5~ $\mu$m radius of rotation. 

    The amplitude spectrum of the sensor output $V_\mathrm{real}$, together with the corresponding \textit{t}-statistic (\textit{t}-score) and signal-to-background ratio (SBR) spectra (see \textit{Methods}), exhibited a single peak at $\sim$2.2~Hz in the first case, Fig.~\ref{fig:5}a. In the second case (Fig.~\ref{fig:5}d), the amplitude and \textit{t}-score spectra exhibited multiple peaks at $\sim$0.33 and 0.86~Hz, as well as $\sim$3.06, 5.23, 8.53, 10.01, 11.94 and 14.71~Hz, whereas the SBR spectrum showed that only the first two frequency peaks had significant power above the background. In the latter experiment, the IC surface was more densely covered with rotating markers, and some of the peaks between 8-10~Hz also came up in the spectra of neighbouring electrode sensors. Therefore extra peaks can be attributed to other markers of motor rotation inside and outside of the field of view and focus, see Supplementary Video~1 and Supplementary Fig.~15. Furthermore, the optically and electrically measured frequencies agree; any differences align with whether the damaging blue-light optical measurement preceded or followed the impedance measurement, Supplementary Fig.~14 and \textit{Supplementary Text}. Lastly, Fig.~\ref{fig:5}f and Supplementary Fig.~16 show a good fit between the DC-corrected, mean-filtered time-domain signal of the cluster in Fig.~\ref{fig:5}d-f and 0.33 and 0.86~Hz components. See \textit{Methods} for details on data analysis. Other examples of rotating markers we detected are shown in Supplementary Fig.~17. The maximum SNR for rotation detection using the IC-based sensors was 15~dB, obtained from the sensor in Fig.~5a–c (see \textit{Methods}). 

\begin{figure*}[!tb] 
    \centering
    \includegraphics[width=0.6\textwidth]{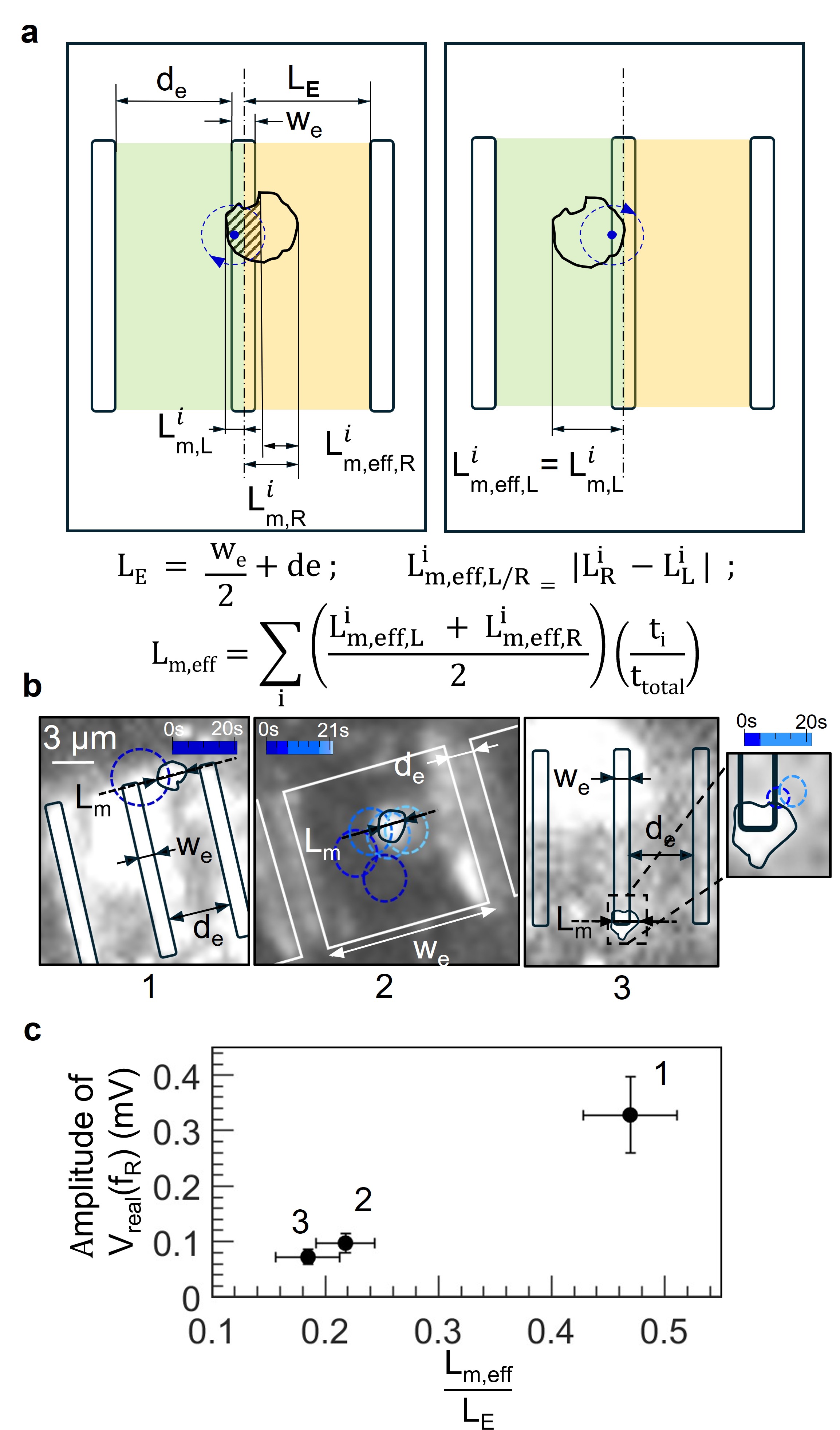} 
    \caption{\textbf{The signal strength from the IC-based sensor increases with effective marker size in the sensing region.} 
    \textbf{a.} Schematic illustration defining $L_{\mathrm{E}}$, $L_{m,\mathrm{eff},R/L}^i$ and $L_{\mathrm{m,eff}}$. All lengths are measured along the electrode-width direction. The regions of the electric field dominated by the left (L) and right (R) electrodes are marked in green and yellow, respectively, each with length $L_{\mathrm{E}}$ (defined below). Here, $w$ is the electrode width and $d$ the interelectrode spacing. The left and right panels show the marker positioned at two points along its trajectory (blue), with maximum overlap with the yellow and green regions, respectively. Left: The marker overlaps with both the yellow region ($L_\mathrm{m,R}^i$) and the green region ($L_\mathrm{m,L}^i$), where $L_\mathrm{m,eff,R}^i$ is defined as the difference between these two overlaps. Right: The marker overlaps only with the green region, such that $L_\mathrm{m,eff,L}^i = L_\mathrm{m,L}^i$. $L_\mathrm{m,eff}$ is defined below as the time-weighted average over all trajectories.
    \textbf{b.} Fluorescence images of three rotating markers on three electrode sensors (1-3), traversing their central electrode. The rotation trajectory is shown in blue, with the colour scale indicating temporal progression. Geometric parameters $w$, $d$, and marker length ($L_\mathrm{m}$) are indicated. 
    \textbf{c.} Signal amplitudes of the three rotating markers in (b), plotted against the ratio of effective marker length to the electric field length, $L_\mathrm{m,eff}/L_\mathrm{E}$. Signal amplitudes were extracted from the amplitude spectra of the impedance data at the identified peak rotation frequency. Error bars represent the standard deviation across measurements in both $x$ and $y$.  
    }
    \label{fig:6}  
    \end{figure*}

    To further confirm successful detection of motor rotation, we examined whether the signal amplitude varied with the size and relative positioning of the marker over the central electrode in the expected manner. As the marker traverses around the central electrode, it generates a signal by displacing the conductive solution and thereby redistributing the electric field from either of the side electrodes to the electrode in the centre. Thus, larger markers or greater overlap with the strong electric field zones, from either of the side electrodes, produce correspondingly larger signals. Approximating the system as one-dimensional along the electrode width (for sensors with the same electrode length), each side electrode dominates the electric field in the region from the inner edge of that electrode to the centre of the centre electrode (shaded regions in Fig.~\ref{fig:6}a, marked with $L_\mathrm{E}$). The signal amplitude is determined by the maximum effective length of the marker, $L_{\mathrm{m,eff}}$, averaged across all trajectories, with each trajectory weighted by the fraction of total time spent in that trajectory. 
    For each trajectory $i$, the maximum effective length is defined as  $L_{\mathrm{m,eff,R/L}}^i = \lvert L_\mathrm{m,L}^i - L_\mathrm{m,R}^i \rvert$. Consequently, different electrode sensor configurations, rotation radii, and marker sizes produce different  $L_{\mathrm{m,eff}}$, as depicted in Fig.~\ref{fig:6}b for markers from Fig.~\ref{fig:5} (here 1 and 3, respectively) and Supplementary Fig.~17a (here 2). Fig.~\ref{fig:6}c shows the expected increase in the signal amplitude with increasing normalised effective marker length ($L_{\mathrm{m,eff}}/L_{\mathrm{E}}$), for the three configurations in Fig.~\ref{fig:6}b. Configuration 3 uses a higher electrode simulation voltage than 1 and 2 (see \textit{Methods}), which here shifts point 3 higher, further confirming the overall trend.

\begin{figure}[htbp] 
    \centering
    \includegraphics[width=0.5\textwidth]{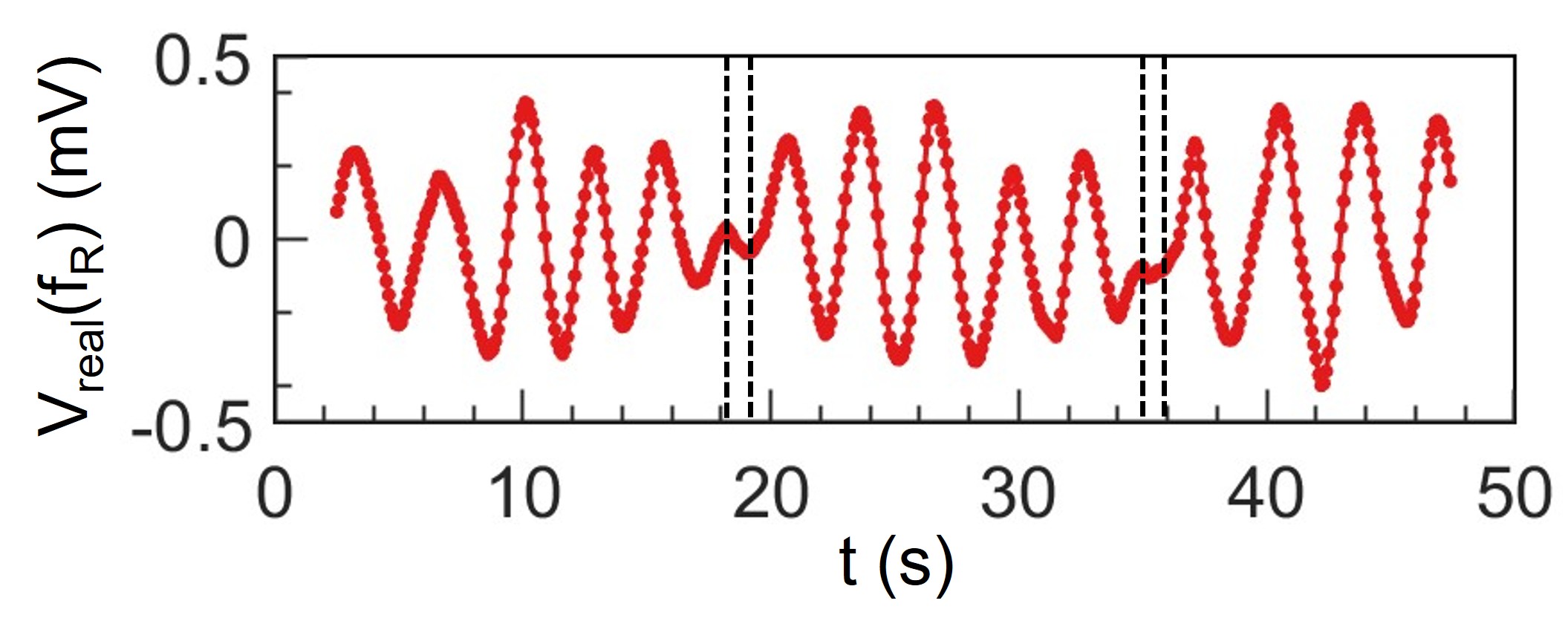} 
    \caption{\textbf{Changes in the flagellar motor rotational direction can be detected by the IC-based sensor.}
   Time-domain electrode sensor output ($V_\mathrm{real}$) from the experiments in Fig.~5d–f (the same experiments is also configuration 1 in Fig.~\ref{fig:6}b–c), after DC offset correction and synchronous (lock-in) detection at 0.33~Hz used to extract the amplitude and phase of the signal component at that rotation frequency (\textit{Methods}). Vertical dashed lines mark phase discontinuities; each pair indicates the start and end of a pause or reversal in the direction of phase evolution.}
    \label{fig:7}  
    \end{figure}

 Finally, to confirm we can detect a change in rotational direction of the motor, we tracked the phase of the marker 1 in Fig.~\ref{fig:6}. To extract the phase of the detected rotation frequency (0.33~Hz) the signal amplitude must exceed the background noise within the measurement bandwidth around that frequency. Thus, we used sliding windows of at least 1.5 signal periods ($\sim$5~s) that give a measurement bandwidth of ~0.2~Hz, Fig.~\ref{fig:7} (see also \textit{Methods}). The amplitude and phase of the detected signal evolves sinusoidally in time except at two time points at which the phase evolution pauses or shifts. These pauses/shifts likely correspond to changes in rotational direction. Flagellar motors of our wild-type \textit{E.\ coli} strain predominantly rotate counter-clockwise, and intermittently switch to clockwise rotation every $\sim$15~s in media without a carbon source \cite{Rosko2017}. The observed phase discontinuities in Fig.~\ref{fig:7} are consistent with these time scales (occurring every 14.5~s). 

\begin{figure}[htbp] 
    \centering
    \includegraphics[width=0.48\textwidth]{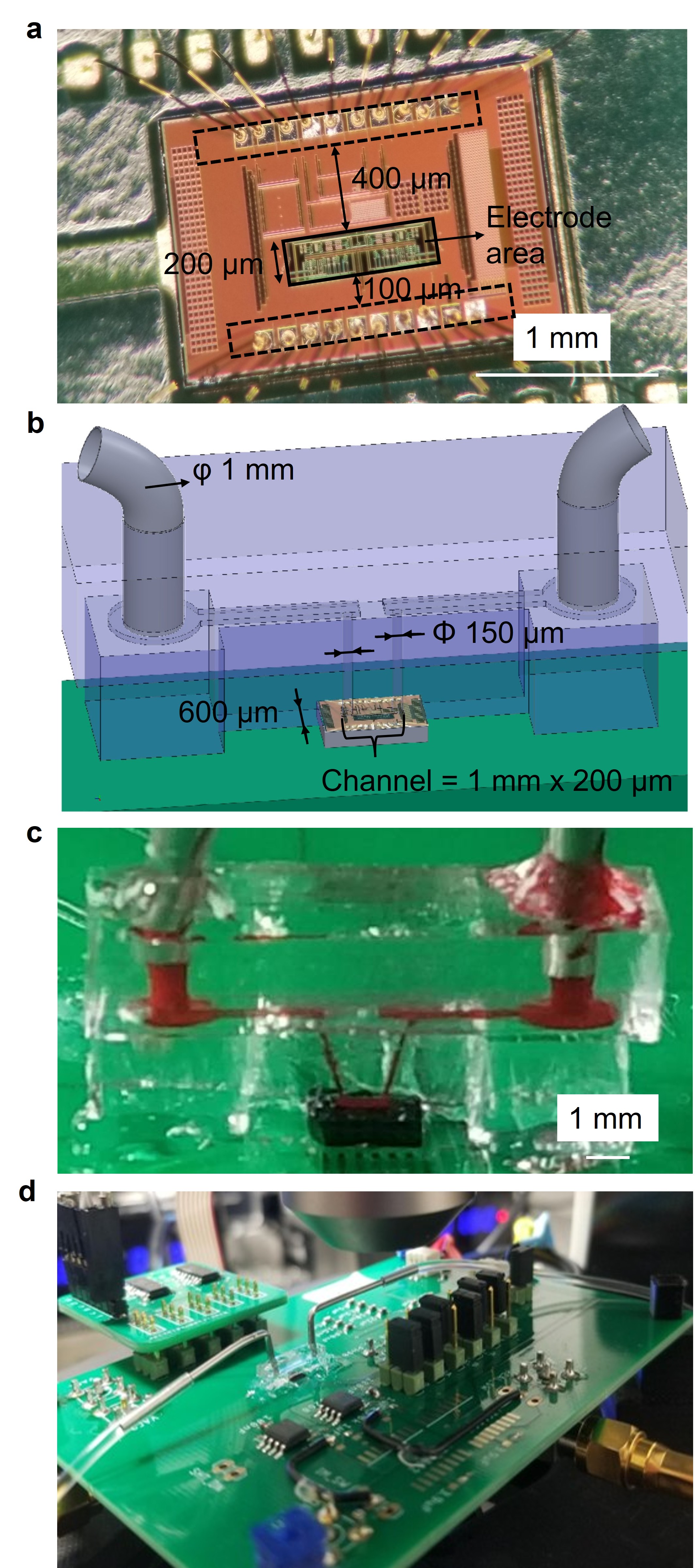} 
    \caption{\textbf{Design, fabrication and integration of a microfluidic channel with the IC sensor.} \textbf{a.} The IC surface shown in Fig.~4a is reproduced here with the physical constraints on the size of the microfluidic device indicated. Spacings of 100 and 400~$\mu$m separate the electrode sensing area from the wire bonds. \textbf{b.} CAD design of the two-layer microfluidic device. The bottom layer, in contact with the IC surface, is 600~$\mu$m wide and contains a microfluidic channel 1~mm long and 200~$\mu$m wide, with two vertical conduits (150~$\mu$m diameter) at either end connecting it to the top layer. The top layer connects each conduit to a millimetre-scale reservoir on either side, into which a 1~mm-diameter tube or needle can be inserted. The microfluidic device is fabricated from PDMS, with the red dye indicating the path of the fluid flow, \textbf{c}, and integrated with the PCB setup placed under a microscope for imaging, \textbf{d.}}
    \label{fig:8}  
    \end{figure}

     \textbf{A microfluidic device was integrated with the IC for controlled delivery of liquids to the sensing area.} A microfluidic device enables the controlled delivery and removal of bacteria, markers, nutrients, and analytes, which will likely be required for the envisioned biosensing applications \cite{Luka2015,Shang2022}. It also maintains a stable liquid volume above the sensors, preventing evaporation-induced DC shifts observed in Supplementary Fig.~13. Finally, by confining the electric field to a reduced height above the electrodes and thus enhancing the sensitivity to surface-localized phenomena, the device could improve sensor SNR \cite{Gawad2001, Warren2023}. Given these advantages, we designed a microfluidic channel above the sensor electrodes, subject to the following constraints imposed by the IC surface: (i) liquids must be delivered to the $\sim(750 \times 200)~\mu\text{m}^2$ electrode sensor area while shielding the surrounding electronics, (ii) the device must fit within the $\sim700~\mu$m-wide space between the two wire-bonded edges, with only a 100~$\mu$m clearance between the channel and the wire bonds on one side, and (iii) the device must interface with millimetre-scale fluid-handling components, such as tubing and syringes, Fig. \ref{fig:8}a. These requirements motivated a two-layer design, in which microscale and millimetre-scale features are separated in the bottom and top layer, respectively, Fig. \ref{fig:8}b.

     The microfluidic device was fabricated from Polydimethylsiloxane (PDMS) due to its biocompatibility, optical transparency, and suitability for rapid prototyping \cite{RajM2020,Sia2003,Morbioli2020}. To fabricate a 600~$\mu$m wide bottom layer of the PDMS channel, and fit it within the above mentioned 100~$\mu$m clearance, we designed a custom three-axis alignment and cutting tool described further in \textit{Methods}, Supplementary Fig.~18-19 and Supplementary Video~2. Together with the fabrication procedures described in \textit{Methods}, this enabled the integration of a microfluidic device on the IC, including reliable formation of vertical fluidic interconnects and leak-free bonding without channel obstruction. The channel height was set to 100~$\mu$m to confine the sensing volume above the electrodes (see \textit{Methods}). Figure \ref{fig:8}c and d show one such channel with a red dye delivered via millimetre-scale tubing and syringe-pumps marking the liquid path. Using the channel, bacteria-bead suspensions can be  brought in contact with the electrode sensors on the PCB, while remaining optically accessible for real-time microscopy, Fig. \ref{fig:8}d.
     
    \section*{Discussion}
We have presented a method and a device that together constitute a novel bioelectrical interface able to detect the rotation of a single bacterial flagellar motor. Several studies, including our own, proposed to utilise the high-sensitivity (nM) and speed (100s of ms) of the bacterial chemotactic network for environmental sensing \cite{Atkinson2022, IndiaWebb2023}. Because the network output changes the frequency of the flagellar motor rotational direction, our architecture forms the basis of a reconfigurable class of whole-cell based biosensors, where inputs are engineered as needed to detect a given analyte, but the detection of the output remains the same technology. 
     
   The maximum SNR obtained was 15~dB, indicating that the periodic rotation signal could be clearly detected from the background noise. Detectable rotation frequencies were confined to below 3~Hz, from large bead clusters on unsheared flagella. Several reasons likely explain why we did not detect faster rotating markers, despite the fact that such higher frequencies could improve SNR by shifting the signal away from dominant 1/f noise. Faster-rotating markers are smaller, because they impose lower drag onto the motor (that operate at low Reynolds number\cite{Sowa2008}), which produces weaker signals. Furthermore, such smaller and fast rotating markers are harder to image optically. Lastly, detecting changes in rotational direction requires higher SNR, because it relies on extraction of phase at the rotation frequency.

    The electrode sensors of the IC do not operate in perfect differential mode due to process asymmetries (in electrode dimensions and interelectrode spacing)  \cite{Pelgrom1989}, and settling of non-rotating bacteria and beads over the sensors during the experiments. The former can be rectified by calibrating the input amplitude difference applied to the two side electrode sensors to produce a zero sensor output prior to the experiment. To rectify the latter, future experiments can pattern the IC surface with molecular adhesives for specific attachment of the cells. For instance, positively charged polymers, such as poly-L-lysine (as used to prepare the bead assay in Fig.~\ref{fig:2}) or polyethyleneimine may be deposited onto the IC surface in defined patterns through microcontact stamping or printing, with surrounding areas coated with an antiadhesive polymer, such as polyethylene glycol to prevent unwanted attachment \cite{Arnfinnsdottir2015,Wang2019}. Alternatives to these molecular adhesives include using antibodies specific to membrane-bound proteins expressed on the bacteria outer membrane \cite{Rozhok2005} or a biotin-streptavidin conjugation to biotinylated outer membrane proteins \cite{Cerf2008}. The IC surface would be prepared with these compounds prior to introducing bacteria through the microfluidic channel. The non-specific adhesion can also be controlled by identifying the optimal incubation period of the cells and/or the markers. Taken together these interventions can enable precise positioning of the bacteria and perfect differential mode of IC operation, and thus improved SNR.

   To increase the operation frequency and thereby improve SNR, our IC integrates the electrode sensors with the amplifier and processing circuitry that follows immediately. In the current implementation, however, the excitation circuitry and digitization remain off chip, with external interconnects introducing parasitic variability and noise. Future versions could integrate these functions on the chip, reducing parasitic loading and moving the platform towards a fully self-contained system. The architecture is also well suited to multiplexing. Because the sensing interface is defined lithographically and read out electrically, it can in principle be scaled to dense arrays with thousands of individually addressable electrodes. Such scaling would enable parallel monitoring of multiple bacterial strains, analytes and environmental conditions on a single chip. Power consumption is expected to depend primarily on the number of active channels, operating frequency, required noise performance and extent of on-chip processing. Although higher bandwidth and lower noise generally require increased analog bias current, power could be reduced through duty-cycled or multiplexed readout, lower-voltage circuit design and application-specific digital processing. Together, these advances provide a route to compact, battery-powered bioelectronic sensors for field deployment.

Lastly, while the frequency of the rotational direction changes of the flagellar motor scales with the environmental analyte concentrations, in \textit{E. coli} the motor is powered by the electrochemical gradient of protons, the so-called proton motive force (PMF) \cite{Lo2024, Biquet-bisquert2021}. In most conditions, the motor speed is linearly proportional to the PMF \cite{Krasnopeeva2026}. Because PMF is one of the main energy sources in bacteria \cite{Lo2024}, and hard to measure otherwise \cite{Lo2024, Bisquert2024}, this linear relationship has been used to infer the PMF changes in a wide range of conditions \cite{Krasnopeeva2019, Terradot2024, Bisquert2024, Krasnopeeva2026}. Our device, by automatically providing motor speed information in electronic form for potentially hundreds of thousands of individual cells, can be a useful PMF sensor  \cite{Krasnopeeva2021}.

    \section*{Methods}
    \subsection*{Bacterial strains}
    Experiments with the cleanroom fabricated four-electrode sensors are conducted with $E. coli$ strain SYC12 (background strain RP437) \cite{Hata2020}, while those with the IC with the EK01 harbouring pwR20 plasmid  (background strain MG1655) \cite{Krasnopeeva2019, Pilizota2012}.
    Both strains have wild-type chemotactic machinery \cite{Sourjik2002, Schulmeister2011} and carry a chromosomal genetic mutation ($\Delta$fliC::fliC$^{st}$) enabling expression of so-called sticky flagellar filament \cite{Turner2000}. Plasmid pWR20 encodes an enhanced green fluorescent protein (eGFP) and kanamycin resistance \cite{Pilizota2012}. 

    \subsection*{Bacterial growth and preparation}
    For experiments with cleanroom-fabricated electrode sensors, SYC12 cells were grown overnight to stationary phase in Miller Lysogeny Broth (LB). The culture was then diluted 100-fold into tryptone broth (1\% w/v tryptone, 0.5~mM NaCl) and incubated at 30$\,^{\circ}\mathrm{C}$ with shaking at 250~rpm until an optical density (OD$_{600}$) of 0.4--0.6 was reached. The culture was subsequently grown overnight (12--16 h) at room temperature without shaking to a final OD$_{600}$ of 0.6--0.8. For experiments with the IC-based electrode sensors, EK01+pwR20 cells were inoculated from frozen glycerol stock at 10$^5$ dilution into 3-(N-morpholino)propanesulfonic acid (MOPS)-buffered medium, prepared from 1$\times$ MOPS mixture containing a MOPS buffer(MOPS, tricine, FeSO$_4$·7H$_2$O adjusted with KOH) and a salt mixture(NH$_4$Cl, K$_2$SO$_4$, CaCl$_2$·2H$_2$O, MgCl$_2$, and NaCl), supplemented with 1$\times$ defined amino acid mixture (EZ supplement \cite{Brouwers2020}), 1.32~mM K$_2$HPO$_4$ and 0.3$\%$ glucose. The culture was grown at 37$\,^{\circ}\mathrm{C}$ with 220~rpm shaking until OD$_{600}$ = 0.3.  
    In both cases, after harvest the cells were washed twice by centrifugation (2~min at 8000~g) and resuspended in Motility Buffer (MB: 10~mM KPO$_4$, 0.1~mM EDTA, pH 7.0). For experiments with SYC12, MB was supplemented with 1~mM L-methionine (MB$_{Met}$) and cells were concentrated to OD$_{600}$=1, and for those with EK01, with 130~mM NaCl (MB$_\mathrm{NaCl}$) and cells were concentrated to OD$_{600}$ = 0.6. To prepare the bacteria-bead suspension for the IC, we next mixed  the bacterial suspension with fluorescent 0.5~$\mu$m beads (Fluoresbrite YG Microspheres, Polysciences, USA) at 1:1 volume ratio. Beads were diluted in MB$_\mathrm{NaCl}$ to a concentration of $3.64 \times 10^9$~ beads/ml before mixing. The bacteria-bead mix was incubated for 10-15~min to allow the beads to attach to bacterial flagella, and subsequently centrifuged at 2000~g for 1~min to remove unattached beads in suspension. The pellet containing bacteria and bead clusters was resuspended in MB$_\mathrm{NaCl}$ at OD of 0.6. This bacteria-bead suspension was added to the custom fluidic chamber constructed around the IC and covered with an $\sim 5 \times 5$~mm coverslip.

    SYC12 in MB$_{Met}$ was introduced into the microfluidic channel under the cleanroom-fabricated four-electrode sensor after it was treated with 0.01$\%$ (w/v) poly-L-lysine and rinsed with MB$_{Met}$. The culture was incubated for 30~min to immobilize the bacteria on the channel surface, and unattached cells were removed by flushing with fresh MB$_{Met}$. The 2~$\mu$m in diameter beads (Polysciences Inc., USA), diluted to 0.125$\%$ (w/v) in MB$_{Met}$, were introduced next and allowed to attached to sticky flagella.

    \subsection*{Preparation of fluorescent beads for testing of cleanroom fabricated microelectrode sensors}
    To test the detection of single beads with the clean room fabricated microelectrodes we used passive siphon-driven flow that moved 2~$\mu$m diameter fluorescent polystyrene beads (Bangs Laboratories, USA) over the electrode array with a volumetric flow rate of 0.5 - 1.0 $\mu$L/min. Beads were washed in MB$_{Met}$ then diluted to a concentration of 0.02\% w/v in MB$_{Met}$ before being introduced to the flowcell.
    
     \subsection*{Microelectrode sensor fabrication and integration with electronics}
    
    The four-electrode sensor was fabricated in a cleanroom on a fused silica wafer substrate. The electrodes, consisting of a Ti/Au metal stack (thickness 5~nm/145~nm), were patterned by a standard lift-off process \cite{Zajdel2017}. The sense electrodes measured 4~$\mu$m in width, while the current-injection electrodes measured 100~$\mu$m x 100~$\mu$m. A polyimide layer was spin coated over them and patterned using an aluminium hard mask and O$_2$ plasma, in order to insulate parts of the sense electrodes, leaving behind a 4~$\mu$m x 4 ~$\mu$m area each. It also provided electrical insulation for the conductive traces connected to the electrodes. A conducting polymer Poly-3,4-ethylenedioxythiophene:poly-styrenesulfonate (PEDOT:PSS) was electropolymerised on the current-injection electrodes to reduce the $C_\mathrm{DL}$ on these electrodes (see \textit{Supplementary Text}). The resulting die, containing a single four-electrode sensor, was affixed over a hole drilled in a PCB. The die and the PCB were electrically connected by aluminium wire bonding that was protected with light-curable epoxy. The PCB consisted of an instrumentation amplifier (Analog Devices AD8221, USA), which provided a high input-impedance and unity-gain interface to measure the voltage across the sense electrodes and transfer to the impedance analyser (Keysight E4980AL, USA). The PCB also contains conductive traces that route current from the impedance analyser to the current-injection electrodes (see \textit{Supplementary Text} and Supplementary Fig.~1 for additional details).

    The IC incorporating the microelectrode sensor array and integrated electronics was designed using the Cadence Virtuoso platform and fabricated by Taiwan Semiconductor Manufacturing Company Ltd (TSMC) using a standard 0.18-$\mu$m complementary metal–oxide–semiconductor (CMOS) process. See \textit{Supplementary Text} and Supplementary Fig.~7 for details of the electronic circuits incorporated on the IC. The fabricated die was wire-bonded to the PCB using gold wire bonds, and the bonds were mechanically protected with epoxy (AITechnology, USA). See Supplementary Fig.~10-11 for details of the PCB. 

     \subsection*{Integration of electrode sensors with fluidic channel/chamber}

   The fluidic channel was integrated beneath the cleanroom-fabricated four-electrode sensor using a glass slide and either vacuum grease or PDMS seals. Inlet and outlet holes were drilled into the PCB on either side of the electrode sensor, and fitted with silicone tubing, with the electrode sensor facing downwards. For bacterial flagellar motor measurements, a Parafilm-stencil defined vacuum grease spacer was used to form the channel. The stencil defined both the footprint area and height (0.13~mm) of the grease, which was then compressed with the PCB to form the channel (see Supplementary Fig.~3c). Assuming volume conservation, its final height (H) was estimated from the change in area as $H = \frac{A_{original}}{A_{final}} \times 0.13 \mathrm{mm} \approx 40~\mu$m.
   The slide was clamped to a fixed mount while the PCB was positioned using a three-axis micromanipulator, enabling lateral adjustment without breaking the seal. For single polystyrene bead flow detection, a PDMS flowcell was used instead, providing a well-defined channel height. I was cast on an SU-8 master (SU-8 2010, MicroChem), which ranged between 15-20~$\mu$m for the given device, and was deposited by one coating on a silicon substrate and lithographically patterning channel geometries. A stylus profilometer was used to confirm channel thickness of the mould (Dektak 3030 Profiler, Veeco Instruments, USA). PDMS was mixed to a 5:1 base to curing agent ratio (Sylgard 184, Dow Corning, USA), cast to a thickness of 10~mm, degassed in a vacuum desiccator, then cured at $90^\circ$C for 60~min in a laboratory oven. A finished PDMS channel was carefully peeled from the mould, then lightly clamped into place over the cleanroom-fabricated electrode array. The PDMS did not allow relative movement between the electrode sensor and the channel, which was sufficient since precise electrode sensor alignment was not required in this experiment.

     The fluidic chamber was built around the IC using UV-curable acrylic adhesive (Loctite AA3526, Henkel Adhesives, Germany). The wire bonds and electronics area of the IC were first covered layer by layer, curing each layer with a handheld UV curing pen. The same procedure was then used to build the millimetre-scale chamber capable of holding 0.5–5~$\mu$L of solution by forming a wall of adhesive. The entire device was subsequently placed in a UV light box for a few minutes to ensure complete curing before use. The chamber was closed with an approximately 5~mm $\times$ 5~mm piece of glass cut from a 0.13~mm thick microscope cover glass (VWR International) using a diamond pen. The glass was cleaned with ethanol and isopropanol (IPA) prior to use.
     
    The PDMS microfluidic devices for the IC-based sensors was made from two layers moulded from a custom made SU-8 (SU-8 3035, MicroChem/Kayaku Advanced Materials) based master. The master was prepared by spin-coating a 100~$\mu$m-thick SU-8 layer (deposited in three sequential coatings) on a silicon substrate and lithographically patterning masters for both layers on the same wafer. See \textit{Supplementary Text} and Supplementary Fig.~20 for details. PDMS 5:1 base to curing agent (Sylgard 184, Dow Corning, USA) was cast to a thickness of 1.5–2~mm, degassed in a vacuum desiccator, and partially cured at $65~^\circ \mathrm{C}$ for 1~h in a standard laboratory oven (Raven Incubator, LTE Scientific, UK) before carefully peeling the PDMS from the master. 
    
    Vertical fluidic interconnects were formed under stereoscopic inspection. In the bottom-layer devices, a 30 gauge (G) needle was used to punch holes through both ends of the channel at a 45$^{\circ}$ angle, and away from the channel and
    nearest device edge. This kept the conduits within the 600~$\mu$m-wide device area. In the top-layer devices, a 23G needle was used to punch holes into the reservoirs at either end. Punched material in the vertical conduits was cleared using 34G and 26G needles as appropriate. Individual PDMS layers were cut to micrometre precision using a custom-built micrometer-controlled cutting tool
    (Supplementary Fig.~18). See Supplementary Video~2 for the procedure used to form vertical conduits and cut the individual layers.

    The PDMS layers were sonicated in isopropanol (IPA) for 30~min, fully cured at $65~^\circ \mathrm{C}$ overnight, and inspected under the stereoscope before use. The two layers were aligned and bonded under the stereoscope after oxygen plasma treatment
    (PDC-002, Harrick Plasma, USA). The assembled device was aligned to the IC surface using a custom alignment tool and bonded with a thin, uniform layer of 1-2~$\mu$L silicone adhesive (MED1-4013, NuSil Technology, USA), which was applied to the device underside (see Supplementary Fig.~19). The device was held under compression over the IC surface overnight (using the Z-control of the alignment tool) until the silicone fully cured.

    The completed microfluidic device was tested for leakage and flow using water. A 21G bent needle connected to PTFE tubing (1/16" OD, Darwin Microfluidics, France) was inserted into the top-layer reservoirs, and water was flown using a syringe. Devices were stored with water filled in them, with the water syringes attached to prevent drying and clogging prior to use.
     
    \subsection*{Data acquisition} 
    \subsubsection*{Optical imaging of bacterial assays}

    The cleanroom-fabricated electrode sensor with the corresponding channel containing the bead assay was imaged from above through the transparent silica substrate, using a 20x long working distance objective (20x M Plan APO, NA 0.42, WD = 20~mm, Mitutoyo, Japan). A CCD camera (Flea Series, Point Grey, Canada) attached to the objective was used to record their rotation at 40~frames per second (fps) with 20~ms exposure time. See Supplementary Fig.~3b for the imaging setup. The electrode sensor and fluorescent beads used for testing the capabilities of the cleanroom fabricated microelectrode sensors were imaged on an inverted Olympus IX81 fluorescence microscope using a Xe lamp light source and the appropriate filter set for Alexa Fluor 647 dye with a 50X long working distance
    objective (M Plan APO, NA 0.55, WD = 13 mm, Mitutoyo, Japan).
    
    Imaging of the electrode sensor area on the IC-based sensors was performed using white-light reflection microscopy and a 20× objective (CFI Achro LWD DL, NA 0.4, Nikon, Japan) on an upright microscope (Eclipse E200, Nikon, Japan). Fluorescence imaging of the markers was carried out using a mercury lamp (HBO 50W AC L1, OSRAM, Germany) with a filter cube set (excitation filter: ET470/40x, emission filter: ET525/50m, dichroic: 59022bs; Chroma Technology Corp., USA). Images were recorded at 20~fps for experiments of Fig.~5a-c and 10~fps for those depicted in Fig.~5d-f, using a ProSilica GigE camera (Allied Vision, Germany) and StreamPix software. Exposure time was 5~ms for experiments of Fig.~5a-c and 10~ms for those experiments of Fig.~5d-f. Gain was set to 24 in both cases. The intensity of the blue light was $\sim$127~mW/cm$^2$, measured using a power metre placed in focus under the objective where the beam diameter was 1$\pm$0.5~mm.

    To image the IC-based sensor surface with rotating markers through the microfluidic channel ($\approx$4~mm of PDMS device height and $\approx$2~mm needle height) a long working distance objective (M Plan Apo 20x, WD=20~mm, Mitutoyo, Japan) was used in a custom-built microscope. The microscope was built in upright geometry. The optical components: camera (WAT-910X, Watec, France), tube-lens, objective and sample (PCB) were aligned vertically from top to bottom using Thorlabs (USA) and custom machined components. A blue-LED light source (470nm LUXEON Rebel LED, Luxeon Star LEDs, Canada) driven by a current driver (FlexBlock A011, LED Dynamics, USA) was used to excite the markers on the IC surface and the green fluorescence emitted was collected by the objective, passed through an emission filter (ET525/50m, Chroma Technology Corp., USA) and recorded by the camera.
    
     \subsubsection*{Impedance measurements}
    
    All impedance measurements on cleanroom fabricated four-electrode sensors were performed using a Keysight Technologies E4980AL LCR meter with 1~Vpp 10~kHz sine wave stimulus applied to the current-injection electrodes. The LCR meter was controlled via a USB to GPIB interface (82357B, Keysight Technologies, USA) and a custom MATLAB script. Measurements were taken continuously with output sampled at an average rate of 33 samples per second (although actual rate varied, see \textit{Supplementary Text}).

    For impedance measurements on the IC, as shown in the setup of Fig.~3e, 10~MHz sine-wave electrode stimulation signals V$_{ac}$ and -V$_{ac}$ were supplied by one signal generator. Square wave demodulation signals  $SQ\_0^\circ$ and $SQ\_90^\circ$ were provided by another signal generator of the same kind (both Keysight 33500B, Keysight Technologies, USA). Several electrode stimulation sine-wave amplitudes were tested. The experiment of Fig.~5a-c (repeated as Fig.~6 sensor 3) used amplitude 900~mVpp, while those of Fig.~5d-f (repeated as Fig.~6 sensor 1) used 100~mVpp. The experiments shown in Supplementary Fig.~17a (repeated as Fig.~6 sensor 2) also used inputs of amplitude 100~mVpp. 
    
    The oscillator of the demodulation generator was synchronized to that of the stimulation generator. The phase relationships between the signals were set on the generators and verified using a 20 MHz oscilloscope (ISO-TECH ISR-622, RS Components Ltd., UK). Any phase shifts detected were corrected manually. The phase relationships set in this manner remained stable for at least 60~min. Where experiments exceeded this duration, phase shifts were checked and reset in the SET step of Fig.~4c. 
    
    In the setup of Fig.~3e, a LabVIEW program was used to record $V_{real}$ and $V_{im}$ for 60~s at 10~kHz, and to select electrodes. The program sent signals to the PCB (Supplementary Fig.~11) via a DAQ board (NI PCIe card 6343), which interfaced with it using a BNC 2110 terminal block (National Instruments, USA). The gains of the current amplifiers and the transimpedance amplifier were configured to 0 and 132~dB, respectively. The cutoff frequency of the low pass filter in the demodulation circuit was 14.6~kHz.

    For the electrodes that had a marker rotating over them, at least 10 control impedance measurements were made under identical conditions except that the bacteria-bead solution was replaced with just MB$_\mathrm{NaCl}$.
    
    \subsection*{Data Analysis}
    All analysis was implemented in MATLAB (MathWorks Inc., USA) unless otherwise specified.
    
    \subsubsection*{Optical image analysis}

    In Fig.~2c, the changes in optical intensity $\Delta F$ were obtained by baseline correcting $F$, which is the sum of pixel intensities within the sensing area (white square). The baseline was calculated as the mean of all $F$ points that visually appeared to belong to the flat portion of the trace.
    
     Analysis of optical images given in Fig.~2d and Fig.~5b,c were performed as follows. Each image in the recorded sequence was cropped to manually selected region of interest (ROI) containing only the rotating marker. The brightest pixel in the ROI was identified to provide an approximate bead location. Pixels with intensities above a fixed threshold  within a circular neighbourhood corresponding to the bead radius were then segmented as bead pixels. The bead position was determined as the centroid of the segmented bead pixels $X + iY = \frac{1}{N}\sum x_i + iy_i$, where $N$ is the number of segmented bead pixels and $x_i + iy_i$ is the complex position of each segmented bead pixel $i$. The sequence of bead coordinates obtained for each frame gave the bead trajectory over time. The rotational speed over time was obtained as the frequency corresponding to the maximum magnitude of the single-sided Fourier spectrum, computed in sliding windows with flattop window function. The window length $w$ was chosen to balance frequency resolution $1/w$ and temporal tracking of the signal speed variation. Specifically, $w$ should be sufficient to resolve rotation frequency $w \gg 1/f_\mathrm{R}$ while being shorter than the average timescale of speed variation $w \leq 1/f_\mathrm{speed var}$. The time step was chosen as $ts <  w/10$ for smooth temporal sampling. Accordingly, in Fig.~2d $w=$3~s and $ts=$0.1~s, whereas in Fig.~5c $w=$2~s and $ts=$0.05~s to follow the respective speed variations while maintaining required frequency resolution. Similarly, the mean and standard deviation of optically observed speed in Supplementary Fig.~17 was calculated using $w$=2~s and $ts$=0.1~s in (a), and $w$=5~s and $ts$=0.1~s in (b), the latter to resolve the slower 0.33~Hz rotation. In Fig.~5c, speeds obtained during periods when the bead appeared to have paused rotation, as seen visually, were attributed to diffusive motion and excluded (greyed area) from the calculation of average speed.
    
    \subsubsection*{Electrical data analysis}
    
    For the bead pass detection check on cleanroom fabricated sensors of Fig.~2c, $\Delta Z$ was obtained from measured $Z$ by removing high-frequency noise and linear baseline drift as follows. The signal was passed through a zero-phase finite-response (FIR), low-pass filter with 10~Hz cut-off frequency. Next, it was sectioned into 1~s windows (corresponding to the periodicity of bead flow through the sensing area), and the $Z$ baseline in each of these sections was calculated as the mean of all points outside $\pm 0.2$~s of the maximum $Z$ value. A linear fit of these baseline values was used to obtain the slope of the baseline drift, which was then subtracted from $Z$. The baseline and noise were then defined as the mean and standard deviation, respectively, of all points that visually appeared to belong to the flat portion of the trace (using a threshold). $\Delta Z$ was calculated as $Z-\text{baseline}$. SNR (in dB) for the detected $\Delta Z$ peaks was calculated as $\text{SNR}_\mathrm{cleanroom,flow,dB} = 20 \log_{10} \left( \frac{\Delta Z_\mathrm{peak}}{\sigma_\mathrm{Noise}} \right)$, where $\sigma_\mathrm{noise}$ is the standard deviation of the baseline as described above.

     For the motor-rotation detection experiments with IC-based sensors, the amplitude spectrum $A(f)$ of the sensor output $V_\mathrm{real}$ in Fig.~5a,d and Supplementary Fig.~17, was computed from the one-sided Fourier transform of the full duration of $V_\mathrm{real}$ measurement (51~s in Fig.~5a and 50~s in Fig.~5d). The signal was multiplied by a flattop window function ($W$) and normalised by $\sum W$ to obtain accurate amplitude estimates. The amplitude spectra of controls were calculated in a similar manner for all control measurements ($N = 11$ for Fig.~5a and $N = 15$ for Fig.~5d), and averaged to obtain their mean spectrum $\overline{A}_\mathrm{C}(f)$ and the corresponding twice-standard-deviation envelope $2\sigma_{A,\mathrm{C}}(f)$. These were then used to calculate the \textit{t}-score and SBR spectra in order to identify signal peaks that are statistically significant with respect to background variability, and have significant contrast with respect to control background, respectively. They were calculated from the power spectra of the signal ($\tilde{P}_\mathrm{S}(f)$) and control measurements (average $\tilde{\overline{P}}_\mathrm{C}(f)$ and standard-deviation $\sigma_{P,\mathrm{C}}(f)$) after log-frequency smoothing with a 1/12-octave window (\textit{Supplementary Text}), as follows:
    \begin{equation}
        P(f) = A(f)^2,
    \end{equation}
    \begin{equation}
        t\text{-score}(f) = \frac{\tilde{P}_{\mathrm{S}}(f) - \tilde{\overline{P}}_{\mathrm{C}}(f)}{\tilde{\sigma}_{P,\mathrm{C}}(f) \, \sqrt{N}},
    \end{equation}
    \begin{equation}
        \mathrm{SBR}(f) = \frac{\tilde{P}_{\mathrm{S}}(f)}{\tilde{\overline{P}}_{\mathrm{C}}(f)}
    \end{equation}.

    The maximum SNR (in dB) obtained from the IC-based sensors was calculated from the power spectra for the data in Fig.~5a at the peak frequency of 2.2~Hz, with log-frequency smoothing, as:
    \begin{equation}
        \mathrm{SNR}_\mathrm{IC,rot,dB} = 10 \log_{10} \left( \frac{\tilde{P}_{\mathrm{S}}(f_{\mathrm{peak}}) - \tilde{\overline{P}}_{\mathrm{C}}(f_{\mathrm{peak}})}{\tilde{\sigma}_{P,\mathrm{C}}(f_{\mathrm{peak}})} \right)
    \end{equation}

    For the motor-rotation detection experiments with cleanroom-fabricated sensors, the amplitude spectrum of the full 60~s impedance measurement in Supplementary Fig.~5, obtained following the procedure used for IC-based sensors, exhibited no clear peak. Given that temporal variability in the signal can result in time-averaging of rotational speed changes in the full-time spectrum, we also calculated time-resolved peak frequencies in sensor output in Fig.~2d. This was done using sliding-window spectral analysis with a flattop window. A time window of 3~s was chosen to match the optically observed signal variability ($\approx 2-4$~s) with a time step of 0.1~s, corresponding to sufficient frequency resolution (0.3~Hz) to distinguish the $\approx 5$~Hz rotation speed.

    For the motor-rotation detection experiments with the IC-based sensors, the amplitude spectra from the full measurement duration had sufficient SNR to resolve the rotation speed (Fig.~5a,d and Supplementary Fig.~17). To check whether the time-resolved peak frequencies from electrical data could provide information on signal variability, we performed similar sliding, flattop window spectral analysis. To determine the smallest window that produces clearly identifiable spectral peaks, we increased it from 5~s to full duration ($\approx 50$~s) in steps of 10~s.  A clear peak was observed only for the sensor with the motor rotating at 2.2~Hz, and only with a minimum window length of 40~s. To account for the control background characterised by $1/f$ noise,  \textit{t}-score spectrum was calculated in each window. Accordingly, the time-resolved frequencies in Fig.~5d were obtained from the peak of the \textit{t}-score spectrum calculated in sliding windows of 40~s with a time step of 0.5~s.

    The mean filtered data in Fig.~5f was obtained by calculating the mean of the DC-corrected IC sensor output, V$_{\mathrm{real}}$, in  sliding windows of 0.2~s. This procedure acts as a low-pass filter with $f_{\mathrm{cutoff}} \sim 5$~Hz. DC corrections were applied by identifying piecewise constant offsets and linear drifts in the signal (see Supplementary Fig.~21).  
    The mean-filtered data were then manually segmented into time intervals for fitting, which were chosen so that the signal amplitude in each segment was approximately constant. The peak frequencies observed in Fig.~5d in a narrow window around 0.33 and 0.86~Hz were included in the fit. For each segment, a least-squares fit of the sum of two sinusoids with frequencies varied in a window around 0.33~Hz (0.29--0.37~Hz) and 0.86~Hz (0.82--0.90~Hz) was computed. The fitting procedure solved for the amplitude, phase, and frequency of the two sinusoids within each segment. In Supplementary Fig.~16, fits using fixed sliding windows and fixed frequencies of 0.33~Hz and 0.86~Hz were evaluated by performing a least-squares fit of the sum of the two sinusoids in each window, estimating their amplitudes and phases.
    
    In Fig.~7, the signal at the frequency of motor rotation was obtained as follows. First, the time-domain electrical signal output was corrected for DC offsets as shown in Supplementary Fig.~21. Synchronous detection of the signal component at the rotation frequency, $S = A\sin(2\pi f_\mathrm{R} t+\phi)$, was then performed in sliding windows of duration $w > 1/f_\mathrm{R}$ (here, $w = 5$~s) with a time step $\Delta t < w/10$ (here, $\Delta t = 0.1$~s). Within each window containing $N$ points, the in-phase ($X_\mathrm{sine}$) and quadrature ($X_\mathrm{cosine}$) components of $S$ at $f_\mathrm{R}$ were calculated as:
    \begin{equation}
        X_\mathrm{sine} = \frac{2}{N}\sum S * \sin(2\pi f_\mathrm{R} t)
    \end{equation}
    \begin{equation}
        X_\mathrm{cosine} = \frac{2}{N}\sum S * \cos(2\pi f_\mathrm{R} t) 
    \end{equation}
    The amplitude and phase of the signal at $f_\mathrm{R}$ in each of these windows were then calculated as:
    \begin{equation}
        A = \sqrt{X_\mathrm{sine}^2 + X_\mathrm{cosine}^2}
    \end{equation}
    \begin{equation}
        \phi =\tan^{-1}\left(\frac{X_\mathrm{sine}}{X_\mathrm{cosine}}\right)
    \end{equation}

    \section*{Author Contributions}
    T.P., M.M., T.Z., A.J., So.S. and J.F.  designed the research. T.Z. fabricated the cleanroom-based sensors and performed experiments using them, with assistance from M.L, B.R. and T.P. T.Z. and A.J. analysed the data obtained with the cleanroom-based sensor. St.S. and M.M. designed the IC. So.S., A.J., M.M., and T.P. integrated the IC with the PCB electronics and designed the data acquisition protocols with this sensor. A.J., M.M., So. S. and T.P. designed the IC-based sensor experimental setup, including the microfluidic channel. A.J. and T.P. designed the bacterial-bead solution protocol. A.J. performed experiments with the IC-based sensor and A.J. and T.P. analysed this data. A.J. and T.P. wrote the manuscript. All authors commented on the manuscript.
    
    \section*{Competing Interests Statement} Amritha Janardanan, Teuta Pilizota and James Flewellen are founders of the company K\=ahu SiliconBio Ltd that aims to develop biosenosors for a range of analytes in liquid based on the technology described in the publication. Amritha Janardanan, Teuta Pilizota, James Flewellen, Soner Sonmezoglu and Michel Maharbiz are inventors on the patent related to the technology: WO2025153807. Stefano Sonedda and Tom Zajdel are contributors on the same patent.    
    
    \section*{Acknowledgements}
    This work was supported by the Engineering and Physical Sciences Research Council (EPSRC) established career fellowship to T.P. (EP/V03264X/1 and EP/V03264X/2), and Office of Naval Research Global X Award to T.P. and M.M. (GRANT12420502). Fabrication of microfluidic device masters and some IC post-processing was carried out at the Scottish Microelectronics Centre (SMC). We thank Peter Lomax, Graham Wood, and Camelia Dunare at SMC, as well as Nadanai Laohakunakorn, for their advice and assistance with these processes. FIB–SEM was performed at the University of Edinburgh Cryo FIB-SEM Facility. We thank Fraser Laidlaw for performing this. We also thank all members of the Pilizota Lab for helpful discussions and feedback. 
	
\newpage

\bibliographystyle{unsrtnat}
\bibliography{refs}

\appendix
\setcounter{figure}{0}
\renewcommand{\thefigure}{\arabic{figure}}
\captionsetup[figure]{labelfont=bf,labelsep=period,name={Supplementary Fig.}}

    \onecolumn
    
    \begin{center}
    {\LARGE \textbf{Supplementary Information}}
    \end{center}
    
    \vspace{1em}
    
    \section*{Supplementary Text}

    \subsection*{Calculation of impedance spectrum for a representative electrode sensor pair} The impedance $Z$ between the electrodes  of a two-electrode sensor was calculated for each input frequency $f$, between 1~kHz and 10~GHz, with 100~Hz resolution. For electrodes as described in Fig.~1b, each of the circuit elements were evaluated as: 
    \begin{equation}
        R_\mathrm{sol} = \frac{\rho k_\mathrm{cell}}{l}
    \label{eq:Rsol}
    \end{equation}
     \begin{equation}
          C_\mathrm{sol} = \frac{\epsilon_\mathrm{r}\epsilon_\mathrm{0}l}{k_\mathrm{cell}}
    \label{eq:Csol}
    \end{equation}
    \begin{equation}
          C_\mathrm{DL} \approx \frac{\epsilon_\mathrm{r}\epsilon_\mathrm{0}A}{\lambda_\mathrm{D}}
    \label{eq:CDL}
    \end{equation}
  
    where $\rho$ is the resistivity of the solution, $\epsilon_{r}$ and $\epsilon_{0}$ are the relative and absolute permittivities of the solution and vacuum, respectively, and $l$ and $A$ are the length and area of the electrodes. Here, $k_\mathrm{cell}$ is a dimensionless 2D proportionality constant between the resistance and resistivity of the cell, and $\lambda_\mathrm{D}$ is the Debye length. To facilitate computation of $k_\mathrm{cell}$ for coplanar electrodes, we follow conformal mapping explained in Ref.~\cite{Linderholm2005}. Thus, the domain comprising the coplanar electrodes and its curved electric field is mapped into a coordinate system where they become parallel electrodes with straight field lines.
    
    In Fig.~1c, the spectrum of a representative two-electrode sensor with $l \times w - d = (10 \times 0.75 - 4)\,\mu\mathrm{m}$ was calculated. The solution above and between the electrodes was assumed to be 10~mM potassium phosphate buffer with $\rho$ = 6.6~$\Omega$m, $\epsilon_\mathrm{r}$ = 78.4 and $\lambda_\mathrm{D}$ = 0.25~nm. A MATLAB program provided by Linderholm (\cite{Linderholm2005}) was used to obtain $k_\mathrm{cell} \approx$ 2.05 for the chosen electrode width, interelectrode spacing and channel height (assumed to be 100~$\mu$m).
    $Z$ was then calculated, based on the model in Fig.~1b, as
    \begin{equation}
          Z = x + \frac{2}{i\,2\pi f C_\mathrm{DL}}
    \label{eq:Z}
    \end{equation}
    where,
     \begin{equation}
          x = \frac{R_\mathrm{sol}}{i\,2\pi f C_\mathrm{sol}R_\mathrm{sol}+1}.
    \label{eq:x}
    \end{equation}
    The dashed vertical lines in Fig.~1c correspond to the characteristic transition frequencies calculated as $f_\mathrm{low} = 2/(2\pi R_\mathrm{sol}C_\mathrm{DL})$ and $f_\mathrm{high} = 1/(2\pi R_\mathrm{sol}C_\mathrm{sol})$.
    \subsection*{Operation of cleanroom fabricated microelectrodes at 10~kHz was achieved using four-point measurements and conducting polymer coating}
    \label{note:PCB_4ppointelectrode}
    The equivalent circuit model of the PCB-sensor-solution interface is provided in Supplementary Fig.~\ref{fig:SI_PCBSchematic_4electrode}b. The amplifier on the PCB provides a large input impedance of 1~G$\Omega$ to the sense electrodes, and feeds the difference in voltage between them to the analyser \cite{Linderholm2006}. The parasitic capacitances between circuit traces from the sense and current-injection electrodes on the PCB are $C_\mathrm{p}\sim$10~pF, while the input
    capacitance of the amplifier is $C_\mathrm{in}=$10~pF.  
    In order to accurately measure $R_\mathrm{sol}$ between the sense electrodes, the capacitive impedances, $|Z_{C_\mathrm{in}}|$ and $|Z_{C_\mathrm{p}}|$ must be significantly larger than $R_\mathrm{sol}$ (typically $\gtrsim 10\,R_\mathrm{sol}$). In Motility Buffer, $R_\mathrm{sol} \sim 20~\mathrm{k}\Omega$ (see \textit{Methods}), which sets an upper limit on the operating frequency of $f_\mathrm{max} \approx \frac{1}{2\pi \cdot 10 R_\mathrm{sol} C_\mathrm{p}} \approx 80~\mathrm{kHz}$.
    From Fig.~1b, this is a frequency below which $C_\mathrm{DL}$ dominates impedance between the two-electrode sensor. 

    To be able to measure solution impedance changes below this frequency, a four-point measurement that separates the current-injection and voltage sensing functions was needed. Further, to reduce $C_\mathrm{DL}$, a conducting polymer poly-3,4-ethylenedioxythiophene:poly-styrene sulfonate (PEDOT:PSS) was electro-polymerized onto the current-injection electrodes. It forms a porous, high–effective-surface-area coating. The electro-polymerization solution was prepared by dissolving 10~mM of the monomer 3,4-ethylenedioxythiophene (EDOT) in deionized water with PSS added at a 1.5:1 mass ratio. A droplet of this solution was deposited onto each electrode, and a current of 50~nA applied between it and a tungsten counter electrode for 30~s (using a Keithley 2400 SourceMeter). The potential varied from 0.8 to 1.2~V in the 30~s period. Each 100~$\mu$m × 100~$\mu$µm current-injection electrode was coated individually. 
    
    Before PEDOT:PSS deposition, the impedance spectrum measured between the sense electrodes in 10~mM KCl solution exhibited a fairly constant phase angle of $-70^\circ$ to $-90^\circ$ between 1 and 100~kHz input frequencies, indicating dominance of the interface impedances. After deposition the interfacial impedance was effectively removed, the electrode $R_\mathrm{sol}$ was $\sim$20~k$\Omega$ between 5 to 50~kHz, with the phase angle reduced to a range from $20^\circ$ to $0^\circ$, Supplementary Fig.~2. 

    \subsection*{Bead distance from the four-electrode sensor influences the detected signal magnitude.}   
    Peak heights in $\Delta F$ correlate with the corresponding bead positions in Supplementary Fig.~4 (blue arrows). Reductions in $\Delta F$ peak magnitude occur when the bead only partially overlaps the sensing area and/or when the bead height above the sense electrodes changes. Variations in bead height are reflected as changes in the apparent bead radius in the optical images due to defocus. The bead height is estimated to be within 10~$\mu$m, i.e. within the limit of our long-working distance objective (see \textit{Methods}). The corresponding $\Delta Z$ peaks follow a similar magnitude trend to $\Delta F$, but are broader, suggesting that the impedance sensing region extends beyond the area between the sense electrodes along the flow path.
    
  Small-magnitude $\Delta Z$ peaks (red arrows) could be caused by one of the following: (i) the bead passing over the nitride insulation covering the sense electrodes (6, 8 and 13), which does not fully suppress the electrical response, (ii) the beads are not too far in $z$ direction to give a weak signal despite not being visible in the field of view (1), and similarly (iii) beads are outside of the designated sensing region in $xy$ direction but still produce a weak signal (14 and 16). 

    \subsection*{Impedance measurements on cleanroom fabricated four-electrode sensors}
    The sampling rate of the impedance analyser varied between 15-60~Hz from one sampled point to another. We suspect several causes, e.g. the Serial-GPIB adapter that ran through a MATLAB Toolbox sending SCPI commands. Given the expected rotation rate of the 2~$\mu$m bead and 60~Hz Mains, such varied sampling rate can cause aliasing of the signal. The solution was to either identify the exact cause(s) or introduce a low-pass anti-aliasing filter. However, we believed that this alone would not have enabled motor rotation detection, because from the signals we obtained with below described considerations taken into account, we observed little to no correlation between optical signal and measured $\Delta Z$. This is why we did not pursue it, and instead moved to the IC approach.

    For the analysis we selected a bead that rotated at 4.95$\pm$1.66~Hz (based on optical detection), which ensured the 15-60~Hz analyser sampling rate did not alias it. Next, we simulated the 60~Hz Mains signal as a 60~Hz sinusoid with arbitrary amplitude and phase and sampled at the same sampling intervals as the impedance data. To check for aliased frequencies from 60~Hz Mains interfering in the measured impedance, both signals were resampled at a constant rate of 33~Hz to obtain their amplitude spectra over the full 60~s measurements. Supplementary Fig.~\ref{fig:SI_4pointelectrode_fullsepctra} shows strong coincident peaks at 8.5-9.5~Hz and smaller ones at 12-13, 4-4.5 and between 1-2.5~Hz, pointing to the aliasing of Mains frequency. However, no stronger peak was observed at the rotation speed of 4.95$\pm$1.66~Hz.
    \label{note:aliasedf}

    \subsection*{Electronic circuit on IC}
    \label{note:ICdesign}
    The signal processing electronic circuits on the  ASIC is shown in Supplementary Fig.~7. The differential current flowing into the central electrode of the sensor is fed to a current pre-amplifier followed by a transimpedance amplifier, that amplify and transform the current to a voltage difference. The current pre-amplifier uses matched NMOS-PMOS transistor pairs in the feedback loop and output of an opamp, such that all transistors have the same operating point \cite{Ferrari2009}. The output  consists of $N$ copies of the transistor-pair in the feedback loop, stacked in parallel, so the output current is amplified by a factor $N$ irrespective of the current flowing in and any geometrical dependence of the transistor parameters. Two stages of these have been cascaded. The overall gain of the current amplifier can be set by digital control signals ranging from 40 to 62~dB; the option to bypass the entire current preamplifier has also been included. 
    
    The next component is a two-stage transimpedance amplifier. The architecture shown in Supplementary Fig.~7 keeps the power requirement to a minimum while providing a high gain at frequencies in the order of 1-10~MHz \cite{Ciofi2006}. The gain of this amplifier is $\frac{C2}{C1}R$. The value of R has been made programmable by digital logic, for a transimpedance gain of 120 to 132~dB.

    The differential current flowing into the middle electrode is a sinusoid at the operating frequency ($f$), amplitude modulated by the rotating signal of frequency ($f_R$). The demodulation circuit to extract the rotation signal is implemented as a switching multiplier followed by a low pass filter. To obtain the real and imaginary part of the impedance, $\Delta V$ signal is split into 2 paths, one which is multiplied by a square wave in phase with the input signal  and the other by a signal at $90^\circ$ to it. The square wave signal drives the transmission gate switches of the operational amplifier circuit to change its configuration from voltage buffer (gain=1) to inverting amplifier (gain=-1). The low pass filter is implemented using $G_m$-C architecture where an operational transconductance amplifier (OTA) and capacitors $C_1$ and $C_2$ determine the cut-off frequency $\omega_t = \frac{G_m}{\sqrt{C_1 C_2}} = 14.6~\mathrm{kHz}$.

    \subsection*{Calculation of impedance spectrum of the IC-based sensors with passivation}
    
    In the equivalent circuit model, the passivation layer changes the impedance between the two electrodes, $C_\mathrm{pas}$ in Fig.~3d. The circuit elements $R_\mathrm{sol}$ and $C_\mathrm{sol}$ remain as evaluated as in Supplementary Eq.~\ref{eq:Rsol} and \ref{eq:Csol}, whereas $C_\mathrm{pas}$ is given as:
    \begin{equation}
          \frac{1}{C_\mathrm{pas}} = \frac{1}{C_\mathrm{SiO_2}} + \frac{1}{C_\mathrm{Si_3N_4}}
    \label{eq:Cpas}
    \end{equation}
    where
    \begin{equation}
          C_\mathrm{SiO_2/Si_3N_4} = \frac{\epsilon_\mathrm{r-SiO_2/Si_3N_4}\epsilon_{0}A}{d_\mathrm{SiO_2/Si_3N_4}}
    \label{eq:Coxide/nitride}
    \end{equation}
    $d_\mathrm{SiO_2}$ and $d_\mathrm{Si_3N_4}$ are the thicknesses of SiO$_2$ and Si$_3$N$_4$ over the centre of the electrode, and  $\epsilon_\mathrm{r-SiO_2}$= 3.9 and $\epsilon_\mathrm{r-Si_3N_4}$ = 7 the relative dielectric constants of the two materials. Here, $C_\mathrm{DL}$ in Fig.~1b is in series with $C_\mathrm{pas}$ and neglected in the model because $C_\mathrm{pas} \ll C_\mathrm{DL}$. For example, Supplementary Fig.~9a evaluates the impedance spectrum for the $l \times w - d = (10 \times 0.75 - 4)\,\mu$m  electrode pair using Supplementary Eq.~\ref{eq:Z} with $C_\mathrm{DL}$ replaced by $C_\mathrm{pas}$ and $d_\mathrm{SiO_2/Si_3N_4}$ from Supplementary Fig.~8b. This was done for 10~mM phosphate buffer with $\rho$ = 6.6~$\Omega$m, $\epsilon_\mathrm{r}$ = 78.4, and for 0.13~M NaCl added to the buffer with $\rho$ = 0.69~$\Omega$m, $\epsilon_\mathrm{r}$ = 75.
    As before $k_{cell} = 2.05$ was obtained using the MATLAB program provided by Linderholm \cite{Linderholm2005} for this electrode configuration and channel height 2~mm. The result shows negligible variation over this salt concentration range. The electrode–electrolyte coupling is dominated by the passivation capacitance. At operating frequency of 10~MHz, the ratio of conductive to capacitive (displacement) current in the solution is given by $\sigma/(\omega \epsilon)$ i.e., $1/(\rho)(\mathrm{2\pi} f\epsilon_\mathrm{r} \epsilon_\mathrm{0})$, which is $\sim 3$ for 10~mM phosphate buffer and $\sim 30$ with 0.13~M NaCl added to the buffer. Thus, the current transport in the solution is predominantly resistive in the presence of 0.13~M salt, and has some reactive contribution without salt. Consequently, perturbations to the conductive and dielectric properties of the solution, such as displacement of electrolyte by the rotating marker, lead to changes in the effective impedance. These changes are transduced through the capacitive interface and detected as variations in the measured current, Supplementary Fig.~9b. 

    \subsection*{Analysing effect of blue light exposure on marker rotation speed in optical and electrical measurements}
    On exposure to light, the motor speed slows down exponentially, with the exponential factor increasing linearly with incident light power \cite{Krasnopeeva2019}. The intensity of the blue light we used, 450-490~nm, incident on the IC was measured to be between 105 and 157~mW/cm$^2$ (see \textit{Methods}). If we use the results from \cite{Krasnopeeva2019}, the average power intensity of 131~mW/cm$^2$ gives an exponential factor of 0.00131/s.
    For the signal in Fig.~5a-c, where the optical measurement precedes the electrical, and the bacteria were exposed to blue light for 48~s in between, the observed speed of (2.3$\pm$1.08)~Hz would reduce to $2.3\pm1.08 \times e^{(-0.00131\times48)} = (2.12\pm1.01)$~Hz, which matches the reduced speed of 2.1~Hz observed from the electrical measurements.
    For the signal in Fig.5d-f, where the electrical measurement precedes the optical, and the bacteria is exposed to blue light for $\sim$1.5~min in between, the electrically detected speed of rotation of 0.33~Hz would reduce to $0.33 \times e^{(-0.00131\times90)} = 0.29$~Hz, which again matches the reduced speeds of (0.27$\pm$0.14)~Hz observed in the optical recording. Supplementary Fig.~4 depicts the specific time intervals of these two experiments.

    \subsection*{SU-8 master fabrication}
    To fabricate the master for the IC micrfoludic channel we used $\sim$100~$\mu$m of SU-8 (SU-8 3035, MicroChem/Kayaku Advanced Materials, USA) to coat the 4-inch single-side polished silicon substrate (Inseto Ltd., UK). The wafer was stripped of its native oxide using 4:1 buffered hydrofluoric acid (BHF) for 10~s and further cleaned in an oxygen plasma barrel asher for 15~min. The resist was spin coated and baked sequentially in three layers of $\sim$33~$\mu$m, as processing thicker layers at once produced cracks. Each layer was spin-coated using a three-step program: 500~rpm for 10~s (spread), 3000~rpm for 35~s (thinning), followed by a controlled deceleration to 0~rpm over 30~s, all with acceleration 100~rpm/s. Spin coating of each layer was followed by a soft bake at $65~^\circ\mathrm{C}$ for 3~min and $95~^\circ\mathrm{C}$ for 11~min, using spacer pins for the first 1~min each to reduce the initial heating rate. All three layers were exposed after completion of the multilayer stack using a maskless lithography tool (direct write, ML3, Durham Magneto Optics Ltd, UK).
    
    The designs for the two layers shown in Supplementary Fig.~20 were made with slightly varying device and channel widths to account for variations in alignment and cutting accuracy. Designs were created in CAD software (CleWin) and exported as layout files onto the lithography tool's software for further edits. To identify exposure parameters yielding optimal dimensional accuracy a dose test was performed by exposing the resist to a dose of 1300~mJ/cm$^2$ with dose factor 0.6 to 1.5, in steps of 0.1, and a focus shift between -20 to 10, in steps of 10. Following exposure, the wafer underwent a post-exposure bake at $65~^\circ\mathrm{C}$ for 17~h to minimise internal stress and prevent cracking in thick films. It was then developed in Propylene glycol monomethyl ether acetate (PGMEA) for 6~min, rinsed in isopropanol, and dried under nitrogen before inspection by optical microscopy for defects (e.g., cracks, incomplete development). Next, the wafer was hard-baked at 110$\degree$C for 15~min, and an SEM was used to evaluate the feature sizes. This procedure was repeated with the optimised exposure parameters and final designs on a fresh substrate. A profilometer was used to confirm a feature height of $\sim$100~$\mu$m. Finally, the wafer was placed in a vacuum desiccator with a vial of trichloro(1H,1H,2H,2H-perfluorooctyl)silane for 3~h. The silanization allows for easy removal of PDMS moulds from the master.
    \label{note:SU8master}

    \subsection*{Log-frequency smoothing of P(f) in calculating \textit{t}-score and SBR}
    In the amplitude spectra of IC-based sensor output signals and controls in Fig.~5a and d, and Supplementary Fig.~16, an apparent increase in deviation of the signal spectra from controls is observed with increasing frequency. This arises from a reduction in the variance of the control mean $\mu$ and standard deviation $\sigma$ at higher frequencies due to averaging, which can spuriously inflate \textit{t}-scores and SBR. To correct for this, we equalise variance across frequencies by applying log-frequency smoothing to both signal and control spectra \cite{Mann1996,Donoghue2020}, as described in \textit{Methods}.
    Frequencies are re-binned into bands with constant relative bandwidth (logarithmic spacing), whose widths increase geometrically by a factor of $2^{(1/k)}$ (i.e., $1/k$-octave spacing), and the average power is computed within each band. This ensures smoothing over proportionally similar frequency ranges, stabilising variance estimates and reducing high-frequency noise. Larger $k$ improves variance equalisation at the cost of frequency resolution; we chose $k = 1/12$, corresponding to a resolution of $0.06 \times f_\mathrm{centre}$. Log-frequency smoothing was applied to the signal and and each control power spectrum to obtain $\tilde{P}\mathrm{S}(f)$, $\tilde{\overline{P}}\mathrm{C}(f)$, and $\tilde{\sigma}_\mathrm{P,C}(f)$, which were used for subsequent \textit{t}-score and SBR calculations.
    \label{note:Logfsmoothing_Tscore}

   \section*{Supplementary Figures}
    \addcontentsline{toc}{section}{Supplementary Figures}

    \begin{figure}[htp!]
    \centering    \includegraphics[width=1\textwidth]{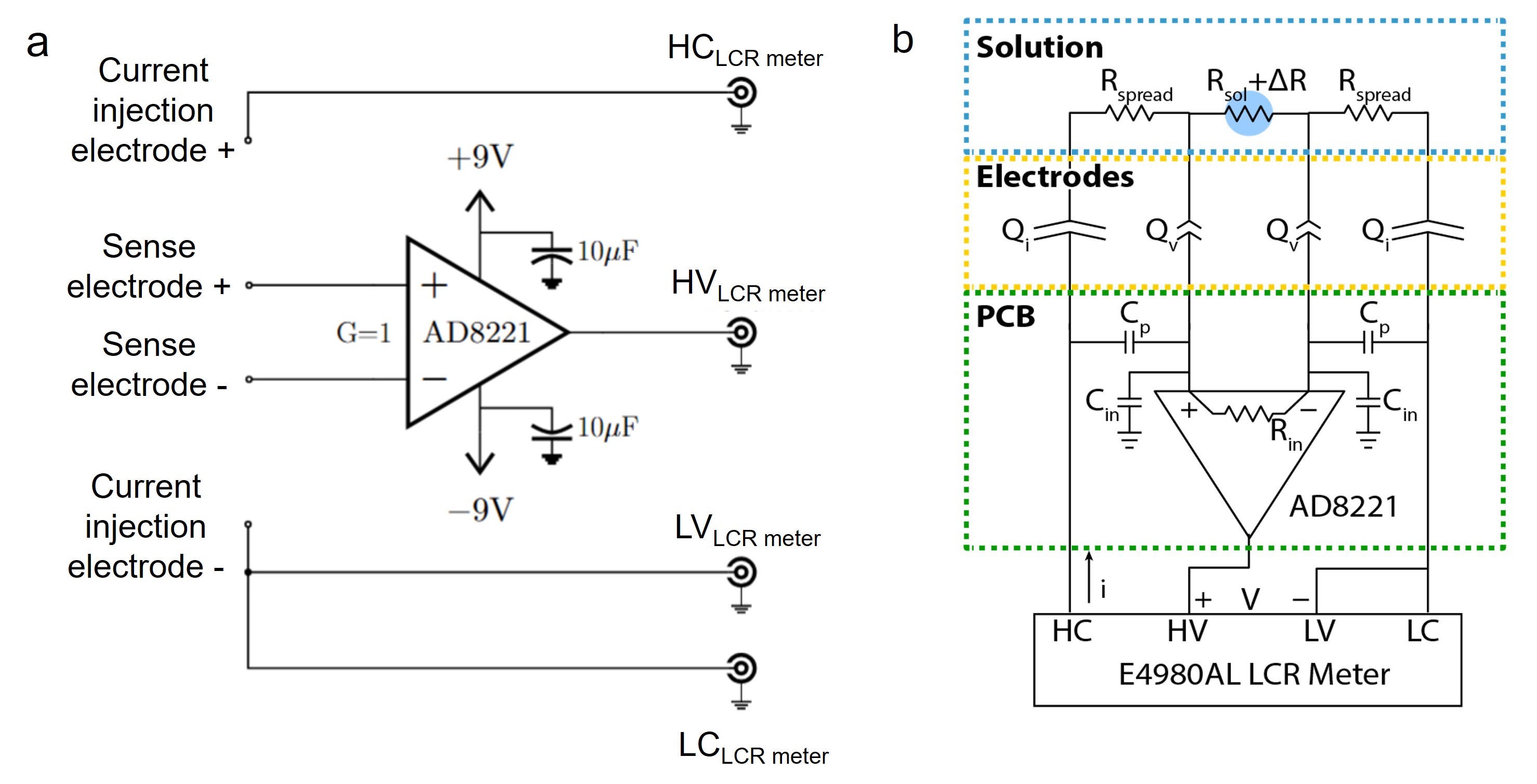} 
    \caption{\textbf{PCB schematic and equivalent circuit model of cleanroom fabricated sensor. a.} The PCB schematic shows the pin connections to the current-injection and sense electrodes. The sense electrodes are connected to an instrumentation amplifier (AD8221) operated at unity gain. The amplifier transfers the voltage difference between the sense electrodes to the coaxial output port, which connects to the LCR meter (impedance analyzer) input port that measures this voltage (HV). The current-injection electrodes are directly connected to coaxial ports linked to the high and low current-injection terminals of the LCR meter (HC and LC, respectively).
    \textbf{b.} Equivalent circuit model of the four-point impedance measurements. In solution, the impedances between the four electrodes are shown. Here, $R_\mathrm{spread}$ is found between the current-injection and sense electrodes, and $R_\mathrm{sol}$ between the two sense electrodes. The rotating marker causes an incremental change $\Delta R$ in $R_\mathrm{sol}$. 
    The electrode interfacial impedances ($Q_\mathrm{i}$ and $Q_\mathrm{v}$) are dominated by the double-layer capacitances at 10~kHz. The PCB interface comprising the circuit traces and the instrumentation amplifier (AD8221) contribute to the inter-trace parasitic capacitance $C_p$, and amplifier input capacitance $C_\mathrm{in}$, respectively. The impedance analyser (LCR meter E4980AL) applies a constant current $i$ on the current-injection electrodes and measures the voltage difference $V$ between the sense electrodes.}  \label{fig:SI_PCBSchematic_4electrode}
    \end{figure}
    \addcontentsline{toc}{subsection}{Supplementary Fig.~\thefigure: PCB schematic and equivalent circuit model of cleanroom fabricated sensor.}

     \begin{figure}[htp!]
    \centering    \includegraphics[width=1\textwidth]{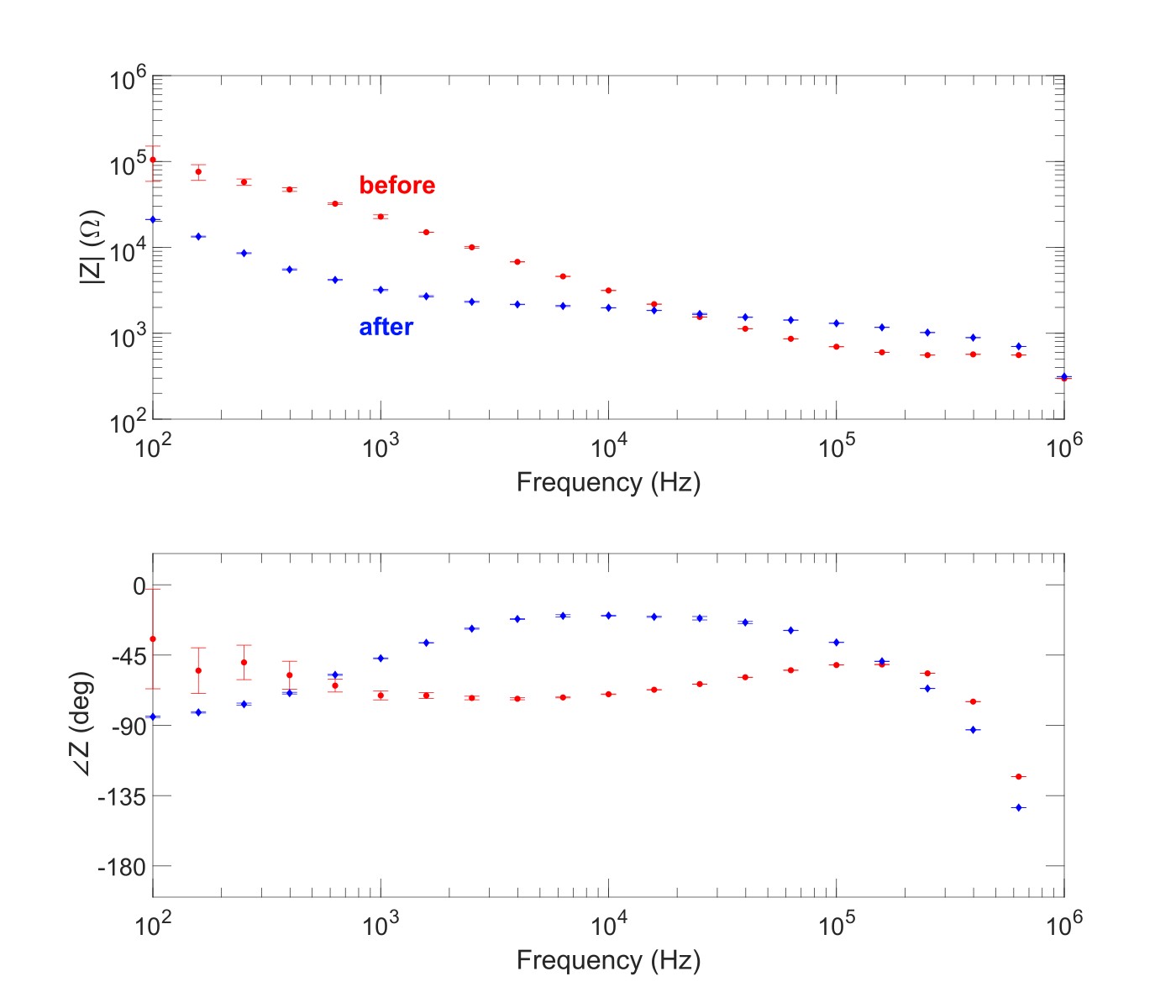} 
    \caption{
    \textbf{Impedance spectrum of a representative four-electrode array before and after PEDOT:PSS deposition.}  
    Magnitude (top) and phase angle (bottom) of the impedance spectrum for a four-electrode sensor immersed in 10~mM KCl before (red
    circles) and after (blue diamonds) PEDOT:PSS deposition. Error bars mark standard deviation across 4 measurements of impedance of a specific array. 
    }
    \label{fig:SI_PEDOTPSS_Zspectrum}
    \end{figure}
    \addcontentsline{toc}{subsection}{Supplementary Fig.~\thefigure: Impedance spectrum of four-electrode array before and after PEDOT:PSS deposition.}

    \begin{figure}[htp!]
    \centering
    \includegraphics[width=0.7\textwidth]{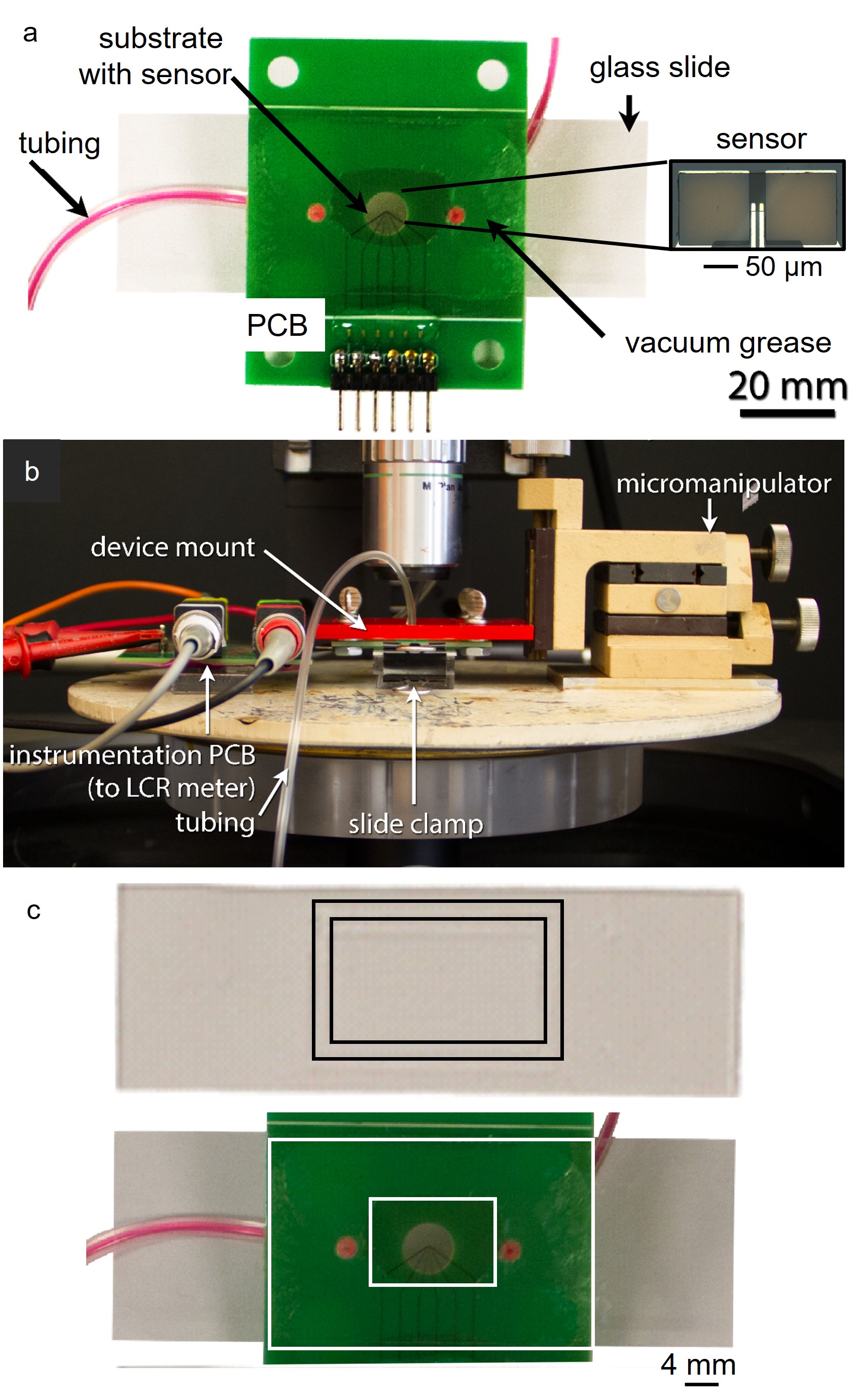} 
    \caption{
    \textbf{Cleanroom fabricated sensor device and measurement setup.} 
    \textbf{a.} PCB (green) with a circular hole at the centre that holds a transparent substrate with a four-electrode sensor (zoomed image) beneath it. A glass slide (white) forms a channel beneath the electrodes, separated by a vacuum-grease spacer that seals the channel while allowing relative motion between the two components. Tubing carrying red dye highlights fluid flow into and out of the channel. 
    \textbf{b.} Measurement setup was on a probe station microscope. The device mount (red) held the PCB shown in (a) This PCB connects to an instrumentation PCB described in Supplementary Fig.~1, which in turn connects to the impedance analyser (LCR meter). The glass slide on the underside of the device is secured to the base using a slide clamp. The micromanipulator connects to the mount, and helps position the PCB (with electrodes) with respect to the glass slide (with immobilised bacteria). The electrodes and bacteria were imaged through the objective from above. Tubing is connected to the device as described in (a). 
     \textbf{c.} Vacuum grease is deposited on the glass slide in a stencil-defined rectangular region shown in the top, with outer dimensions ($29 \times 18$)~mm and inner dimensions ($25 \times 14$)~mm. After compression by the PCB, the grease spreads to an approximately rectangular area of outer dimensions ($43 \times 25$)~mm and inner dimensions ($19 \times 12$)~mm as shown at the bottom. This change in footprint is used to estimate an average channel height of 40~$\mu$m (\textit{Methods}).
     }  
    \label{fig:SI_Setup_4electrode}
    \end{figure}
    \addcontentsline{toc}{subsection}{Supplementary Fig.~\thefigure: Cleanroom fabricated sensor device and measurement setup.}

    \begin{figure}[htp!]
    \centering
    \includegraphics[width=0.8\textwidth]{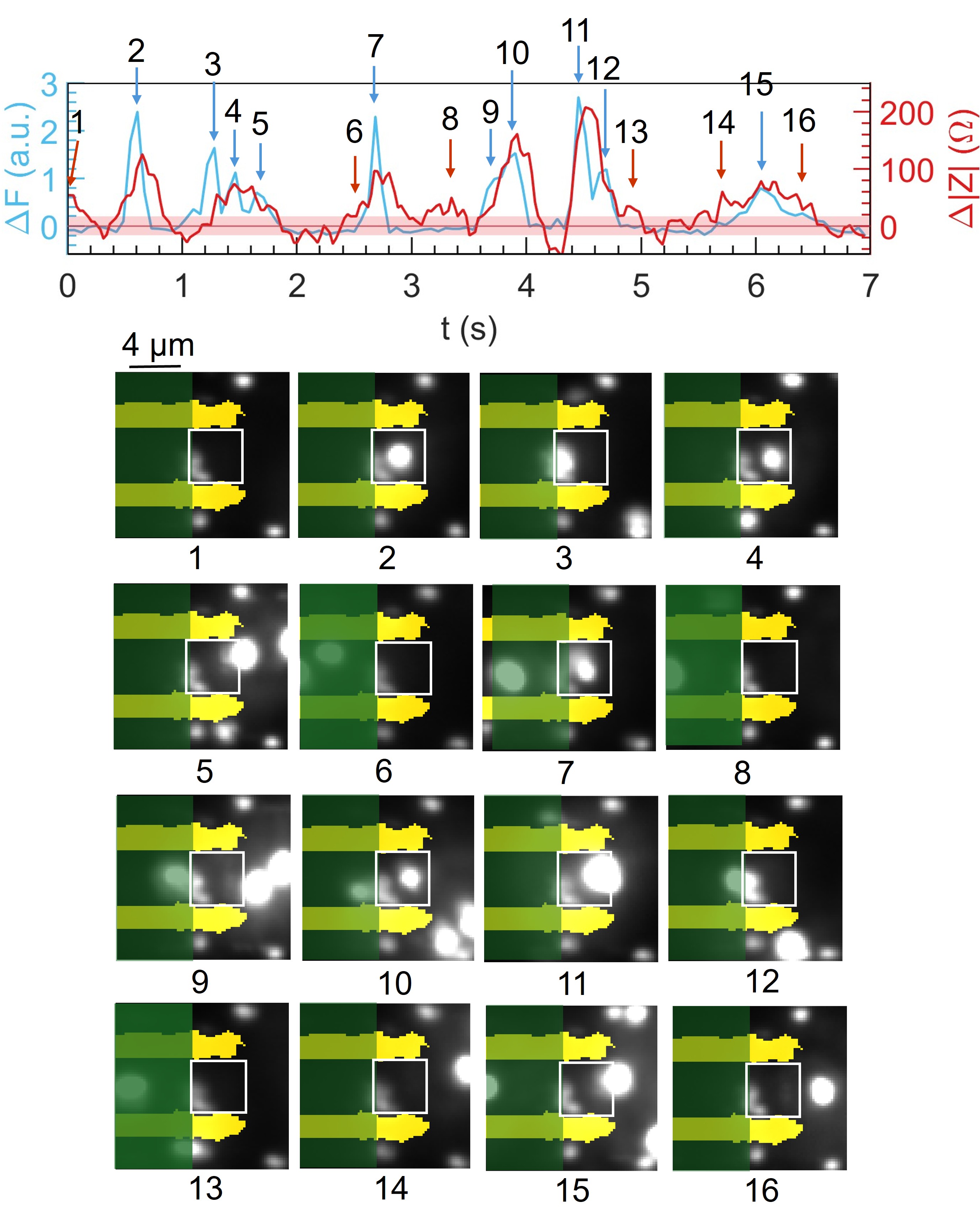} 
    \caption{
    \textbf{Bead positions relative to the four-electrode sensor and corresponding electrical signals.}
    The plot from Fig.~2c is reproduced for the time interval 0–7~s (top) and annotated with arrows and numbers corresponding to the optical images recorded simultaneously (bottom). The  baselines for $\Delta F$ and $\Delta Z$ (see \textit{Methods} for baseline calculation) are aligned and the red shaded area marks the standard deviation of the impedance data baseline. 
    The blue arrows mark peaks in $\Delta F$, with the corresponding bead positions shown in the optical images. 
    Red arrows indicate time instances where no clear $\Delta F$ peak is seen, but a small magnitude peak is seen in $\Delta Z$.  
    }
    \label{fig:SI_beadposition_detected}
    \end{figure}
     \addcontentsline{toc}{subsection}{Supplementary Fig.~\thefigure: Bead positions relative to the four-electrode sensor and corresponding electrical signals.}

     \begin{figure}[htp!]
    \centering
    \includegraphics[width=0.9\textwidth]{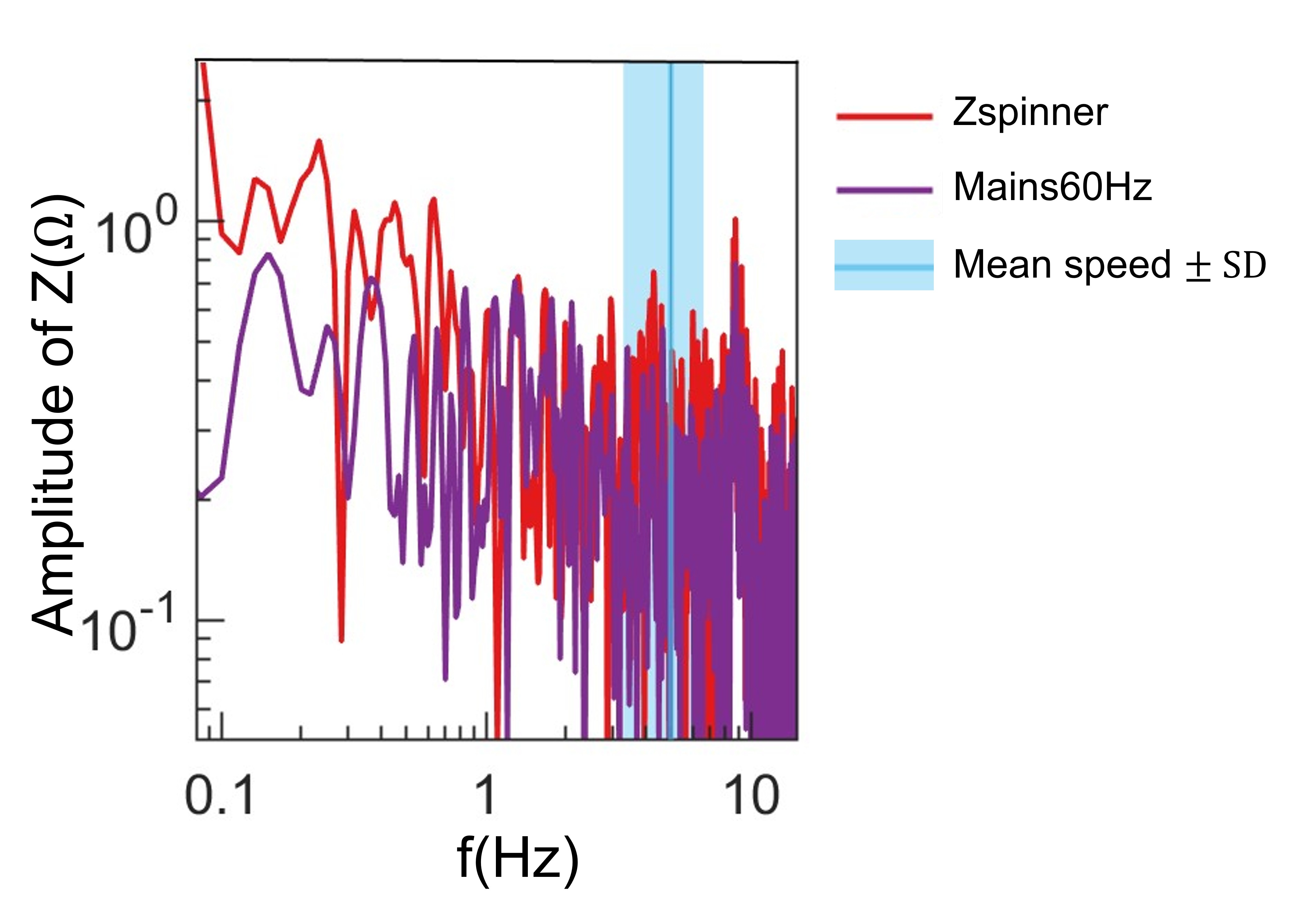}
    \caption{
    \textbf{Amplitude spectrum of the full 60~s impedance measurement from the cleanroom-fabricated sensor.}  
    Amplitude spectrum of the impedance data shown in Fig.~2d (Zspinner) for the full 60~s measurement, along with a simulated 60~Hz sinusoid sampled at the same time points (Mains60Hz). The vertical blue line and shaded region indicate the mean and standard deviation of simultaneously optically observed marker rotation speed.  
    See Supplementary text for analysis of aliased peaks and comparison with marker rotation frequency. 
     }    \label{fig:SI_4pointelectrode_fullsepctra}
    \end{figure}
     \addcontentsline{toc}{subsection}{Supplementary Fig.~\thefigure: Amplitude spectra of measured impedance from the cleanroom-fabricated sensor.}

    \begin{figure}[htp!]
    \centering
    \includegraphics[width=1\textwidth]{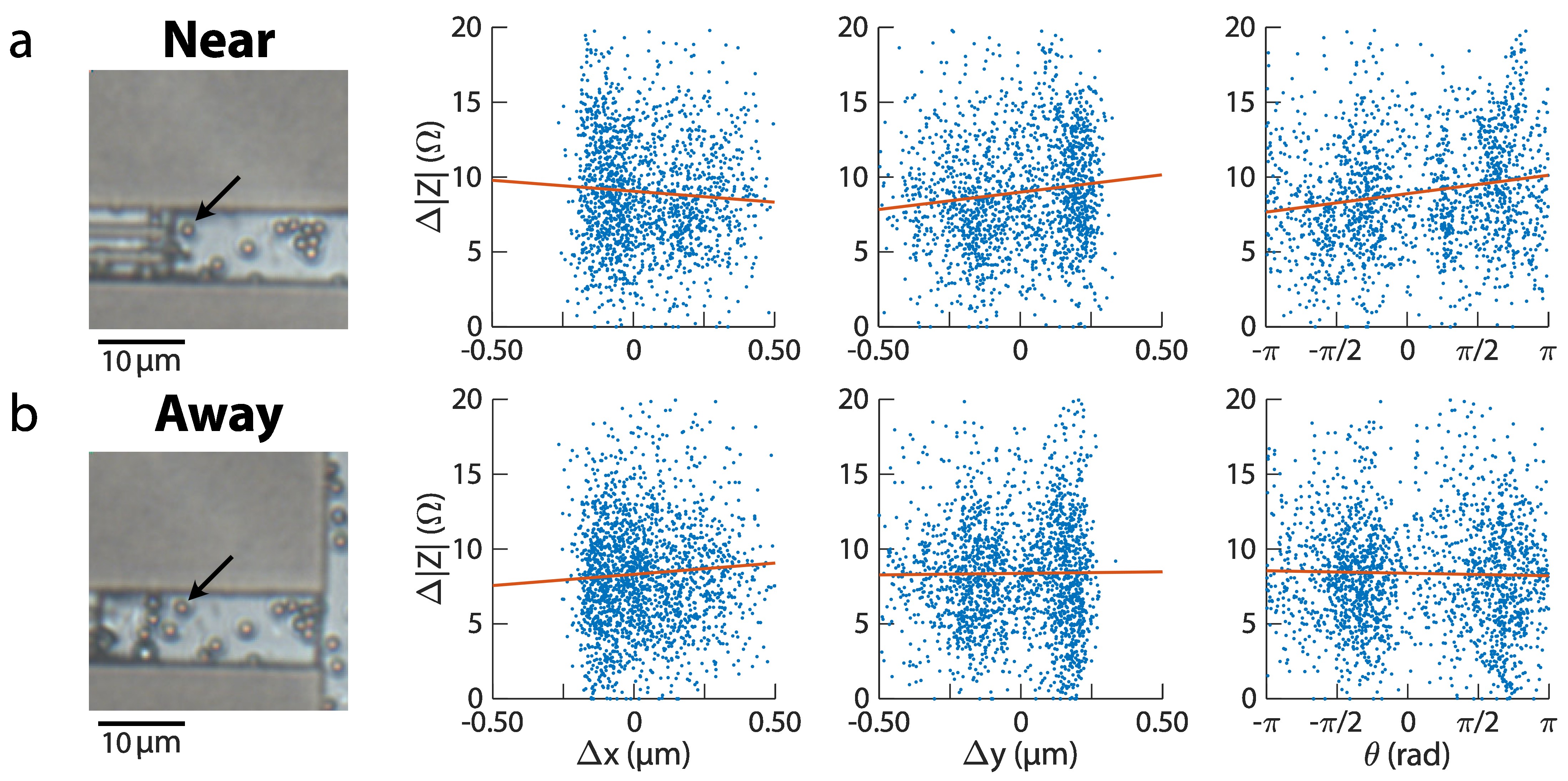} 
    \caption{
    \textbf{No correlation between the displacement of the rotating marker from its median and the change in impedance was observed.} Left: Optical images of electrodes with rotating markers indicated by arrows. 
    Right: Scatter plots comparing impedance changes with x-displacement ($\Delta x$) and y-displacement ($\Delta y$) from the median bead position, and rotation angle ($\theta = \tan^{-1}{\left(\Delta y/\Delta x\right)}$) for a marker near the sense electrodes, \textbf{a} and away from it, \textbf{b}. The red lines in the plots are linear least-squares fits to the data. Each individual dot represents the position of the bead in one individual frame, mapped to the change in impedance measured from the baseline at that time $\Delta Z$.
    }    \label{fig:SI_4pointelectrode_nearvsfar}
    \end{figure}
     \addcontentsline{toc}{subsection}{Supplementary Fig.~\thefigure: Weak correlation between rotating marker position and change in impedance.}

     \begin{figure}[htp!]
    \centering    \includegraphics[width=1\textwidth]{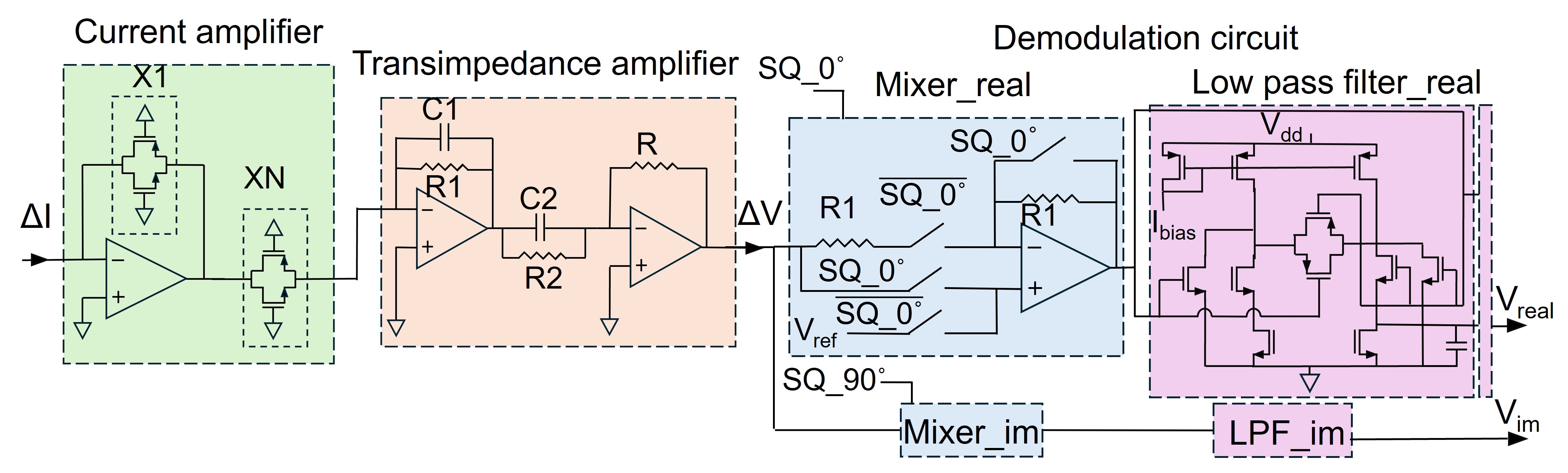} 
    \caption{
    \textbf{Electronic circuits that process the sensor output on the IC.}
    The current amplifier circuit consisted of an operational amplifier with matched NMOS-PMOS that amplifies the signal by factor N with very low noise. The transimpedance amplifier consisted of two operational-amplifier stages with $C_1R_1 = C_2R_2$ that convert current to voltage with an overall gain of $\frac{C2}{C1}R$.
    Next, the differential voltage was processed along two paths---one was multiplied by an in sync (phase and $f$) square wave, while the other was multiplied by a square wave of input frequency that is $90^{o}$ out of phase with it, both using a switching mixer circuit (blue). Each of these outputs is low-pass filtered (pink) to obtain voltages that represent the real and imaginary impedance, respectively. The low-pass filters along each path are implemented as two cascaded OTA–C stages (shown in pink boxes; the narrow pink box represents a repetition of the first stage). Each stage consists of a low-transconductance ($G_m$) operational transconductance amplifier (OTA) and a capacitor ($C$).
    }
    \label{fig:SI_IC_processingcircuits}
    \end{figure}
    \addcontentsline{toc}{subsection}{Supplementary Fig.~\thefigure: Electronic circuits processing sensor output on the IC.}

    \begin{figure}[htp!]
    \centering
    \includegraphics[width=1\textwidth]{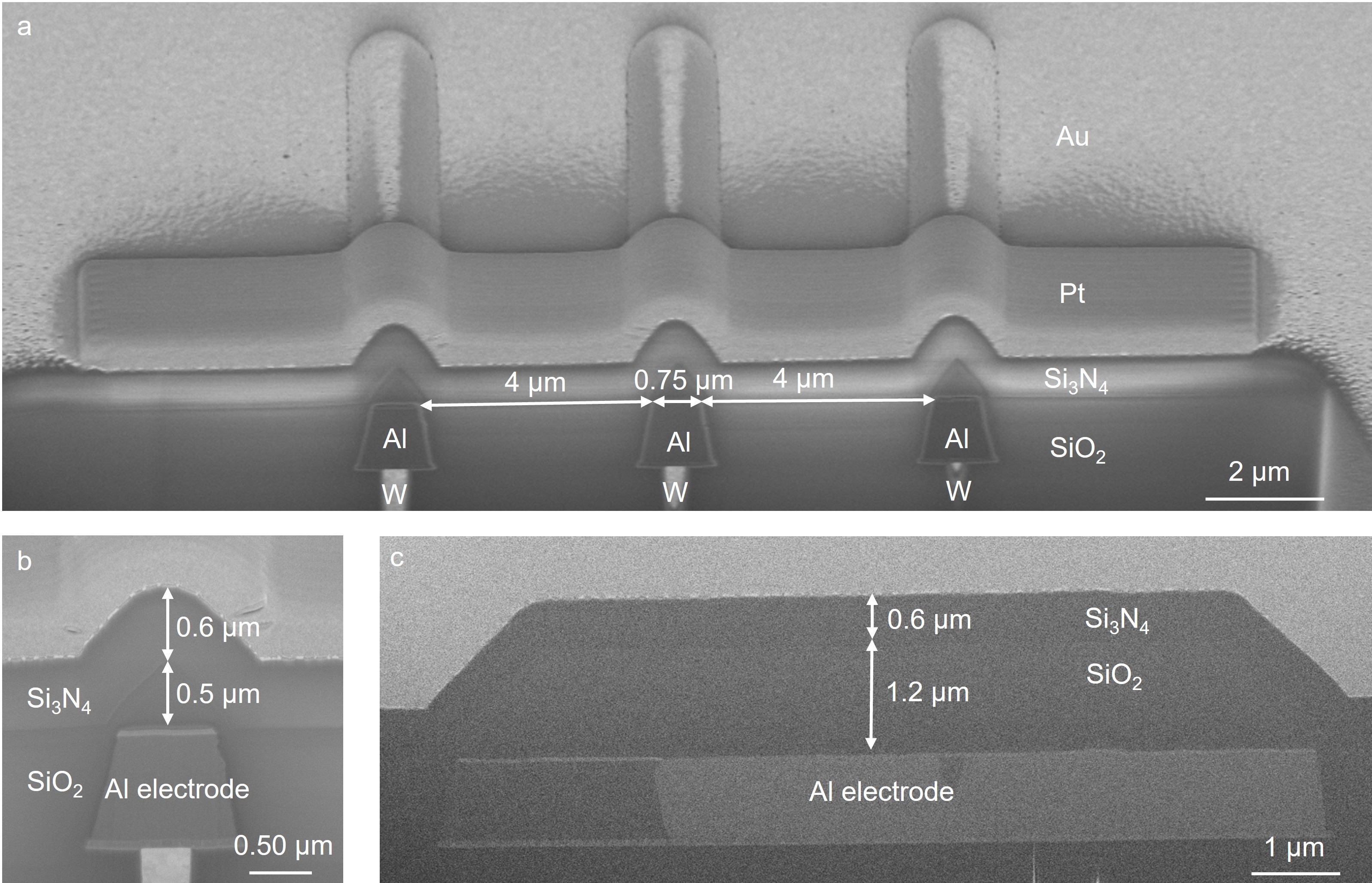} 
    \caption{
    \textbf{Near-surface cross-section of the IC.}
    \textbf{a.} Scanning electron microscopy (SEM) image of a focused ion beam (FIB)-milled near-surface cross-section through an electrode sensor with $l \times w - d = (10 \times 0.75 - 4)~\mu \mathrm{m}$ on the IC. The three electrodes of the sensor are visible as three ridged lines on the surface. A thin coating of gold (Au) and a narrow strip of platinum (Pt) deposited on the surface to reduce charging during imaging are also visible. The cross-section shows the three aluminium (Al) electrodes surrounded by silicon dioxide (SiO${_2}$) passivation, with additional SiO${_2}$ and silicon nitride (Si${_3}$N${_4}$) layers above the electrodes, forming tapered passivation structures above the electrodes. Tungsten (W) interconnects to the electrodes from below are also visible. 
    \textbf{b.} Zoomed SEM image of one of the electrodes in (a), showing the thickness of tapered oxide and nitride passivation layers above the midpoint of the electrode. 
    \textbf{c.} Zoomed SEM image of one electrode of a sensor with dimensions $l \times w - d = (10 \times 10 - 2)~\mu \mathrm{m}$, showing the thickness of tapered oxide and nitride passivation layers above the midpoint of the electrode.
    (a) and (b) show that the passivation thickness varies with electrode width: for narrow electrodes ($w$ = 0.75~$\mu$m), the oxide thickness is smaller, whereas for wider electrodes ($w$ = 10~$\mu$m), the oxide thickness increases.}
    \label{fig:SI_FIB-SEM}
    \end{figure}
     \addcontentsline{toc}{subsection}{Supplementary Fig.~\thefigure: Near-surface cross-section of the IC}

     \begin{figure}[htp!]
    \centering
    \includegraphics[width=0.8\textwidth]{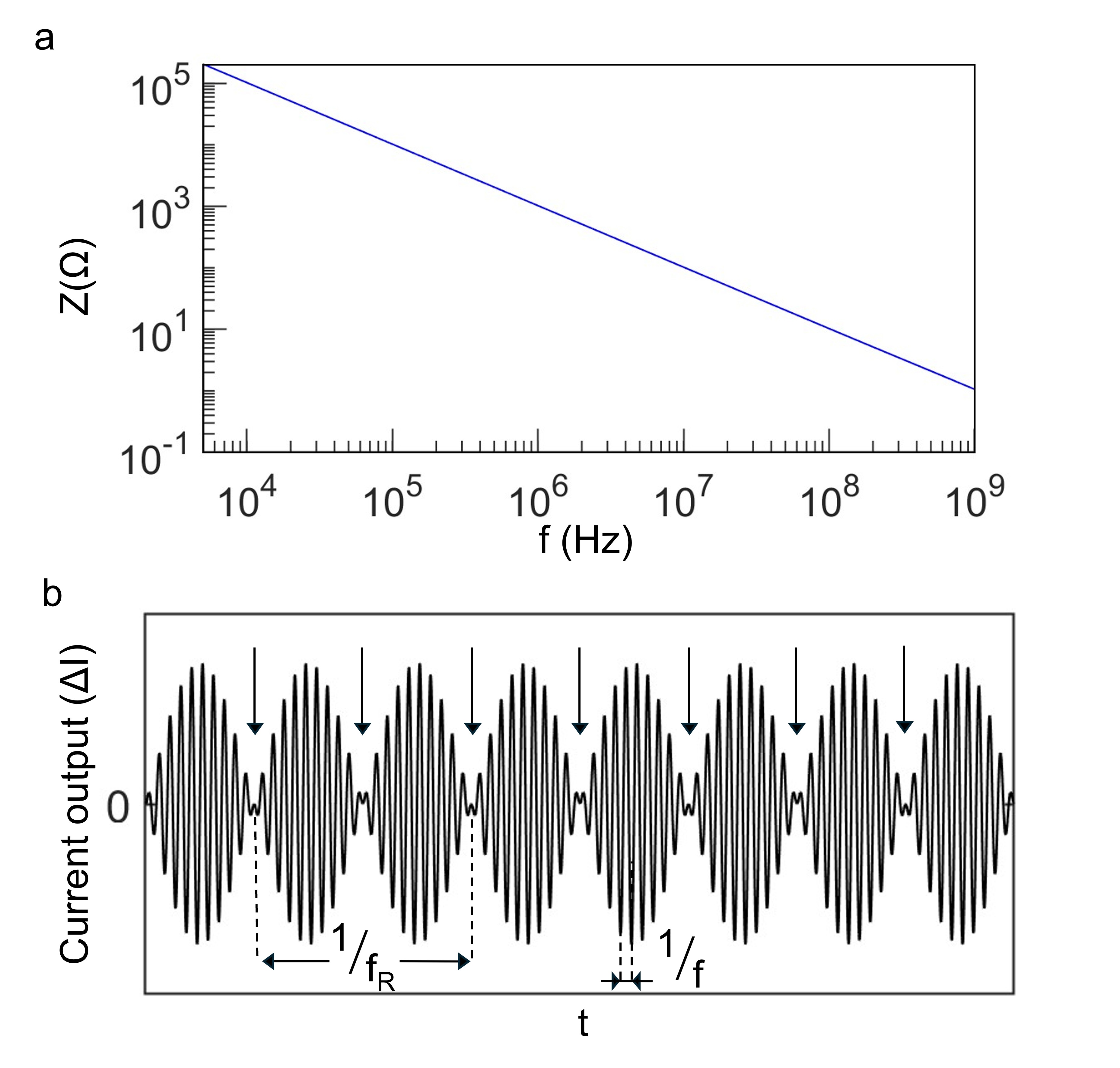}
    \caption{
    \textbf{Impedance spectrum and current output of the IC-based sensor.}
    \textbf{a.} Representative impedance spectrum calculated for one pair of electrodes in an IC-based sensor with dimensions $l \times w - d = (10 \times 0.75 - 4)\,\mu$m in solution. The passivation thicknesses shown in Supplementary Fig.~8b were used, producing this spectrum for solutions containing 0-0.13~M added NaCl in 10~mM phosphate buffer (see \textit{Supplementary Text} for further details). The impedance varies linearly on a log scale with input frequency, indicating dominance of the passivation capacitance across the frequency range $5 \times 10^{3}-10^{9}$~Hz, consistent with the equivalent circuit model in Fig.~3d.  
    \textbf{b.} Schematic of the sensor output current from the central electrode, showing the high-frequency electrode input component $f$, amplitude-modulated by the marker rotation frequency $f_\mathrm{R}$. A $180^\circ$ phase shift is introduced every half-cycle of rotation (i.e., at frequency $2f_\mathrm{R}$, marked with arrows), corresponding to the marker being on either side of the central electrode.
    }
    \label{fig:SI_current_IC}
    \end{figure}
    \addcontentsline{toc}{subsection}{Supplementary Fig.~\thefigure: Schematic representation of current output from the three-electrode IC-based sensor.}

     \begin{figure}[htp!]
    \centering
    \includegraphics[width=1\textwidth]{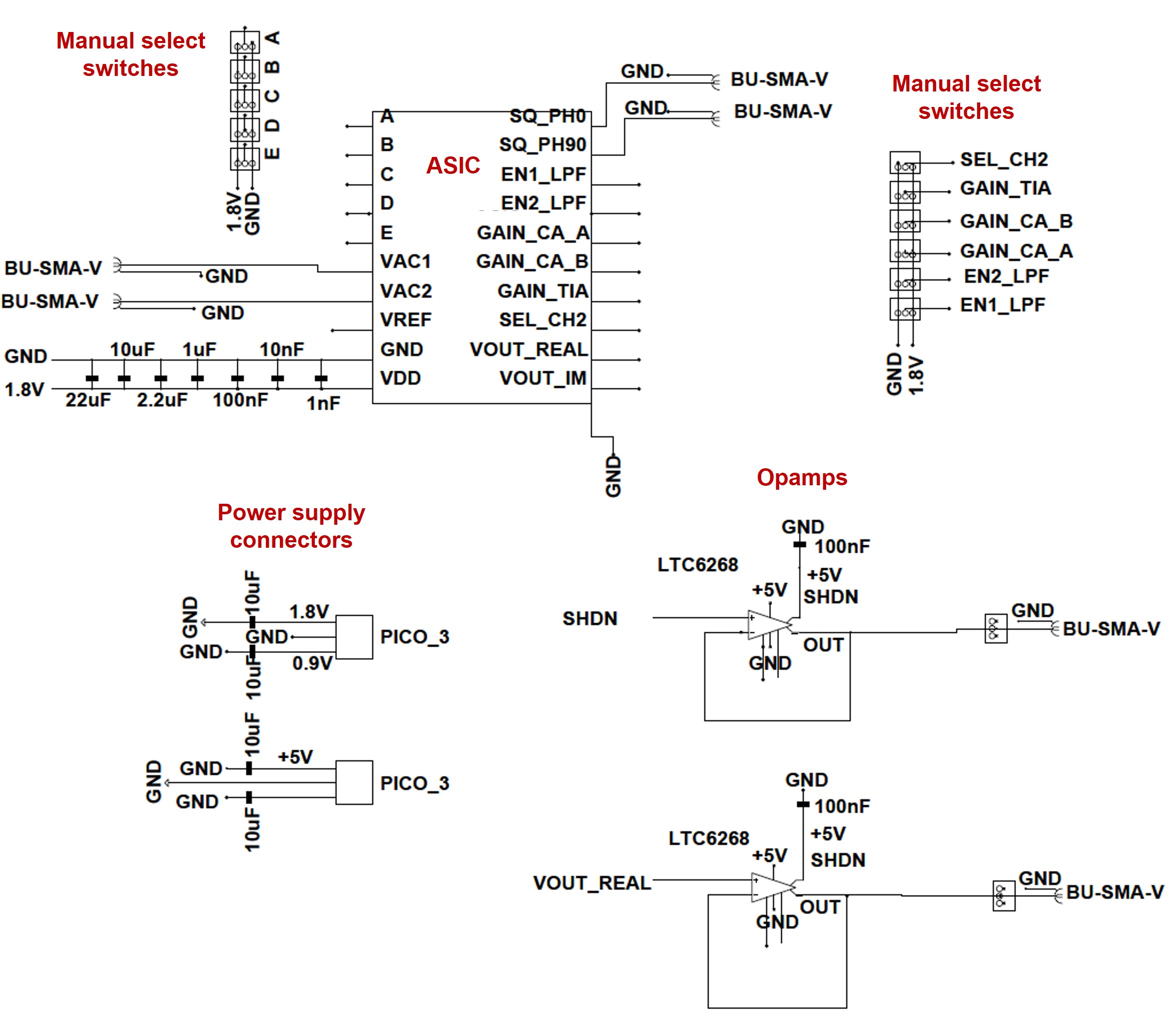} 
    \caption{ 
    \textbf{PCB schematic of the Main PCB for the IC.} 
    Schematic of the ASIC and its pin connections. The ASIC operates on supply voltage VDD = 1.8~V and reference voltage VREF = 0.9~V. The digitally programmable electrode-select pins (A-E) are connected to manual switches that select between VDD and GND. The electrode sensor inputs (VAC1 and VAC2) are routed to SMA connectors to receive signals from a signal generator. Decoupling capacitors of multiple values are placed between VDD and GND to filter unwanted frequency components. Demodulator inputs (SQ\_PH0 and SQ\_PH90) are connected to SMA connectors driven by signal generators. Pins for digitally programming the low-pass filter frequency (EN1\_LPF and EN2\_LPF), current amplifier gains (GAIN\_CA\_A and GAIN\_CA\_B), transimpedance amplifier gain(GAIN\_TIA), and electrode sensor channel (SEL\_CH2) are connected to manual select switches. 
    The ASIC outputs VOUT\_REAL and VOUT\_IM connect to the operational amplifiers (op-amps) shown below.
    The op-amps (LTC6268) are connected as unity-gain buffers before routing the signals to SMA connectors, and operate from +5~V and GND supplies. The power supplies (1.8~V, +5~V and GND) and VREF (0.9~V) connect to the PCB via a compact connector (PICO\_3).
    }
    \label{fig:SI_PCB_IC_Main}
    \end{figure}
    \addcontentsline{toc}{subsection}{Supplementary Fig.~\thefigure: PCB schematic of the Main PCB for the IC}

    \begin{figure}[htp!]
    \centering
    \includegraphics[width=0.95\textwidth]{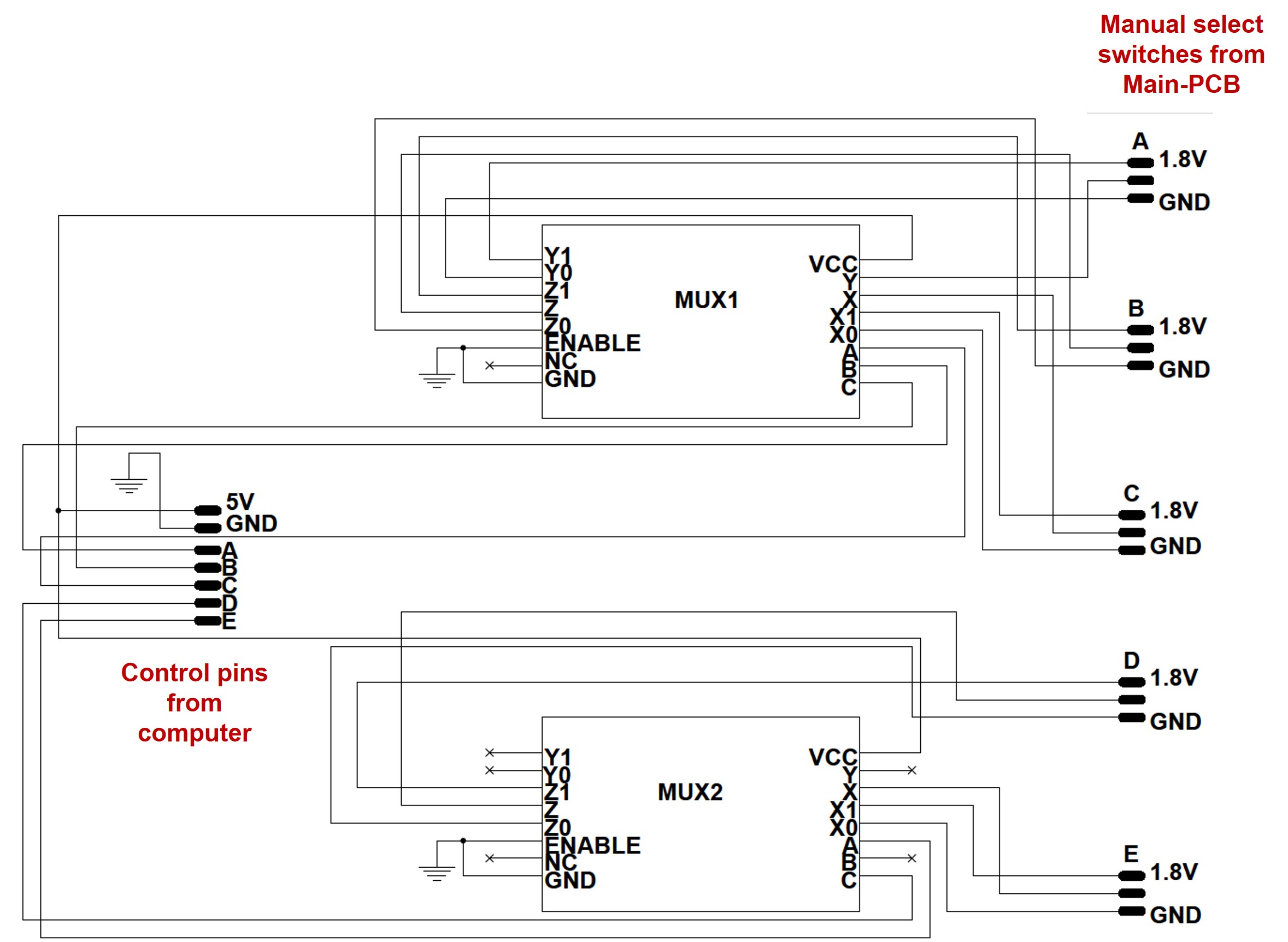} 
    \caption{ 
    \textbf{PCB schematic of the computerised electrode selection addendum board.} 
    The schematic shows the board connecting to the five manual electrode-select switches (A-E, shown on the right) on the Main PCB given in Supplementary Fig.~\ref{fig:SI_PCB_IC_Main}). It contains two multiplexer ICs (MUX1 and MUX2) connected to these switch pins. Each MUX IC contains three MUXs (X,Y and Z), which route the middle pin of each MUX (V) to either of the two inputs voltages (V0 or V1) corresponding to 1.8~V and GND. The selection is controlled by the corresponding seven digital control inputs (A–E, shown on the left) provided by the computer through a DAQ. The computer supplies 5~V and GND to power the MUX ICs, and digital logic signals (1 = 5~V, 0 = GND) to select between the two voltages.
    }
    \label{fig:SI_PCB_IC_addendum}
    \end{figure}
    \addcontentsline{toc}{subsection}{Supplementary Fig.~\thefigure: PCB schematic of the computerised electrode selection board}

    \begin{figure}[htp!]
    \centering
    \includegraphics[width=0.95\textwidth]{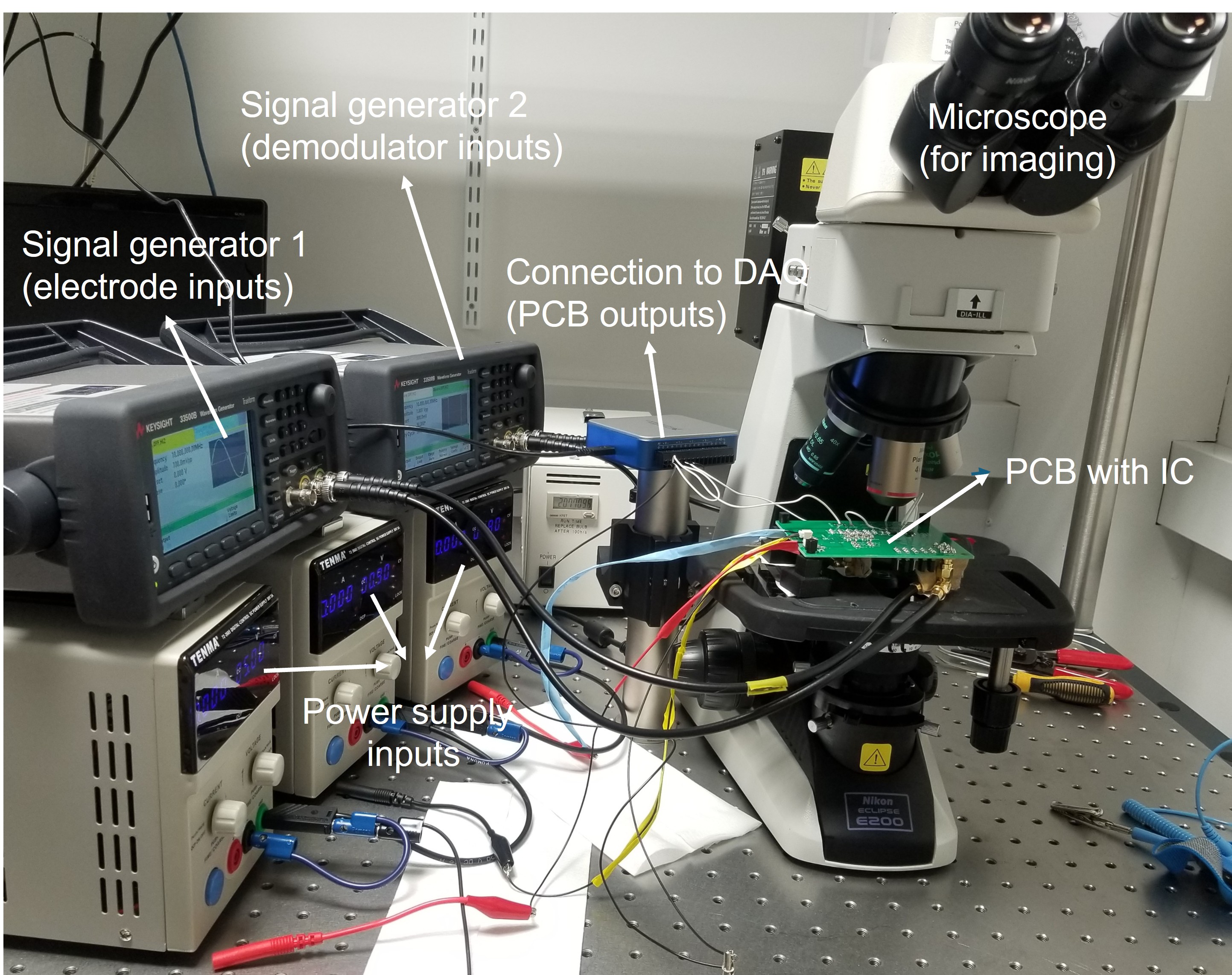} 
    \caption{
    \textbf{Setup for optical and electrical recordings with the IC-based sensor.}
     Photograph of the instrument setup. An upright microscope on an optical table is used to perform the experiments with the IC-based sensor described in the main text. The PCB containing the IC is placed on the custom-designed microscope stage to image and record the rotation markers and electrodes. The signal generators supplying the electrode excitation and demodulation inputs are shown. Three power supplies providing the PCB supply voltages (1.8, 0.9, and 5~V; see also Supplementary Fig.~\ref{fig:SI_PCB_IC_Main}), and the terminal connecting the PCB outputs to the DAQ are visible.
    }
    \label{fig:SI_Setup_IC}
    \end{figure}
    \addcontentsline{toc}{subsection}{Supplementary Fig.~\thefigure: Instrumentation setup for the IC-based sensor.}

\begin{figure}[htp!]
    \centering
    \includegraphics[width=0.9\textwidth]{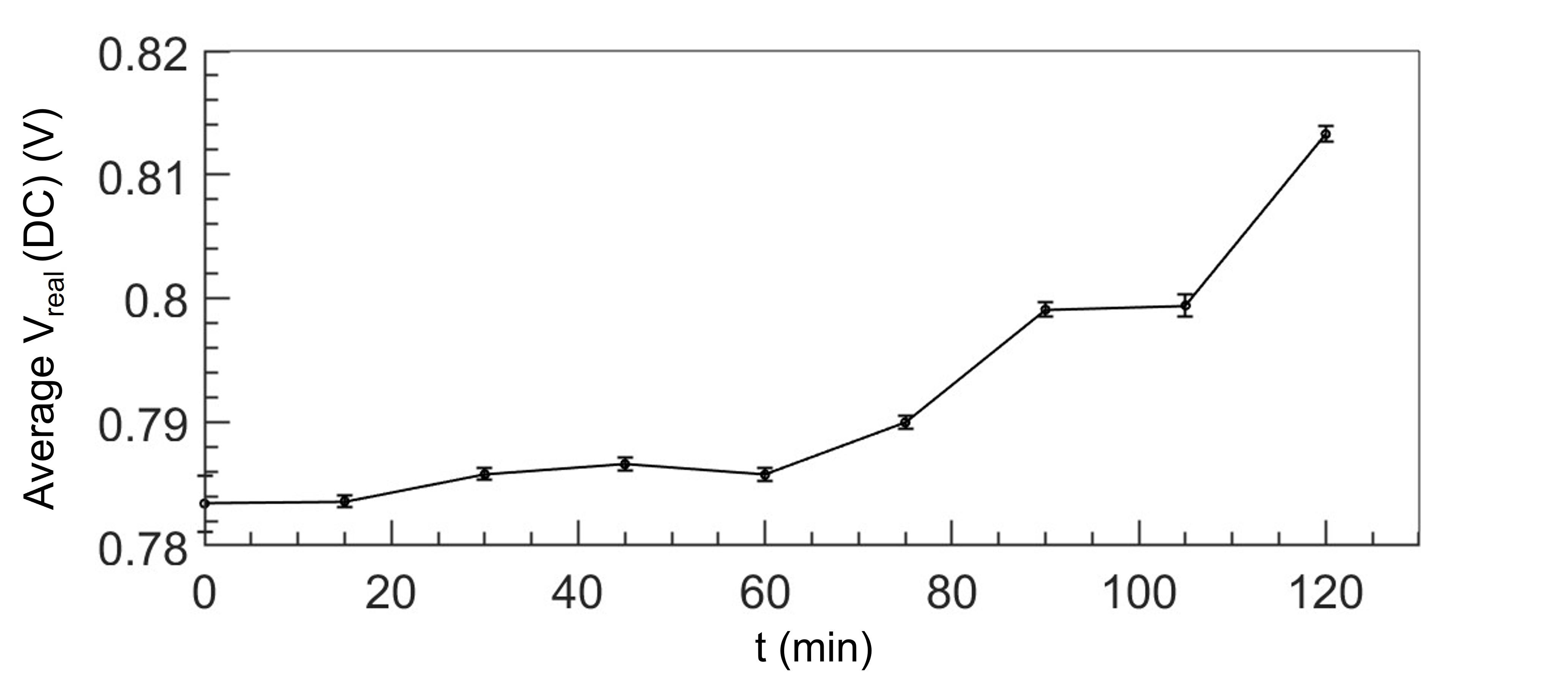} 
    \caption{
    \textbf{Evaporation driven change in the DC voltage output of the IC-based sensor}
    Temporal evolution of the DC sensor voltage V$_\mathrm{real}$ measured every 15~min, with each data point being the mean of a measurement of length $> 60$~s. The DC value remains approximately constant for $\sim$60~min, after which a steep increase is observed. 
    The increase in voltage reflects the effect of surface evaporation, where the liquid loss in the chamber was visible by eye after $\sim$90~min. 
    The error bars correspond to the standard deviation of V$_\mathrm{real}$ during each measurement.
     }
    \label{fig:SI_OutputV_evaporation}
    \end{figure}
    \addcontentsline{toc}{subsection}{Supplementary Fig.~\thefigure: Change in IC-based sensor's DC voltage output with evaporation.}

    \begin{figure}[htp!]
    \centering
    \includegraphics[width=0.95\textwidth]{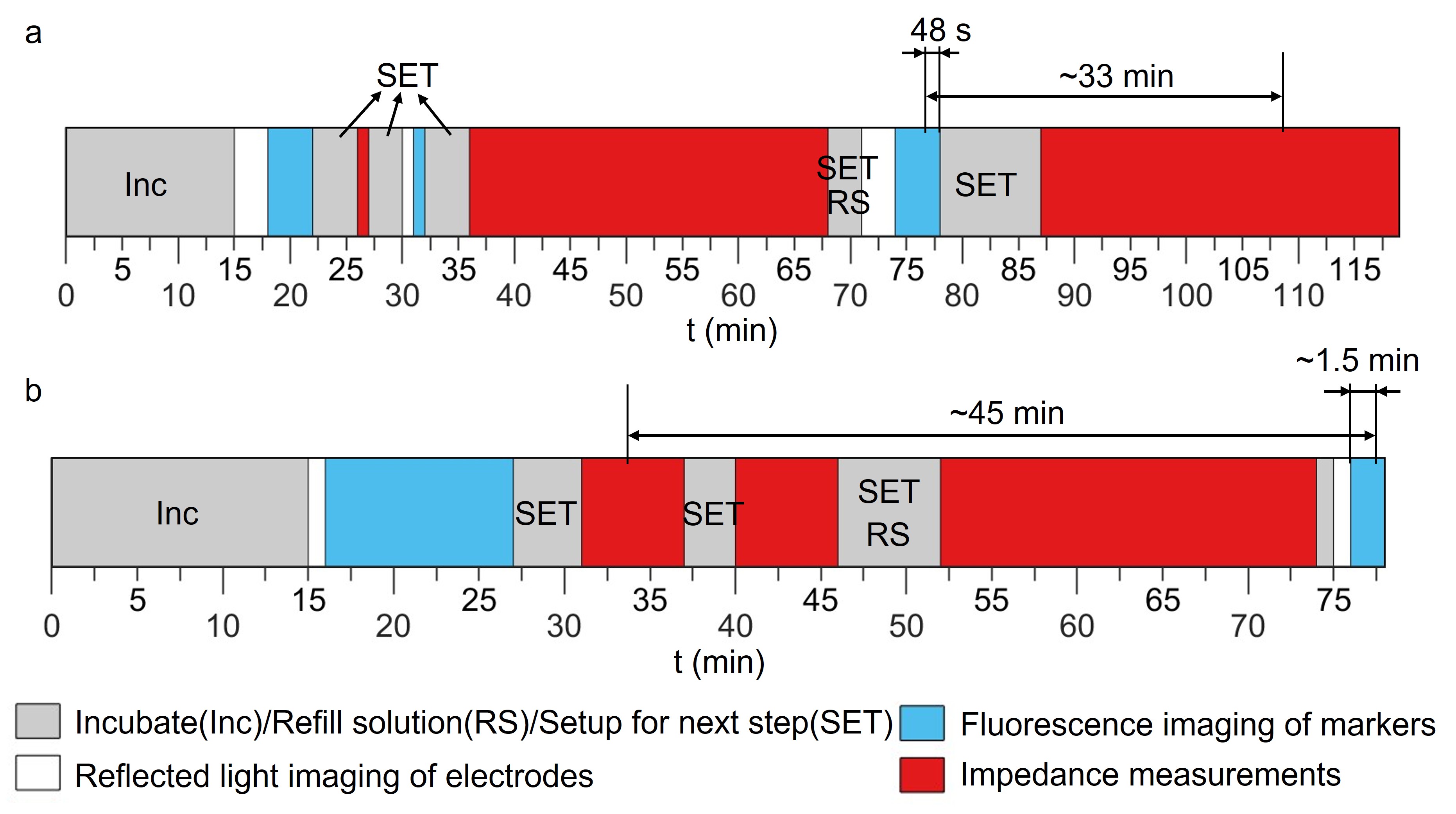} 
    \caption{
    \textbf{Acquisition sequence of experiments in Fig.~5.}
    \textbf{a} and \textbf{b} correspond to experiments in Fig.~5a--c and Fig.~5d--f, respectively. The experimental steps depicted are colour coded as indicated in the figure itself. The longer double-headed arrow indicates the time interval between the impedance and optical measurements reported in Fig.~5, while the shorter one indicates the duration of blue-light exposure between them. 
    }
    \label{fig:SI_Timelineexps_IC}
    \end{figure}
     \addcontentsline{toc}{subsection}{Supplementary Fig.~\thefigure: Timeline of experiments in Fig.~5}

     \begin{figure}[htp!]
    \centering   \includegraphics[width=0.75\textwidth]{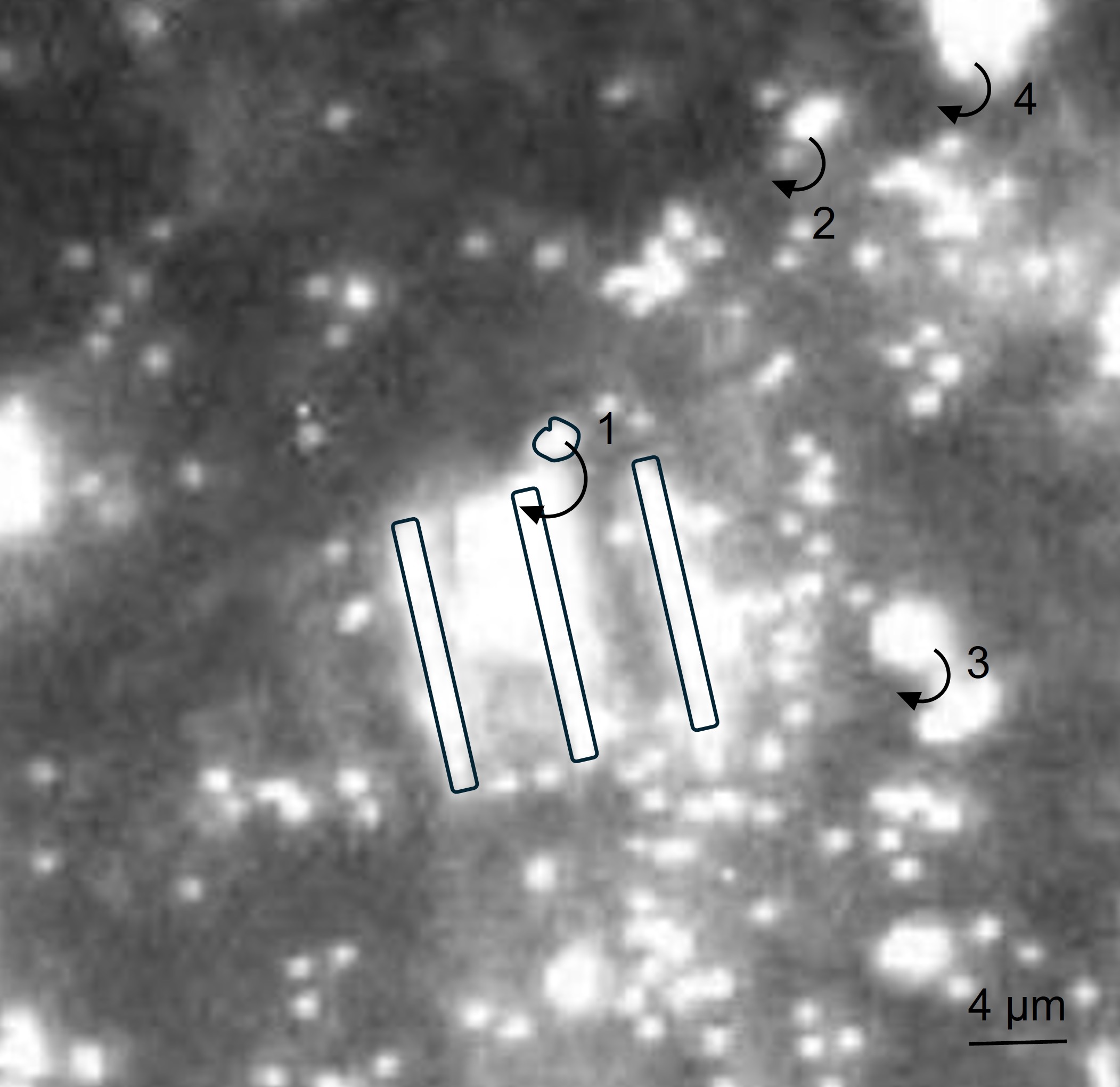} 
    \caption{
    \textbf{Microscopy image of the IC-based sensor used in Fig.~5d–f and its surrounding area.}
    The image reproduces the microscope view shown in Fig.~5e but with a wider field of view, revealing several single beads (bright round spots), bead clusters (larger bright clumps) and bacteria (diffuse bright regions) around the sensor. Four rotating markers that could be clearly identified in the focal plane (numbered rotation symbols) give an indication of the density of rotating markers in this experiment.  
    }
     \label{fig:SI_microscopeview_Fig5dtof}
    \end{figure}
     \addcontentsline{toc}{subsection}{Supplementary Fig.~\thefigure: Microscopy image of the IC-based sensor used in Fig.~5d–f and its surrounding area.}

    \begin{figure}[htp!]
    \centering
    \includegraphics[width=1\textwidth]{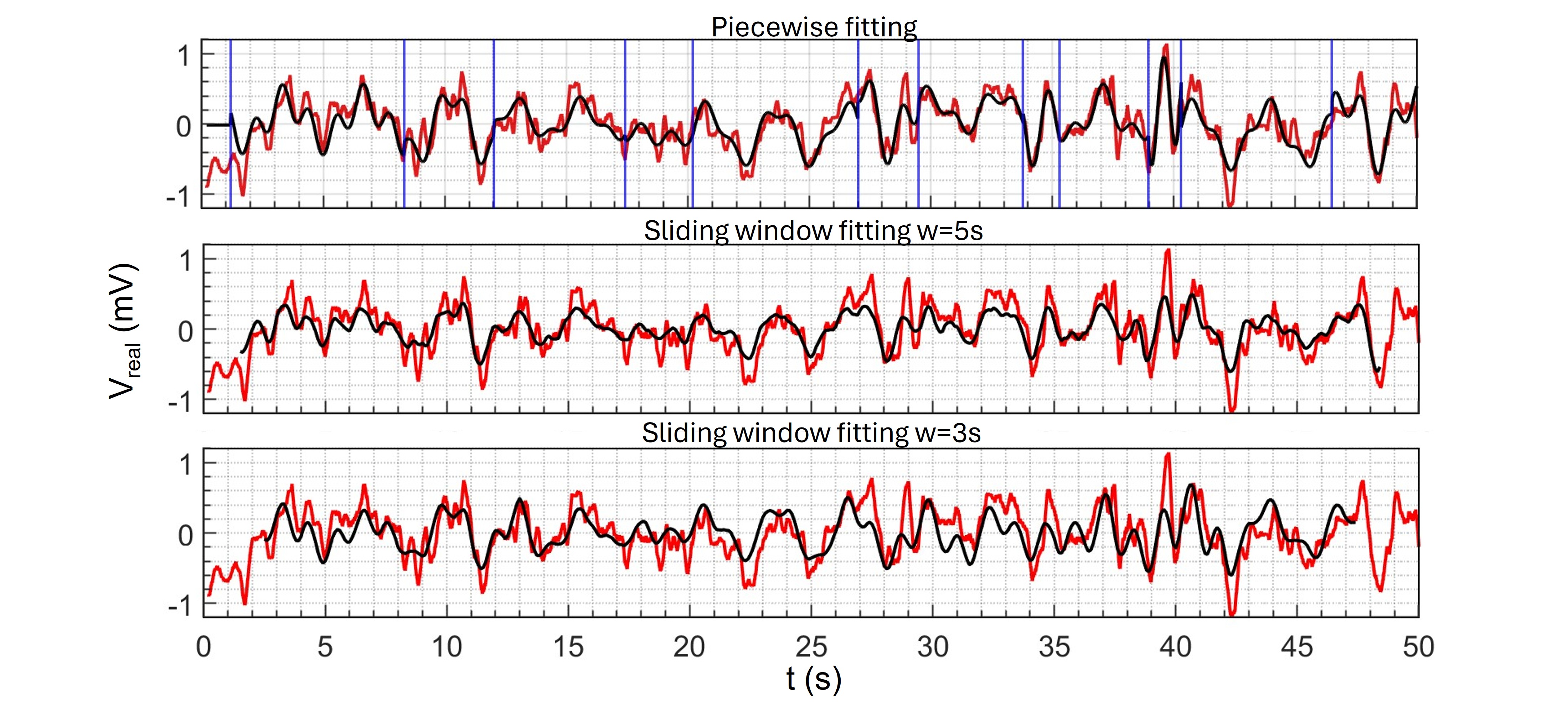} 
    \caption{
    \textbf{Least-squares fits of data in Fig.~5f to a sum of two sinusoids.}  
    Top: Piecewise fits (black) to a sum of two sinusoids performed within manually selected time windows on the mean-filtered sensor output from Fig.~5f (red). The frequency of each sinusoid in the sum was varied in the 0.29--0.37~Hz and 0.82--0.90~Hz intervals, respectively, to obtain the best fit.
    Middle and bottom: comparison of the same fits but now obtained using constant sliding window(s) of 5~s and 3~s, respectively, with a time step of 0.1~s. These fits use a sum of two sinusoids at the fixed frequencies 0.33~Hz and 0.86~Hz. 
    All three methods provide a good fit of frequency. While the piecewise/manual segmentation provides the best agreement in amplitudes, the sliding-window fits are comparable, with the 3~s window performing better (see Methods for details of procedure).}    \label{fig:SI_alternativewindows_LSF_timesignal}
    \end{figure}
     \addcontentsline{toc}{subsection}{Supplementary Fig.~\thefigure: Comparison of manual (piecewise) and fixed-window least-squares fits of Fig.~5f sensor output.}
     
    \begin{figure}[htp!]
    \centering   \includegraphics[width=0.75\textwidth]{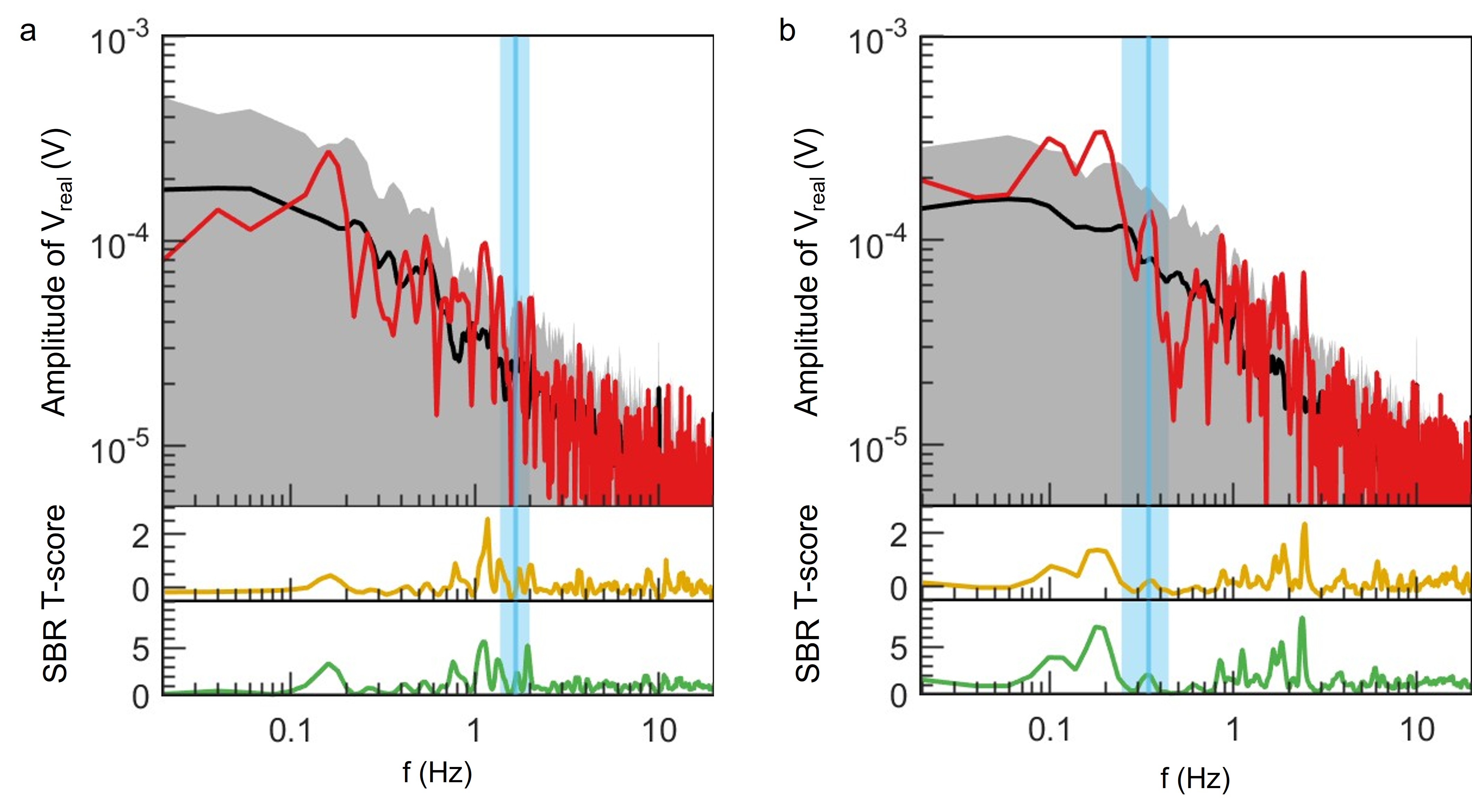} 
    \caption{
    \textbf{More examples of rotation detection by IC-based sensors.} 
    Amplitude, \textit{t}-score, and SBR spectra of two additional sensor outputs, obtained and plotted as in Fig.~5a and d. The detected peak frequencies correspond to the optically observed rotation speeds (blue), with differences between optical and impedance peaks attributed to whether or not the exposure to blue light occurred before or after the impedance measurement (see \textit{Methods}). 
    \textbf{a.} Output from an electrode sensor of $l \times w - d = (10 \times 10 - 2)\,\mu$m, measured 6~min after a rotating bead cluster of size $\sim$2~$\mu$m was observed over the central electrode at 1.68$\pm$0.31~Hz. The spectra show a sensor peak at 1.1~Hz.  
    \textbf{b.} Output from an electrode sensor of $l \times w - d = (10 \times 0.75 - 4)\,\mu$m, measured 8~min after a rotating bead cluster of size $\sim$5~$\mu$m was observed rotating at 0.33~Hz. The spectra show sensor peaks at 0.2~Hz and $\sim$2~Hz. An additional out-of-focus rotating marker a few $\mu$m away from the sensor likely accounts for the peak at $\sim$2~Hz.
    }
    \label{fig:SI_othersignals_IC}
    \end{figure}
     \addcontentsline{toc}{subsection}{Supplementary Fig.~\thefigure: More examples of rotation detection by IC-based sensors.}

    \begin{figure}[htp!]
    \centering
    \includegraphics[width=0.98\textwidth]{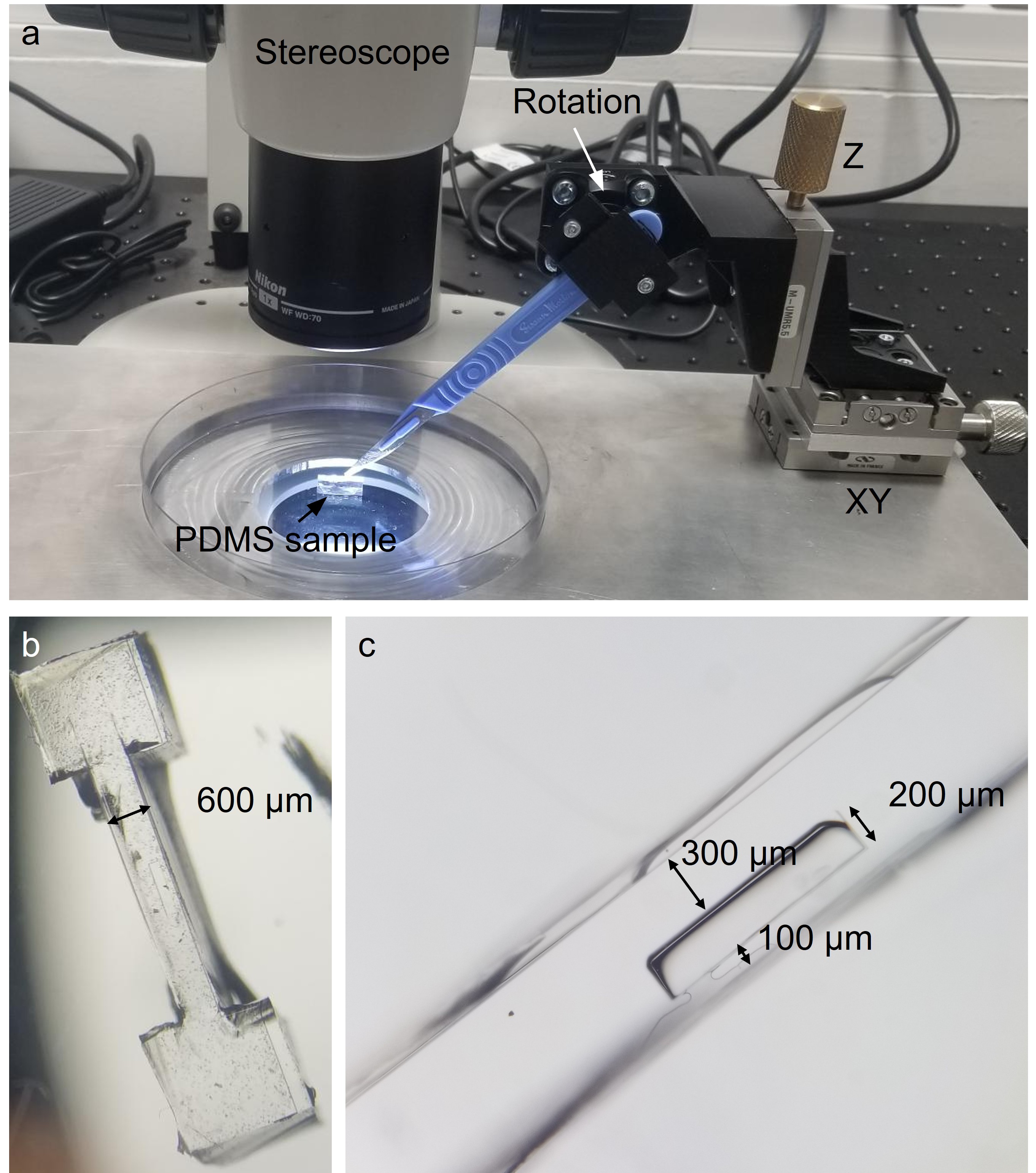} 
    \caption{
    \textbf{Micrometre-precision PDMS cutting.}
    \textbf{a.} The PDMS cutting tool was constructed around a stereoscope. The cutting arm is mounted on a custom designed, mechanically controlled micrometre-precision XYZ translation stage. A rotation stage was attached to the arm to orient the cutting blade (in blue) at the desired angle. Cutting of the channel is performed while it is observed through the stereoscope. 
    \textbf{b.} Image of the bottom layer of the PDMS device described in Fig.~8, showing its 600~$\mu$m width.
    \textbf{c.} Magnified view of the bottom layer in (b), showing the 200~$\mu$m-wide channel and the 100~$\mu$m clearance from the edge.
    }
    \label{fig:SI_PDMScutting}
    \end{figure}
     \addcontentsline{toc}{subsection}{Supplementary Fig.~\thefigure: Micrometre-precision PDMS cutting}
    
    \begin{figure}[htp!]
    \centering
    \includegraphics[width=0.8\textwidth]{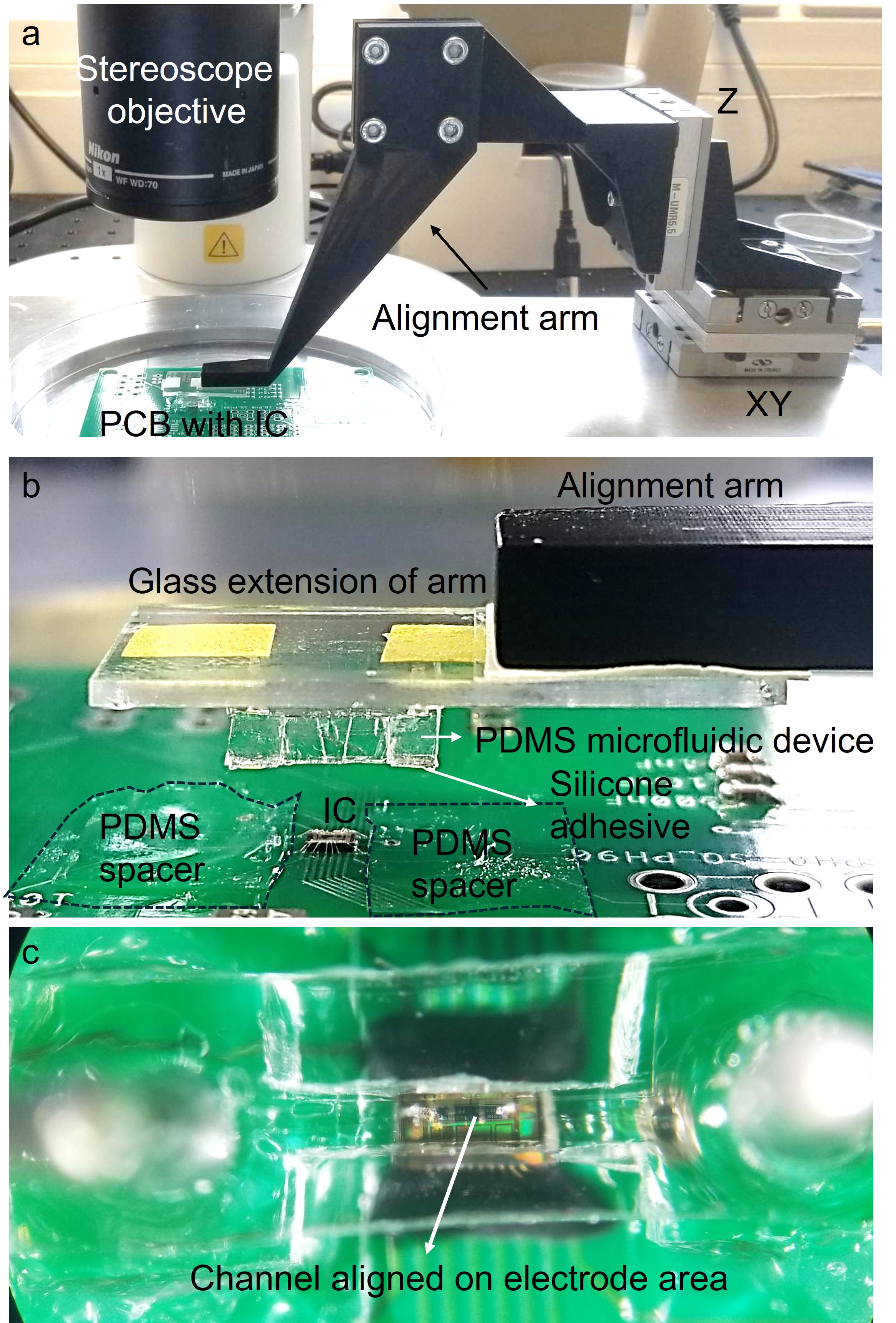} 
    \caption{
    \textbf{Micrometre-precision alignment and bonding of PDMS device on the IC.}
    \textbf{a.} An alignment arm replaced the cutting arm from Fig.~\ref{fig:SI_PDMScutting}a to perform the alignment and bonding of the PDMS microfluidic device onto the IC surface. The XYZ translational stage was again used, and the process was performed under the stereoscope.
    \textbf{b.} The alignment arm was fitted with a transparent glass extension that held the PDMS device, with silicone adhesive applied to its underside. The transparent extension enabled simultaneous visualization of the microfluidic channel and the IC electrode area during alignment onto the IC surface. PDMS spacers were placed on either side of the IC to support the device.    
    \textbf{c.} Top view of the microfluidic device mounted on the IC surface, showing the channel region aligned with micrometre precision to the electrode area on the IC. The whole device fitted well into the space between the wire bonds.
    }
    \label{fig:SI_PDMS_alignandstick}
    \end{figure}
    \addcontentsline{toc}{subsection}{Supplementary Fig.~\thefigure: Micrometre-precision alignment and bonding of PDMS microfluidic device on IC}
     
    \begin{figure}[htp!]
    \centering
    \includegraphics[width=0.95\textwidth]{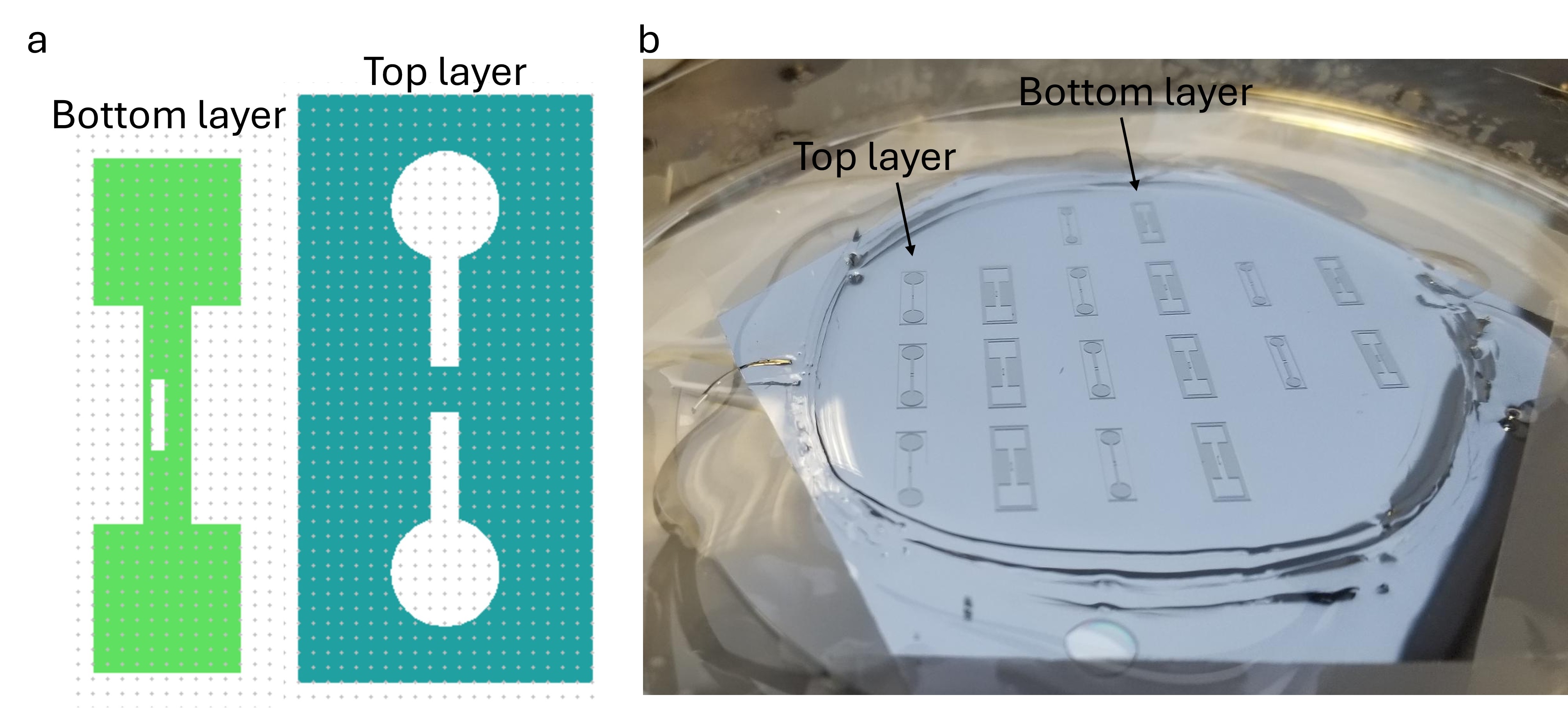} 
    \caption{
    \textbf{SU-8 master for PDMS microfluidic device moulding.}
    \textbf{a} Digital mask designs for the bottom and top layers used to pattern the SU-8 master using a maskless optical lithography system (see \textit{Methods}).
    \textbf{b} Photograph of the SU-8 master on a silicon wafer placed in a Petri dish, with PDMS surrounding the area where devices are cut out. The master contains six variations of top or bottom layer designs (each in a column as pointed above) with different dimensions. The range enabled identification of channel and device dimensions best suited to the IC.
    }
    \label{fig:SI_SU8master_PDMSmould}
    \end{figure}
    \addcontentsline{toc}{subsection}{Supplementary Fig.~\thefigure: SU-8 master for PDMS microfluidic device moulding.}

    \begin{figure}[htp!]
    \centering
    \includegraphics[width=0.95\textwidth]{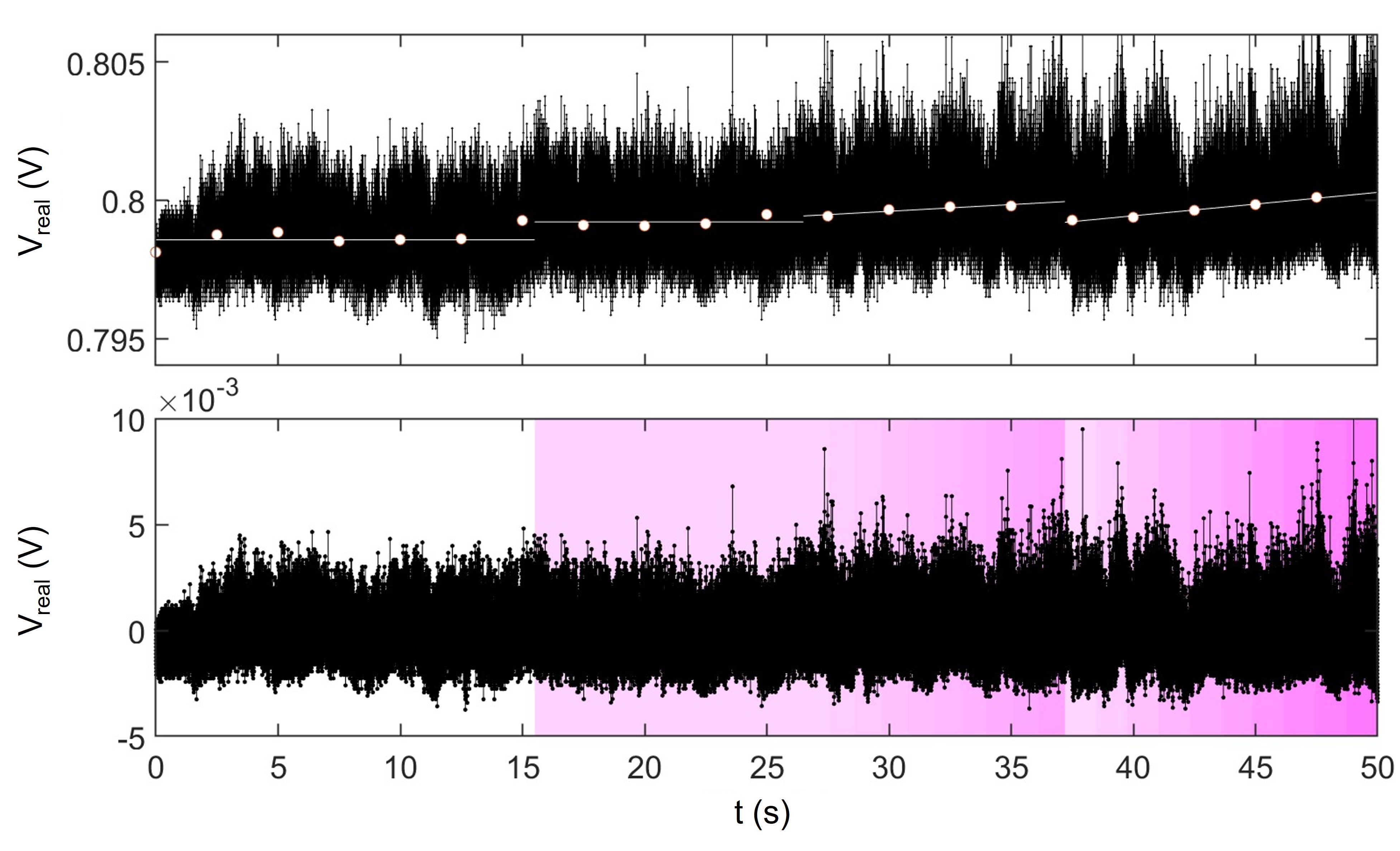} 
    \caption{\textbf{Identification and correction of DC offsets in time-domain data.}
    Top: raw time-domain IC-sensor output (V$_{\mathrm{real}}$) corresponding to the experiment in Fig.~5d-f, with the DC levels calculated in non-overlapping 2.5~s windows (white points). DC level trends longer than twice the time period of the dominant frequency $2/f_{\mathrm{dom}} = 2/0.33$~Hz are identified that exhibit either a constant offset (15–37~s) or linear trends (26–50~s). These correspond to transient passage of bacteria/markers over the sensor and evaporation-induced drift (refer Supplementary Fig.~13), respectively. Linear fits to these regions are used to estimate DC offsets and drift slopes (white lines). 
    Bottom: DC-corrected signal after subtraction of the estimated offsets and trends. The intensity of the pink background indicates the magnitude of the applied DC correction over time.
    }
    \label{fig:SI_DCoffset_correction}
    \end{figure}
    \addcontentsline{toc}{subsection}{Supplementary Fig.~\thefigure: Identification and correction of DC offsets in time-domain data.} 

    \clearpage
    \section*{Supplementary Videos}

    \setcounter{figure}{0}
    \renewcommand{\thefigure}{\arabic{figure}}
    \captionsetup[figure]{labelfont=bf,labelsep=period,name={Supplementary Video}}

    \begin{figure}[htp!]
    \centering
    \includegraphics[width=0.55\textwidth]{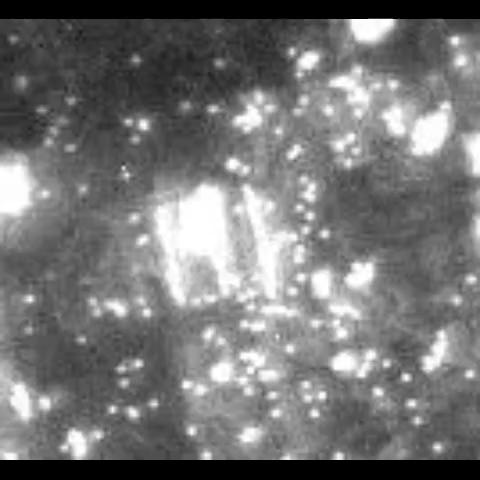} 
    \caption{
    Video showing the expanded field of view of the sensor imaged for the experiments of Fig.~5d--f, also shown in Supplementary Fig.~15. The video shows the rotating markers indicated in Supplementary Fig.~15 and all movements within the field of view. A gentle flow of the solution is visible during imaging due to an increased rate of evaporation, which causes some of the rotating markers to drift. The marker located on the sensor remains well anchored. Some out-of-focus motion is also visible above the sensor.
    }
    \label{fig:SI_Video1_image1}
    \end{figure}
    \addcontentsline{toc}{subsection}{Supplementary Fig.~\thefigure: First frame of Supplementary Video 1.}

    \begin{figure}[htp!]
    \centering
    \includegraphics[width=0.55\textwidth]{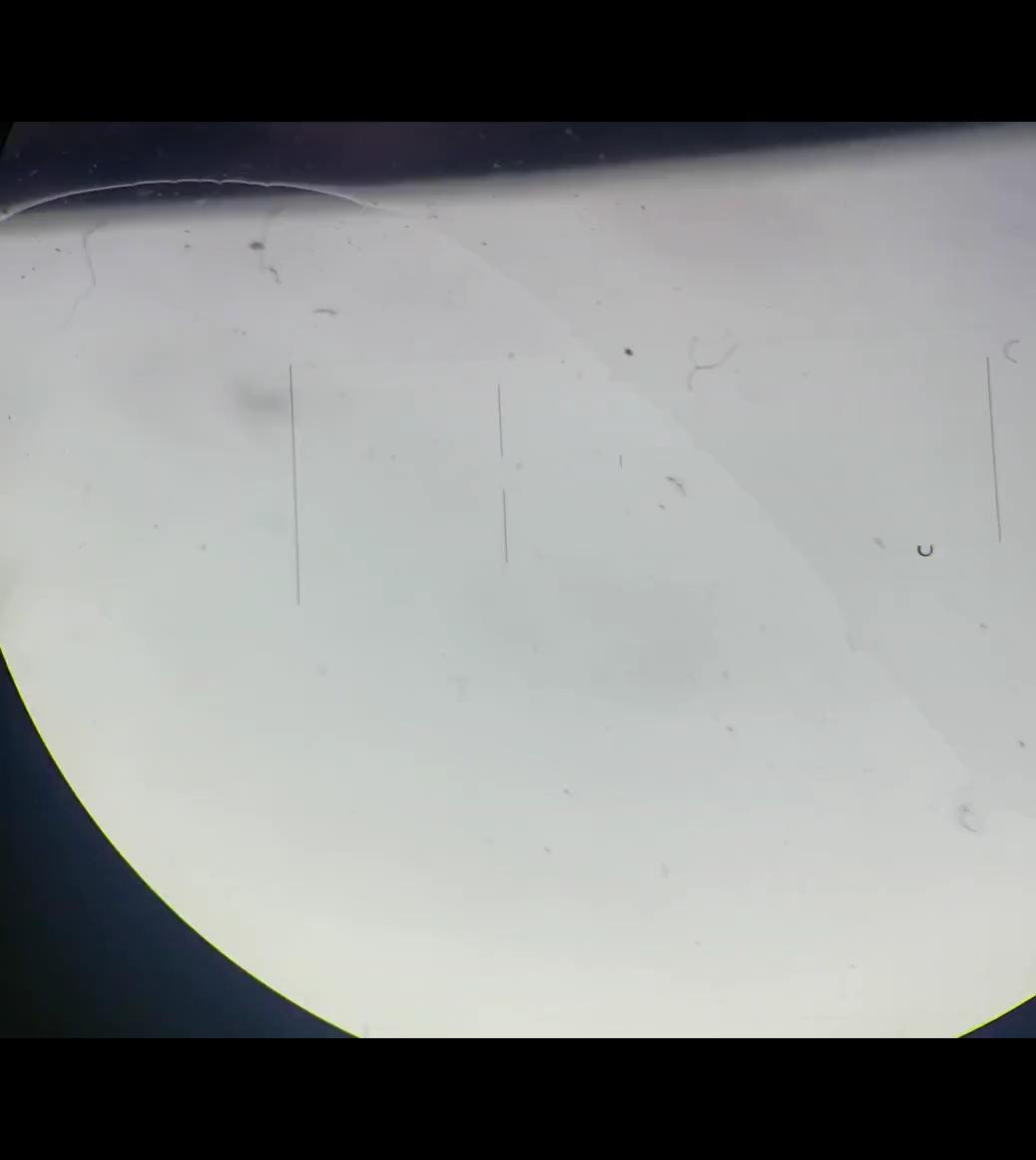} 
    \caption{
    Video showing the procedure used to fabricate the top and bottom PDMS layers of the microfluidic device in Fig.~8. The video first demonstrates the formation of vertical fluidic interconnects in the bottom layer, followed by cutting out the bottom layer using the blade on the micrometer-precision cutting tool, Supplementary Fig.~18, and removal of the punched parts. The top layer is fabricated using a similar procedure, with less stringent requirements on punching and cutting precision. The video also shows the alignment of the two layers prior to bonding.
    }
    \label{fig:SI_Video2_image}
    \end{figure}
    \addcontentsline{toc}{subsection}{Supplementary Fig.~\thefigure: First frame of Supplementary Video 2.}
   
    \addcontentsline{toc}{section}{Supplementary Videos}

    \clearpage
    \addcontentsline{toc}{section}{Supplementary References}

\end{document}